\documentclass[longauth]{aa}
\usepackage[utf8]{inputenc}
\usepackage{graphicx}
\usepackage{txfonts}
\usepackage{subfigure}
\usepackage{natbib}
\usepackage{appendix}
\usepackage{euclid}

\newcommand{\sersics}{S\'ersic }
\newcommand{\bts}{${\rm b}/{\rm t}$ }
\newcommand{\bt}{${\rm b}/{\rm t}$}
\newcommand{\btm}{{\rm b}/{\rm t}}
\newcommand{\sersic}{S\'ersic}

\begin{document}


\title{\Euclid preparation. XXV. The \Euclid Morphology Challenge -- Towards model-fitting photometry for billions of galaxies}



\author{\normalsize Euclid Collaboration: E.~Merlin\orcid{0000-0001-6870-8900}$^{1}$\thanks{\email{emiliano.merlin@inaf.it}}, M.~Castellano\orcid{0000-0001-9875-8263}$^{1}$, H.~Bretonni\`ere\orcid{0000-0001-9935-9109}$^{2,3}$, M.~Huertas-Company\orcid{0000-0002-1416-8483}$^{4,5,6,7}$, U.~Kuchner\orcid{0000-0002-0035-5202}$^{8}$, D.~Tuccillo$^{9}$, F.~Buitrago\orcid{0000-0002-2861-9812}$^{10,11}$, J.~R.~Peterson\orcid{0000-0001-5471-9609}$^{12}$, C.J.~Conselice$^{13}$, F.~Caro$^{1}$
% ...
}

\institute{$^{1}$ INAF-Osservatorio Astronomico di Roma, Via Frascati 33, I-00078 Monteporzio Catone, Italy\\
$^{2}$ Universit\'e Paris-Saclay, CNRS, Institut d'astrophysique spatiale, 91405, Orsay, France\\
$^{3}$  Universit\'e Paris Cit\'e, CNRS, Astroparticule et Cosmologie, F-75013 Paris, France\\
$^{4}$ Instituto de Astrof\'isica de Canarias (IAC); Departamento de Astrof\'isica, Universidad de La Laguna (ULL), E-38200, La Laguna, Tenerife, Spain\\
$^{5}$ Instituto de Astrof\'isica de Canarias, Calle V\'ia L\'actea s/n, E-38204, San Crist\'obal de La Laguna, Tenerife, Spain\\
$^{6}$ Universit\'e Paris-Cit\'e, 5 Rue Thomas Mann, 75013, Paris, France\\
$^{7}$ Universit\'e PSL, Observatoire de Paris, Sorbonne Universit\'e, CNRS, LERMA, F-75014, Paris, France\\
$^{8}$ School of Physics and Astronomy, University of Nottingham, University Park, Nottingham NG7 2RD, UK\\
$^{9}$ Instituto de F\'isica de Cantabria, Edificio Juan Jord\'a, Avenida de los Castros, E-39005 Santander, Spain\\
$^{10}$ Departamento de F\'{i}sica Te\'{o}rica, At\'{o}mica y \'{O}ptica, Universidad de Valladolid, 47011 Valladolid, Spain\\
$^{11}$ Instituto de Astrof\'isica e Ci\^encias do Espa\c{c}o, Faculdade de Ci\^encias, Universidade de Lisboa, Tapada da Ajuda, PT-1349-018 Lisboa, Portugal\\
$^{12}$ Department of Physics and Astronomy, Purdue University, 525 Northwestern Ave., West Lafayette, IN 47907, USA\\
$^{13}$ Jodrell Bank Centre for Astrophysics, Department of Physics and Astronomy, University of Manchester, Oxford Road, Manchester M13 9PL, UK\\
$^{14}$ Universit\"ats-Sternwarte M\"unchen, Fakult\"at f\"ur Physik, Ludwig-Maximilians-Universit\"at M\"unchen, Scheinerstrasse 1, 81679 M\"unchen, Germany\\
$^{15}$ European Southern Observatory, Alonso de Cordova 3107, Casilla 19001, Santiago, Chile\\
$^{16}$ Department of Astronomy, University of Geneva, ch. d\'Ecogia 16, CH-1290 Versoix, Switzerland\\
$^{17}$ Institut d'Astrophysique de Paris, UMR 7095, CNRS, and Sorbonne Universit\'e, 98 bis boulevard Arago, 75014 Paris, France\\
$^{18}$ Canada-France-Hawaii Telescope, 65-1238 Mamalahoa Hwy, Kamuela, HI 96743, USA\\
$^{19}$ Instituto de Matem\'{a}tica Estat\'{i}stica e F\'{i}sica, Universidade Federal do Rio Grande, 96203-900, Rio Grande, RS, Brazil\\
$^{20}$ University of Nottingham, University Park, Nottingham NG7 2RD, UK\\
$^{21}$ Aix-Marseille Univ, CNRS, CNES, LAM, Marseille, France\\
$^{22}$ ICRAR, M468, University of Western Australia, Crawley, WA 6009, Australia\\
$^{23}$ INAF-Osservatorio Astronomico di Capodimonte, Via Moiariello 16, I-80131 Napoli, Italy\\
$^{24}$ Institute of Cosmology and Gravitation, University of Portsmouth, Portsmouth PO1 3FX, UK\\
$^{25}$ Institut f\"ur Theoretische Physik, University of Heidelberg, Philosophenweg 16, 69120 Heidelberg, Germany\\
$^{26}$ INAF-Osservatorio di Astrofisica e Scienza dello Spazio di Bologna, Via Piero Gobetti 93/3, I-40129 Bologna, Italy\\
$^{27}$ Dipartimento di Fisica e Astronomia "Augusto Righi" - Alma Mater Studiorum Universit\`{a} di Bologna, via Piero Gobetti 93/2, I-40129 Bologna, Italy\\
$^{28}$ INFN-Sezione di Bologna, Viale Berti Pichat 6/2, I-40127 Bologna, Italy\\
$^{29}$ Max Planck Institute for Extraterrestrial Physics, Giessenbachstr. 1, D-85748 Garching, Germany\\
$^{30}$ Dipartimento di Fisica, Universit\`{a} di Genova, Via Dodecaneso 33, I-16146, Genova, Italy\\
$^{31}$ INFN-Sezione di Roma Tre, Via della Vasca Navale 84, I-00146, Roma, Italy\\
$^{32}$ Department of Physics "E. Pancini", University Federico II, Via Cinthia 6, I-80126, Napoli, Italy\\
$^{33}$ Dipartimento di Fisica, Universit\'a degli Studi di Torino, Via P. Giuria 1, I-10125 Torino, Italy\\
$^{34}$ INFN-Sezione di Torino, Via P. Giuria 1, I-10125 Torino, Italy\\
$^{35}$ INAF-Osservatorio Astrofisico di Torino, Via Osservatorio 20, I-10025 Pino Torinese (TO), Italy\\
$^{36}$ INAF-IASF Milano, Via Alfonso Corti 12, I-20133 Milano, Italy\\
$^{37}$ Institut de F\'{i}sica d'Altes Energies (IFAE), The Barcelona Institute of Science and Technology, Campus UAB, 08193 Bellaterra (Barcelona), Spain\\
$^{38}$ Port d'Informaci\'{o} Cient\'{i}fica, Campus UAB, C. Albareda s/n, 08193 Bellaterra (Barcelona), Spain\\
$^{39}$ Institut d'Estudis Espacials de Catalunya (IEEC), Carrer Gran Capit\'a 2-4, 08034 Barcelona, Spain\\
$^{40}$ Institute of Space Sciences (ICE, CSIC), Campus UAB, Carrer de Can Magrans, s/n, 08193 Barcelona, Spain\\
$^{41}$ INFN section of Naples, Via Cinthia 6, I-80126, Napoli, Italy\\
$^{42}$ Dipartimento di Fisica e Astronomia "Augusto Righi" - Alma Mater Studiorum Universit\'a di Bologna, Viale Berti Pichat 6/2, I-40127 Bologna, Italy\\
$^{43}$ INAF-Osservatorio Astrofisico di Arcetri, Largo E. Fermi 5, I-50125, Firenze, Italy\\
$^{44}$ Centre National d'Etudes Spatiales, Toulouse, France\\
$^{45}$ Institut national de physique nucl\'eaire et de physique des particules, 3 rue Michel-Ange, 75794 Paris C\'edex 16, France\\
$^{46}$ Institute for Astronomy, University of Edinburgh, Royal Observatory, Blackford Hill, Edinburgh EH9 3HJ, UK\\
$^{47}$ ESAC/ESA, Camino Bajo del Castillo, s/n., Urb. Villafranca del Castillo, 28692 Villanueva de la Ca\~nada, Madrid, Spain\\
$^{48}$ European Space Agency/ESRIN, Largo Galileo Galilei 1, 00044 Frascati, Roma, Italy\\
$^{49}$ Univ Lyon, Univ Claude Bernard Lyon 1, CNRS/IN2P3, IP2I Lyon, UMR 5822, F-69622, Villeurbanne, France\\
$^{50}$ Institute of Physics, Laboratory of Astrophysics, Ecole Polytechnique F\'{e}d\'{e}rale de Lausanne (EPFL), Observatoire de Sauverny, 1290 Versoix, Switzerland\\
$^{51}$ Mullard Space Science Laboratory, University College London, Holmbury St Mary, Dorking, Surrey RH5 6NT, UK\\
$^{52}$ Departamento de F\'isica, Faculdade de Ci\^encias, Universidade de Lisboa, Edif\'icio C8, Campo Grande, PT1749-016 Lisboa, Portugal\\
$^{53}$ Instituto de Astrof\'isica e Ci\^encias do Espa\c{c}o, Faculdade de Ci\^encias, Universidade de Lisboa, Campo Grande, PT-1749-016 Lisboa, Portugal\\
$^{54}$ Department of Physics, Oxford University, Keble Road, Oxford OX1 3RH, UK\\
$^{55}$ INFN-Padova, Via Marzolo 8, I-35131 Padova, Italy\\
$^{56}$ Universit\'e Paris-Saclay, Universit\'e Paris Cit\'e, CEA, CNRS, Astrophysique, Instrumentation et Mod\'elisation Paris-Saclay, 91191 Gif-sur-Yvette, France\\
$^{57}$ INAF-Osservatorio Astronomico di Trieste, Via G. B. Tiepolo 11, I-34143 Trieste, Italy\\
$^{58}$ Istituto Nazionale di Astrofisica (INAF) - Osservatorio di Astrofisica e Scienza dello Spazio (OAS), Via Gobetti 93/3, I-40127 Bologna, Italy\\
$^{59}$ Istituto Nazionale di Fisica Nucleare, Sezione di Bologna, Via Irnerio 46, I-40126 Bologna, Italy\\
$^{60}$ INAF-Osservatorio Astronomico di Padova, Via dell'Osservatorio 5, I-35122 Padova, Italy\\
$^{61}$ Institute of Theoretical Astrophysics, University of Oslo, P.O. Box 1029 Blindern, N-0315 Oslo, Norway\\
$^{62}$ Leiden Observatory, Leiden University, Niels Bohrweg 2, 2333 CA Leiden, The Netherlands\\
$^{63}$ Jet Propulsion Laboratory, California Institute of Technology, 4800 Oak Grove Drive, Pasadena, CA, 91109, USA\\
$^{64}$ von Hoerner \& Sulger GmbH, Schlo{\ss}Platz 8, D-68723 Schwetzingen, Germany\\
$^{65}$ Technical University of Denmark, Elektrovej 327, 2800 Kgs. Lyngby, Denmark\\
$^{66}$ Institut d'Astrophysique de Paris, 98bis Boulevard Arago, F-75014, Paris, France\\
$^{67}$ Max-Planck-Institut f\"ur Astronomie, K\"onigstuhl 17, D-69117 Heidelberg, Germany\\
$^{68}$ Aix-Marseille Univ, CNRS/IN2P3, CPPM, Marseille, France\\
$^{69}$ Universit\'e de Gen\`eve, D\'epartement de Physique Th\'eorique and Centre for Astroparticle Physics, 24 quai Ernest-Ansermet, CH-1211 Gen\`eve 4, Switzerland\\
$^{70}$ Department of Physics and Helsinki Institute of Physics, Gustaf H\"allstr\"omin katu 2, 00014 University of Helsinki, Finland\\
$^{71}$ NOVA optical infrared instrumentation group at ASTRON, Oude Hoogeveensedijk 4, 7991PD, Dwingeloo, The Netherlands\\
$^{72}$ Argelander-Institut f\"ur Astronomie, Universit\"at Bonn, Auf dem H\"ugel 71, 53121 Bonn, Germany\\
$^{73}$ Department of Physics, Institute for Computational Cosmology, Durham University, South Road, DH1 3LE, UK\\
$^{74}$ University of Applied Sciences and Arts of Northwestern Switzerland, School of Engineering, 5210 Windisch, Switzerland\\
$^{75}$ European Space Agency/ESTEC, Keplerlaan 1, 2201 AZ Noordwijk, The Netherlands\\
$^{76}$ Department of Physics and Astronomy, University of Aarhus, Ny Munkegade 120, DK-8000 Aarhus C, Denmark\\
$^{77}$ Centre for Astrophysics, University of Waterloo, Waterloo, Ontario N2L 3G1, Canada\\
$^{78}$ Department of Physics and Astronomy, University of Waterloo, Waterloo, Ontario N2L 3G1, Canada\\
$^{79}$ Perimeter Institute for Theoretical Physics, Waterloo, Ontario N2L 2Y5, Canada\\
$^{80}$ Space Science Data Center, Italian Space Agency, via del Politecnico snc, 00133 Roma, Italy\\
$^{81}$ Institute of Space Science, Bucharest, Ro-077125, Romania\\
$^{82}$ Departamento de Astrof\'{i}sica, Universidad de La Laguna, E-38206, La Laguna, Tenerife, Spain\\
$^{83}$ Dipartimento di Fisica e Astronomia "G.Galilei", Universit\'a di Padova, Via Marzolo 8, I-35131 Padova, Italy\\
$^{84}$ Departamento de F\'isica, FCFM, Universidad de Chile, Blanco Encalada 2008, Santiago, Chile\\
$^{85}$ Centre for Electronic Imaging, Open University, Walton Hall, Milton Keynes, MK7~6AA, UK\\
$^{86}$ AIM, CEA, CNRS, Universit\'{e} Paris-Saclay, Universit\'{e} de Paris, F-91191 Gif-sur-Yvette, France\\
$^{87}$ Centro de Investigaciones Energ\'eticas, Medioambientales y Tecnol\'ogicas (CIEMAT), Avenida Complutense 40, 28040 Madrid, Spain\\
$^{88}$ Universidad Polit\'ecnica de Cartagena, Departamento de Electr\'onica y Tecnolog\'ia de Computadoras, 30202 Cartagena, Spain\\
$^{89}$ Infrared Processing and Analysis Center, California Institute of Technology, Pasadena, CA 91125, USA\\
$^{90}$ INAF-Osservatorio Astronomico di Brera, Via Brera 28, I-20122 Milano, Italy\\
$^{91}$ Junia, EPA department, F 59000 Lille, France\\
$^{92}$ SISSA, International School for Advanced Studies, Via Bonomea 265, I-34136 Trieste TS, Italy\\
$^{93}$ IFPU, Institute for Fundamental Physics of the Universe, via Beirut 2, 34151 Trieste, Italy\\
$^{94}$ INFN, Sezione di Trieste, Via Valerio 2, I-34127 Trieste TS, Italy\\
$^{95}$ Dipartimento di Fisica e Scienze della Terra, Universit\'a degli Studi di Ferrara, Via Giuseppe Saragat 1, I-44122 Ferrara, Italy\\
$^{96}$ Istituto Nazionale di Fisica Nucleare, Sezione di Ferrara, Via Giuseppe Saragat 1, I-44122 Ferrara, Italy\\
$^{97}$ Institut de Physique Th\'eorique, CEA, CNRS, Universit\'e Paris-Saclay F-91191 Gif-sur-Yvette Cedex, France\\
$^{98}$ Dipartimento di Fisica - Sezione di Astronomia, Universit\'a di Trieste, Via Tiepolo 11, I-34131 Trieste, Italy\\
$^{99}$ NASA Ames Research Center, Moffett Field, CA 94035, USA\\
$^{100}$ INAF, Istituto di Radioastronomia, Via Piero Gobetti 101, I-40129 Bologna, Italy\\
$^{101}$ INFN-Bologna, Via Irnerio 46, I-40126 Bologna, Italy\\
$^{102}$ Institut de Recherche en Astrophysique et Plan\'etologie (IRAP), Universit\'e de Toulouse, CNRS, UPS, CNES, 14 Av. Edouard Belin, F-31400 Toulouse, France\\
$^{103}$ Universit\'e C\^{o}te d'Azur, Observatoire de la C\^{o}te d'Azur, CNRS, Laboratoire Lagrange, Bd de l'Observatoire, CS 34229, 06304 Nice cedex 4, France\\
$^{104}$ Institute for Theoretical Particle Physics and Cosmology (TTK), RWTH Aachen University, D-52056 Aachen, Germany\\
$^{105}$ Department of Physics \& Astronomy, University of California Irvine, Irvine CA 92697, USA\\
$^{106}$ University of Lyon, UCB Lyon 1, CNRS/IN2P3, IUF, IP2I Lyon, France\\
$^{107}$ INFN-Sezione di Genova, Via Dodecaneso 33, I-16146, Genova, Italy\\
$^{108}$ INAF-Istituto di Astrofisica e Planetologia Spaziali, via del Fosso del Cavaliere, 100, I-00100 Roma, Italy\\
$^{109}$ Instituto de F\'isica Te\'orica UAM-CSIC, Campus de Cantoblanco, E-28049 Madrid, Spain\\
$^{110}$ Department of Physics, P.O. Box 64, 00014 University of Helsinki, Finland\\
$^{111}$ Ruhr University Bochum, Faculty of Physics and Astronomy, Astronomical Institute (AIRUB), German Centre for Cosmological Lensing (GCCL), 44780 Bochum, Germany\\
$^{112}$ Department of Physics, Lancaster University, Lancaster, LA1 4YB, UK\\
$^{113}$ Universit\'{e} Paris-Saclay, CNRS/IN2P3, IJCLab, 91405 Orsay, France\\
$^{114}$ Department of Physics and Astronomy, University College London, Gower Street, London WC1E 6BT, UK\\
$^{115}$ Astrophysics Group, Blackett Laboratory, Imperial College London, London SW7 2AZ, UK\\
$^{116}$ Univ. Grenoble Alpes, CNRS, Grenoble INP, LPSC-IN2P3, 53, Avenue des Martyrs, 38000, Grenoble, France\\
$^{117}$ Centre de Calcul de l'IN2P3, 21 avenue Pierre de Coubertin F-69627 Villeurbanne Cedex, France\\
$^{118}$ Dipartimento di Fisica, Sapienza Universit\`a di Roma, Piazzale Aldo Moro 2, I-00185 Roma, Italy\\
$^{119}$ Zentrum f\"ur Astronomie, Universit\"at Heidelberg, Philosophenweg 12, D- 69120 Heidelberg, Germany\\
$^{120}$ Department of Mathematics and Physics E. De Giorgi, University of Salento, Via per Arnesano, CP-I93, I-73100, Lecce, Italy\\
$^{121}$ INFN, Sezione di Lecce, Via per Arnesano, CP-193, I-73100, Lecce, Italy\\
$^{122}$ INAF-Sezione di Lecce, c/o Dipartimento Matematica e Fisica, Via per Arnesano, I-73100, Lecce, Italy\\
$^{123}$ Institute for Computational Science, University of Zurich, Winterthurerstrasse 190, 8057 Zurich, Switzerland\\
$^{124}$ Higgs Centre for Theoretical Physics, School of Physics and Astronomy, The University of Edinburgh, Edinburgh EH9 3FD, UK\\
$^{125}$ Universit\'e St Joseph; Faculty of Sciences, Beirut, Lebanon\\
$^{126}$ Department of Astrophysical Sciences, Peyton Hall, Princeton University, Princeton, NJ 08544, USA\\
$^{127}$ Helsinki Institute of Physics, Gustaf H{\"a}llstr{\"o}min katu 2, University of Helsinki, Helsinki, Finland\\
$^{128}$ Kapteyn Astronomical Institute, University of Groningen, PO Box 800, 9700 AV Groningen, The Netherlands\\
$^{129}$ Department of Mathematics and Physics, Roma Tre University, Via della Vasca Navale 84, I-00146 Rome, Italy\\
$^{130}$ Cosmic Dawn Center (DAWN)\\
$^{131}$ Niels Bohr Institute, University of Copenhagen, Jagtvej 128, 2200 Copenhagen, Denmark\\
$^{132}$ Departement of Physics and Astronomy, University of British Columbia, Vancouver, BC V6T 1Z1, Canada}

\date{September 2022}

 \abstract
  {The European Space Agency's \Euclid  mission will provide high-quality imaging for about $1.5$ billion galaxies. A software pipeline to automatically process and analyse such a huge amount of data in real time is being developed by the Science Ground Segment of the Euclid Consortium; this pipeline will include a model-fitting algorithm, which will provide photometric and morphological estimates of paramount importance for the core science goals of the mission and for legacy science.
   The Euclid Morphology Challenge is a comparative investigation of the performance of five model-fitting software packages on simulated \Euclid data, aimed at providing the baseline to identify the best suited algorithm to be implemented in the pipeline. In this paper we describe the simulated data set, and we discuss the photometry results. A companion paper (Euclid Collaboration: Bretonni\`ere et al. 2022) is focused on the structural and morphological estimates.
   We created mock \Euclid images simulating five fields of view of 0.48 deg$^2$ each in the $I_{\scriptscriptstyle\rm E}$ band of the VIS instrument, containing a total of about one and a half million galaxies (of which 350\,000 have nominal signal-to-noise ratio above $5$), each with three realisations of galaxy profiles (single and double \sersic, and `realistic' profiles obtained with a neural network); for one of the fields in the double \sersics realisation, we also simulated images for the three near-infrared $Y_{\scriptscriptstyle\rm E}$, $J_{\scriptscriptstyle\rm E}$ and $H_{\scriptscriptstyle\rm E}$ bands of the NISP-P instrument, and five Rubin/LSST optical complementary bands ($u$, $g$, $r$, $i$, and $z$), which together form a typical data set for a \Euclid observation. The images were simulated at the expected \Euclid Wide Survey depths. To analyse the results we created diagnostic plots and defined metrics to take into account the completeness of the provided catalogues, and the median biases, dispersions, and outlier fractions of their measured flux distributions. 
   Five model-fitting software packages (\texttt{DeepLeGATo}, \texttt{Galapagos-2}, \texttt{Morfometryka}, \texttt{ProFit}, and \texttt{SourceXtractor++}) were compared, all typically providing good results. Of the differences among them, some were  at least partly due to the distinct strategies adopted to perform the measurements. In the best case scenario, the median bias of the measured fluxes in the analytical profile realisations is below 1\% at signal-to-noise ratio above 5 in $I_{\scriptscriptstyle\rm E}$, and above 10 in all the other bands; the dispersion of the distribution is typically comparable to the theoretically expected one, with a small fraction of catastrophic outliers. However, we can expect that real observations will prove to be more demanding, since the results were found to be less accurate on the most realistic realisation.
   We conclude that existing model-fitting software can provide accurate photometric measurements on \Euclid data sets. The results of the challenge are fully available and reproducible through an online plotting tool.}
 
\keywords{
    Galaxies: structure  --
    Galaxies: evolution --
    Cosmology: observations
}

\titlerunning{Euclid Morphology Challenge - Data set and photometry}
\authorrunning{E. Merlin et al.}

\maketitle


\section{Introduction}

The European Space Agency's \Euclid mission \citep[][mission RedBook]{Laureijs2011}, due to start operations in 2023, is designed to provide accurate photometric, spectroscopic and morphological data (in particular cosmic shear and clustering distributions) for billions of galaxies across 15\,000 deg$^2$ of sky, using them as tracers to study the properties of the dark components of the Universe.

To this end, a processing pipeline is being assembled by the Science Ground Segment, a team that is in charge of releasing the data to the community. This pipeline is ready to ingest, process and analyse the raw imaging data from the satellite on a daily basis; optical data from external ground-based instruments \citep[Rubin/LSST, DECAM, CHFT, Pan-STARRS, OmegaCAM, Subaru; see][]{Scaramella2022} will also be used to complement the optical and near-infrared images obtained by the two satellite photometers VIS \citep[observing in $I_{\scriptscriptstyle\rm E}$, a broad optical band; see][]{Cropper2016} and NISP-P \citep[observing in the three near-infrared -- NIR -- bands $Y_{\scriptscriptstyle\rm E}$, $J_{\scriptscriptstyle\rm E}$, and $H_{\scriptscriptstyle\rm E}$;][]{Maciaszek2016,Schirmer2022}, allowing for high-quality photometric redshift estimates. The final step of the image analysis pipeline 
will produce a global catalogue containing all the astrometric, photometric and morphological information about each source detected in the $I_{\scriptscriptstyle\rm E}$ images (plus an additional sample of NIR-detected sources). This catalogue will then be exploited for scientific use by the Euclid Collaboration, and it will also be released to the community for legacy use.

The pipeline currently implements two photometric techniques \citep[aperture and template-fitting, performed with \Euclid-specific versions of two public software tools, \texttt{a-phot} and \texttt{t-phot} respectively; see][]{Merlin2015,Merlin2016a,Merlin2019}, and a module to estimate so-called CAS morphological parameters  \citep[Concentration/Asymmetry/Smoothness: non-parametric morphological features that can be used to distinguish between discs, ellipticals, compact, diffuse, symmetric/asymmetric or clumpy objects by means of a dimensional reduction, see][]{Conselice2003}. However, the pipeline is foreseen to also include a profile model-fitting algorithm. The Euclid Morphology Challenge (EMC) was organized 
with the aim of analysing and comparing the performance of various model-fitting software tools on \Euclid data, in order to establish the foundations for choosing the tools that will be integrated into the official processing pipeline. The final choice will be driven by many factors, including computational performance, robustness of the algorithm, and compatibility with the current version of the pipeline; however, the accuracy of the parameter estimates will of course be the main driver. Therefore, assessing the performance of the different software packages on simulated data, for which the ground truth is known, is a necessary and fundamental step for a sound selection. 
Eight development teams of model-fitting software packages were invited to participate to the challenge, and five provided at least partial results.

In this paper we present the data set created for the EMC, and we discuss the results concerning photometry. In fact, albeit not being the central focus of the challenge, flux measurements obtained via model-fitting techniques will have great relevance, providing a crucial complement to the more straightforward methods already included in the pipeline. A companion paper is dedicated to the analysis of such morphological estimates (Euclid Collaboration: Bretonni\`ere et al. 2022, EMC2022b hereafter). 

This paper is structured as follows. Section~\ref{sim} describes the technique used to create the simulated data set for the Challenge, with some technical details given in Appendix~\ref{AppTec}. In Sect.~\ref{codes} we briefly present the software tools taking part in the Challenge, and in Sect.~\ref{fom} we describe the methods used to analyse and rank the data provided by the participants. The results are then presented in Sect.~\ref{results}, where we investigate the general accuracy of the photometric measurements, and the reliability of the estimated uncertainty budgets, with a further focus on each software package's performance given in Appendix~\ref{AppCodes}. Finally, Sect.~\ref{conclusions} presents a summary of the work and provides conclusions. 

All magnitudes are given in the AB system.

\section{Simulating the \Euclid universe} \label{sim}

Simulated data sets are being produced and used by the Euclid Science Ground Segment to test the full processing pipeline from image reduction to data analysis. These simulations consist of raw single exposures, including observational features and defects, and they must be processed and stacked to reach the nominal depth and be ready for scientific analysis, with background light and defects removed. 
To simplify this complex procedure, and to have all details under control, for the EMC we decided to produce a tailored data set, directly simulating background-subtracted images at the expected nominal depths of the final stacked mosaics in all bands. With this approach, we were also free to try different options, producing simulations with single and double \sersics analytical profiles, and also with realistic morphologies. In this section we explain the procedure we followed to obtain all these simulated data sets.

\subsection{Catalogues and images creation} \label{sims}

We started by creating mock cosmological catalogues with the code \texttt{Egg} \citep[v1.3.1,][]{Schreiber2017}. \texttt{Egg} uses the statistical distributions of real galaxies as detected and classified in the five CANDELS fields \citep{Grogin2011, Koekemoer2011} to build a simulated catalogue of a patch of the sky, complete with the properties of the objects as observed by a chosen set of pass-band filters, with a chosen pixel resolution, and to a chosen limiting magnitude. We refer the reader to the paper describing the code for a detailed description of its workflow; here we provide a short summary. \texttt{Egg} draws redshifts and stellar masses from observed galaxy stellar mass ($M_*$) functions, and subsequently attributes a star-formation rate (SFR) to each galaxy from the observed SFR--$M_*$ main sequence; dust attenuation, optical colours and simple disc-plus-bulge morphologies are obtained from empirical relations established from the high-quality \textit{Hubble} and \textit{Herschel} observations of the CANDELS fields. Random scatter is introduced in each step to reproduce the observed distributions of each parameter. Finally, based on these observables, a suitable panchromatic spectral energy distribution (SED) is selected for each galaxy and synthetic photometry is produced by integrating the redshifted SED over the chosen broad-band filters.
The galaxies are created as two-component objects, with a bulge and a disc both described by a \citet{Sersic1968} profile,
\begin{equation}
    I(r) \propto \exp{\left[ -b_n (r/r_{\rm{e}})^{1/n}\right]},\
\end{equation}
where \sersics indices for the bulge and disc components are $n_{\rm bulge}=4$ and $n_{\rm disc}=1$. The output catalogue contains the physical and observed properties of the galaxies within a field of view (FoV) corresponding to the chosen area; the objects are placed at random positions with a fixed angular two-point correlation function, neglecting large-scale clustering beyond \ang{;3;} (i.e., beyond $\sim$1 Mpc at $z>0.5$).

\begin{figure}[h!]
\centering
\includegraphics[width=0.45\textwidth]{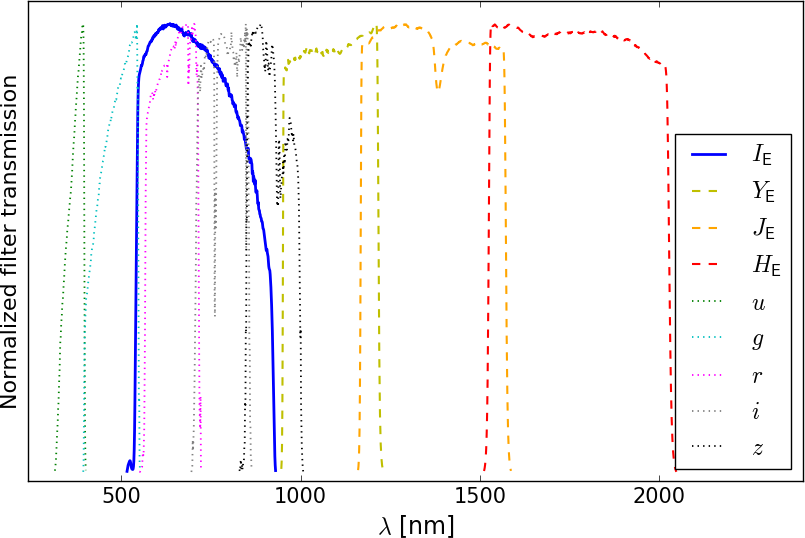}
\caption{Filter transmission curves used to simulate the images (normalised to arbitrary units) in this work. The four filters $I_{\scriptscriptstyle\rm E}$, $Y_{\scriptscriptstyle\rm E}$, $J_{\scriptscriptstyle\rm E}$, and $H_{\scriptscriptstyle\rm E}$ form the \Euclid set, while $u$, $g$, $r$, $i$, and $z$ are external complementary filters (from LSST/VRO). The NIR curves are early estimates of the actual transmission functions, which are described in their latest updated version in \citet{Schirmer2022}; however, since we only use these data to compute the input fluxes, the differences are not relevant for the present work.}
\label{filters}
\end{figure}

We created five catalogues, each one with a size of $0.482$ deg$^2$ (\ang{;41.66;} per side, which is comparable to the typical area on which each single photometric catalogue will be extracted from real data), with limiting magnitude $I_{\scriptscriptstyle\rm E}=27.1$ (the nominal 1$\sigma$ limit as given in the mission RedBook). The total number of simulated galaxies was about $1.5$ million.
We then used an \texttt{Egg} built-in script to obtain all the observational properties of the sources. In particular, for each galaxy the following parameters are given: position of the centroid in pixels; total flux in the simulated band; bulge-to-total flux ratio (\bt); scale length of the bulge and of the disc (defined as the radius at which the component is a factor of $e$ less bright than it is at its center); 
axis ratio for both components; and position angle for both components. For the first of the five fields (F0), we produced nine lists, to include a full multi-wavelength realisation of a Euclidian sky patch: one for each of the four \Euclid bands $I_{\scriptscriptstyle\rm E}$, $Y_{\scriptscriptstyle\rm E}$, $J_{\scriptscriptstyle\rm E}$ and $H_{\scriptscriptstyle\rm E}$, plus five for the Rubin/LSST bands $u$, $g$, $r$, $i$ and $z$. The filter transmission curves are shown in Fig.~\ref{filters}. For the other four fields (F1--4) we only produced the $I_{\scriptscriptstyle\rm E}$ list, since the main purpose of these simulations is the morphological analysis, which with real data will mostly be performed on the $I_{\scriptscriptstyle\rm E}$ images, given that it will be the band with the highest resolution and depth. We point out that in the multi-band realisation the morphological parameters do not change across the spectrum, while total fluxes and \bts do; this information was not explicitly shared with the participants.

We fed these catalogues to \texttt{GalSim} \citep[][v2.2.1]{Rowe2015}, a Python package that produces simulated astronomical images (it is also used for the official \Euclid simulations). We created the images at their expected native pixel scale: $\ang{;;0.1}$ for $I_{\scriptscriptstyle\rm E}$ ($25\,000\times25\,000$ pixels); $\ang{;;0.3}$ for NIR bands; and $\ang{;;0.2}$ for LSST bands. After the procedure described in the following paragraphs, we resampled all the images to the $I_{\scriptscriptstyle\rm E}$ pixel scale, since this is the procedure that will be followed in the real pipeline. 
To simulate the effects of point spread functions (PSFs), we provided \texttt{GalSim} with the Euclid Mission Database models for the $I_{\scriptscriptstyle\rm E}$ and NIR bands (as provided by the corresponding \Euclid working groups; both of them are over-sampled by a factor of 6), while for LSST we provided custom simulated PSFs created using \texttt{PhoSim} \citep{Peterson2015}, at the expected observed pixel scale of $\ang{;;0.2}$ (with no over-sampling). The approximate FWHMs of these PSFs are $\ang{;;0.17}$ ($I_{\scriptscriptstyle\rm E}$), $\ang{;;0.54}$ (NIR), and $\ang{;;1.00}$--$\ang{;;1.12}$ (LSST).

We produced noiseless galactic profiles with \texttt{GalSim}, with pixel values in {\si{\micro}Jy}/pixel. We simulated three different sets of images, all having identical sets of coordinates and total fluxes of the galaxies, which we describe in the following; see also EMC2022b for further details on this.
\begin{itemize}

\item Double \sersics profiles (DS), directly using the standard output of \texttt{Egg}, which consists of a catalogue formatted to be used with \texttt{SkyMaker} \citep{Bertin2009}. In particular, this means that the dimensions of the objects are given as \textit{scale lengths}. On the contrary, \texttt{GalSim} requires half-light radii; the two values coincide for the bulges, while the conversion factor is 1.678 for the discs \citep[see e.g.][]{Graham2005}, so we applied this correction before simulating the images. Also, \texttt{GalSim} requires that the fluxes of the two components are given separately, while \texttt{Egg} outputs a total magnitude and a \bt; therefore, to assign a flux to each component we simply used the relations $f_{\rm bulge}=\btm\; f_{\rm tot}$ and $f_{\rm disc}=(1-\btm)\; f_{\rm tot}$, where $f_{\rm tot}=10^{-0.4(m-{\rm ZP})}$ (where $m$ is the magnitude of the sources as given in the \texttt{Egg} catalogue and ZP is the zero-point of the image, see Sect. \ref{noise}). 

\item Single \sersics profiles (SS), in which galaxies are modeled with a single \sersics index, defined using the \bts values from the \texttt{Egg} catalogue as $n_{\rm total}=(1-\btm)\, n_{\rm disc} + \btm \; n_{\rm bulge} = 3\;\btm +1$. To compute the single effective radius $r_{\rm e,tot}$ from the two values given in the 2-component catalogue, we used the following formula, calibrated empirically to obtain a good visual match between the two realisations: $r_{\rm tot}=[\btm\; r_{\mathrm{e,b}}]^{\alpha}+[(1-\btm)\, r_{\mathrm{e,d}}]^{\alpha}$, where $\alpha=0.8$ if $r_{\mathrm{e,b}}<r_{\mathrm{e,d}}$ (98\% of the cases), and $\alpha=2.0$ otherwise. Finally, position angles and axis ratios already had the same values for bulges and discs, so we simply kept them unchanged.

\item Realistic morphologies (RM), in which galaxy stamps are created by means of a neural network using a variational auto-encoder trained on observed COSMOS galaxies, as described in full detail in \citet{Lanusse2021} and \citet{Bretonniere2022}; each simulated galaxy mimics the properties of its corresponding analytical realisation. In this data set, the biggest and brightest objects ($r_{\rm e}>\ang{;;0.2}$, $I_{\scriptscriptstyle\rm E}<20.5$) are not simulated due to technical limitations; the list of excluded sources was provided to the participants, and accounts for approximately $1\%$ of the total simulated galaxies. Also, the position angles of the galaxies are not constrained to be close to those of the \texttt{Egg} catalogue (and therefore they were not considered in the final analysis of the results).
We point out that  this is the first time that such a demanding test has been performed: the codes must provide an analytical fit on non-analytical shapes for which a ground-truth value is known. This is inherently a very challenging task. Moreover, the method used to create the images is not perfect. The conditioning of the latent space with galaxy morphology is not always exact, and this can introduce a systematic bias with respect to the input values \citep[see the discussion in][]{Bretonniere2022}, although the consistency is fully guaranteed in a statistical sense; some level of scatter remains on a object-by-object basis, meaning that the comparison with the input catalogue must be taken with caution. For more details, see EMC2022b. 
\end{itemize}


\begin{figure}[h!]
\centering
\includegraphics[width=0.45\textwidth]{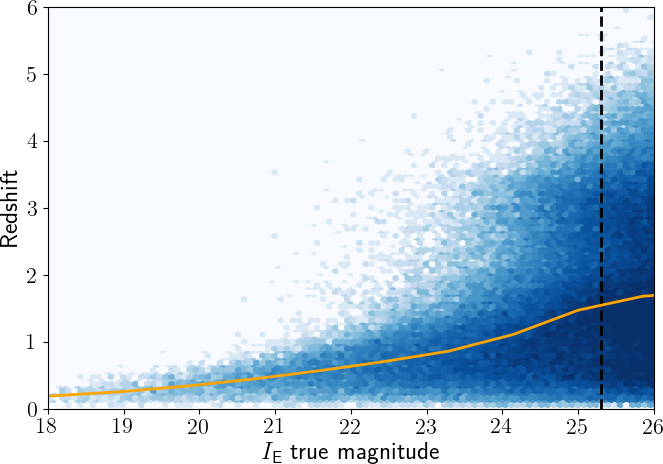}
\caption{Distribution of redshifts as a function of $I_{\scriptscriptstyle\rm E}$ magnitude (input values) in F0. The orange line is the running mean, the vertical dashed line the 5$\sigma$ limit.}
\label{inputcat}
\end{figure}

\begin{figure}[h!]
\centering
\includegraphics[width=0.45\textwidth]{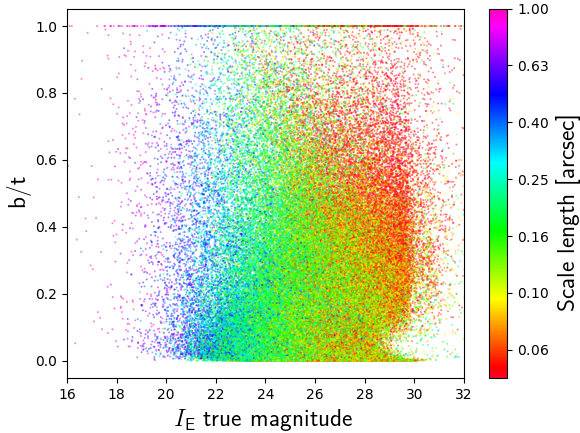}
\caption{Distribution of bulge-to-total ratios as a function of $I_{\scriptscriptstyle\rm E}$ magnitude in the input catalogue of F0; the colours encode the global scale length, defined here as $r = \btm\; r_{\mathrm{e,bulge}}+(1-\btm)\, r_{\mathrm{e,disc}}$.}
\label{inputcat2}
\end{figure}

\begin{figure}[h!]
\centering
\includegraphics[width=0.45\textwidth]{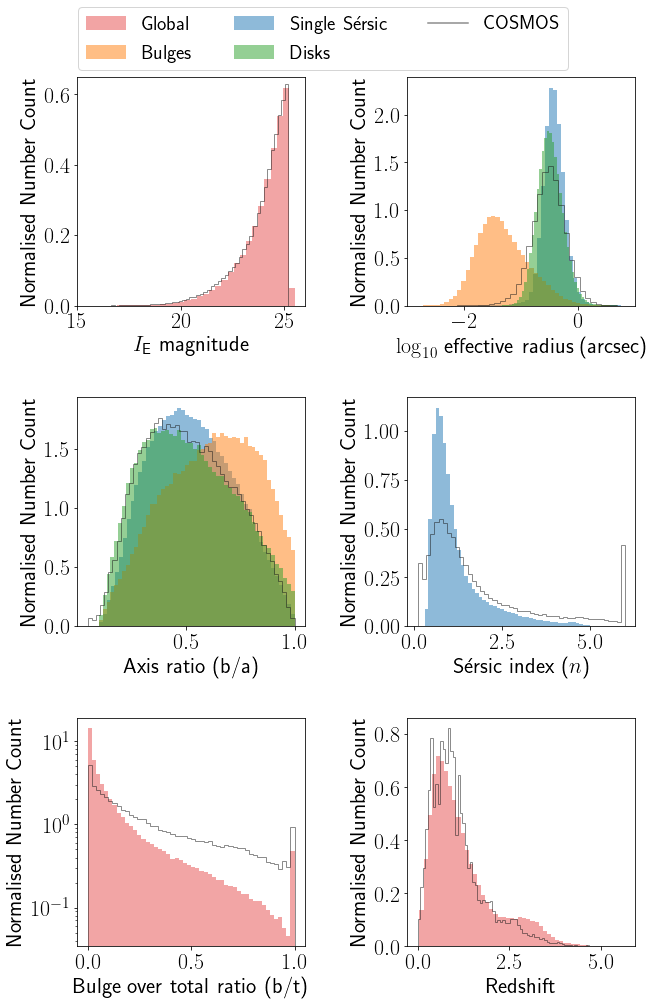}
\caption{Distributions of various input parameters for the $I_{\scriptscriptstyle\rm E}$ band, in all five simulated fields, for all the separate components and realisations in the Challenge, as described in the legend and in the labels of the panels. For comparison, we also show the corresponding parameter distribution in COSMOS.}
\label{distr}
\end{figure}

Figure \ref{inputcat} shows the redshift distribution as a function of  $I_{\scriptscriptstyle\rm E}$ input magnitude. \texttt{Egg} outputs the redshift of each simulated galaxy, and although this information is not explicitly used in the present work, it might nevertheless be useful to have an idea of the global distribution; indeed, most of the analysis and figures will be presented as a function of the input $I_{\scriptscriptstyle\rm E}$ magnitude, which correlates with redshift. For example, looking at the plot one can see that galaxies at $z=3$ typically begin to be detectable at $I_{\scriptscriptstyle\rm E}=23$.
Figure \ref{inputcat2} shows the distribution of simulated galaxies in the magnitude-size-\bts space for the same field (note how there is a non-negligible fraction of bulge-only objects with  $\btm=1$). Finally, in Fig.~\ref{distr} we show the distributions of various input parameters for the $I_{\scriptscriptstyle\rm E}$ band components and realisations (magnitude, effective radius, axis ratio, \sersics index, bulge-to-total ratio and redshift), for the full data set (the five fields), down to the nominal 5$\sigma$ limit $I_{\scriptscriptstyle\rm E}=25.35$ \citep{Laureijs2011}; we also show the COSMOS distributions \citep{Mandelbaum2012} for reference. As expected, simulations and observations agree remarkably well, with the exception of the \bts distribution, which is more skewed towards disc galaxies in the \texttt{Egg} catalogue. This might have some impact on the analysis of the results (see Sect. \ref{results}).

By construction, we could not simulate irregulars, which are estimated to constitute less than $10\%$ of the galaxies at $z<1$, but up to $70\%$ at $z=3$ \citep[e.g. ][, although recent preliminary results from the James Webb Space Telescope seem to indicate a lower number]{Huertas2015}. This is an obvious but unavoidable limitation of this work. The RM realisation can provide a hint about how model-fitting codes can deal with non-analytical shapes.

We then added a field of stars, to include the effects of their presence as contaminants in the fitting procedures. To obtain a realistic distribution, we took their celestial coordinates (converted to pixel positions) from one FoV of the official \Euclid simulations used for Scientific Challenges,\footnote{Scientific Challenges are official Euclidean benchmark tests performed to check and validate the progress of the work in preparation for the launch of the satellite} and simply placed PSF stamps at the positions of the sources, scaling their flux to match the catalogue magnitudes. We excluded very bright stars ($I_{\scriptscriptstyle\rm E}<15$), in order to avoid that large regions of the simulations were affected by their presence, and also because -- given the limited extension of the PSF stamps -- they would saturate creating artificial defects on the images. The fraction of pixels significantly contaminated by stellar light (that is, where the surface brightness from stellar light is more than the 1$\sigma$ surface brightness per pixel) is approximately $1\%$ in the $I_{\scriptscriptstyle\rm E}$ images, $5\%$ in the NIR bands, and from $1\%$ to $7\%$ in LSST bands ($u$ and $z$ respectively).

\subsubsection{Observational noise}\label{noise}

\begin{table}
\label{depths}
\centering       
\begin{tabular}{|l| c| c| c|} 
\hline
Band & $m_{\rm lim}$ & SB$_{\rm bkg}$ & $t_{\rm exp}$ [s] \\
\hline
$I_{\scriptscriptstyle\rm E}$ & 24.6 & 22.33 & 4 $\times$ 590\\
$Y_{\scriptscriptstyle\rm E}$ & 23.0 & 22.10 & 4 $\times$ 88\\
$J_{\scriptscriptstyle\rm E}$ & 23.0 & 22.11 & 4 $\times$ 90\\
$H_{\scriptscriptstyle\rm E}$ & 23.0 & 22.28 & 4 $\times$ 54\\
$u$ & 23.6 & 22.70 & 150 \\
$g$ & 24.5 & 22.00 & 150 \\
$r$ & 23.9 & 20.80 & 150\\
$i$ & 23.6 & 20.30 & 150 \\
$z$ & 23.4 & 19.40 & 150 \\
\hline
\end{tabular}
\caption{Parameters used to simulate the images. $\mbox{m}_{\rm lim}$ is the 10$\sigma$ limiting magnitude within a 2\arcsecond\ aperture, SB$_{\rm bkg}$ is the background surface brightness, and $t_{\rm exp}$ is the total exposure time of the final mosaic (these are not updated to the latest estimates of the actual in-flight values). See text for more details.} 
\end{table}

Once the seed images containing the sources were produced, we proceeded to add simulated observational noise. First of all we replaced the smooth analytical profiles with stochastic realisations from a Poissonian distribution, to simulate the effects of photon shot noise. We used the Python module \texttt{scipy.stats.poisson.rvs} for this purpose. 
We paid particular attention to keep the units of the images always consistent during the whole process: we first converted the noiseless images from {\si{\micro}Jy}/pixel to observational units, using the correct image observational ZP at 1 second, to obtain consistent Poissonian realisations, which depend on the total exposure time. Only after this step did we convert the images back to {\si{\micro}Jy}. To calculate the ZPs we followed the procedure described in \citet{Martinet2019}, which we summarize in Appendix \ref{AppZP}.
The same method was also used to produce an empty sky map containing only Gaussian noise, simulating the observational background at the desired depth. Since the images were simulated with zero background light, the Gaussian noise must have zero mean, and the standard deviation of the pixel values defines the depth of the final simulated image. 

The values used in this procedure are summarized in Table \ref{depths}. The exposure times used in our simulations the for $I_{\scriptscriptstyle\rm E}$ and for the NIR bands were taken from \citet{Laureijs2011}, and it is worth pointing out that the actual in-flight values will be slightly different ($4 \times 560$ seconds for $I_{\scriptscriptstyle\rm E}$ and $4 \times 88$ seconds for all NIR bands). The exposure times for the LSST bands were estimated from simulated data created for one of the internal validation Scientific Challenges, and are representative of an early release (LSST final data after ten years of observations will of course be much deeper). The limiting magnitudes and background surface brightness values, which are needed in the computations, are set to be consistent with the expected values for the \Euclid Wide Survey, and they were taken from a dedicated study by J. C. Cuillandre (priv. comm.); these too are slightly different from the current best estimated values, which can be found in \citet{Scaramella2022}. These small inconsistencies are due to the fact that the present work began before the latest estimates had been made available; however, they have negligible impact on the scientific results of the EMC. It is worth pointing out that the images were produced with homogeneous noise levels, i.e we did not simulate regions of different depths. 
Finally, we summed each image containing the Poissonian realisations of the galaxies and stars with the corresponding `empty' Gaussian sky noise. 

\subsubsection{Rebinning}

As mentioned, we simulated all the images at their native pixel scales ($\ang{;;0.1}$ for $I_{\scriptscriptstyle\rm E}$, $\ang{;;0.3}$ for NIR, and $\ang{;;0.2}$ for LSST), and we then rebinned F0 NIR and LSST images to the $I_{\scriptscriptstyle\rm E}$ pixel scale using \texttt{Swarp} \citep{Bertin2002}. This is consistent with the real pipeline workflow, with some differences: in the pipeline, a newer version of the software named \texttt{Swarp++} is used, and single-exposure images are combined to create the mosaics, allowing information to be gained in the process. 
We also rebinned the PSFs accordingly. All the rebinning processes were performed using the \texttt{BILINEAR} interpolation mode. We note that this resampling procedure introduces artifacts in the noise map (in particular, pixel correlations) that alter the apparent signal-to-noise ratio (S/N) of the map, so the actual uncertainties of the measurements must be computed using a dedicated RMS map, which we discuss next. 

\subsection{RMS maps}

It is common practice to assign uncertainties to the measurements performed on scientific images by means of a weight or an RMS map, which is often obtained from first principles during the data reduction chain. When this is not possible, it can be easily determined, at least to a first approximation, by measuring the RMS of the pixel values in `empty' regions of the (non-rebinned) science frame -- although such a measurement only provides information on the noise due to the unresolved sky background. 
For the sake of the EMC goals, we wanted to factorise out any possible source of complication, and therefore ready-to-use RMS maps were provided to the participants, along with the scientific images. This is also again consistent with the pipeline architecture. 
The procedure to build the RMS map is descibed in Appendix \ref{AppRMS}.

\begin{figure}[h!]
\centering
\includegraphics[width=0.45\textwidth]{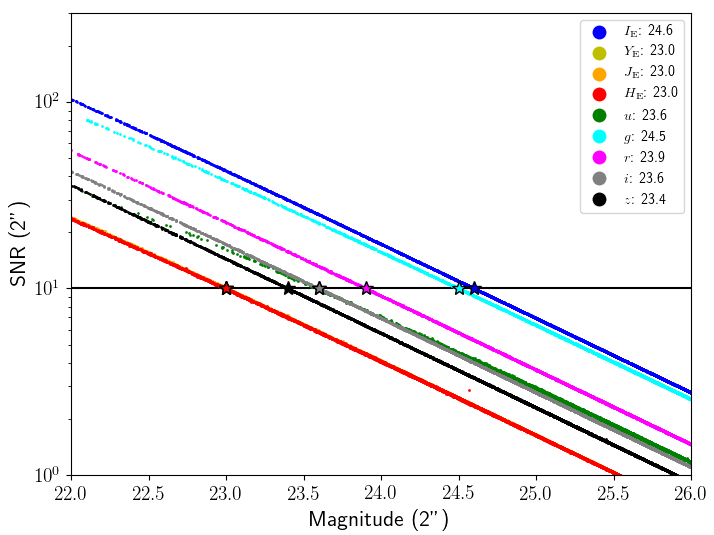}
\caption{Signal-to-noise ratios of the simulated images at the native pixel scales, measured with forced photometry in 2\arcsecond\ diameter apertures at the input positions of the sources. Star symbols show the expected values, which are also reported in the legend.}
\label{snr}
\end{figure}

The RMS maps were produced at the native pixel scales of the scientific images, and we checked that the pre-resampling S/Ns are consistent with the expected values. This is shown in Fig.~\ref{snr}, where we plot the S/N estimated for each simulated source and check that it is equal to the expected value at the limiting magnitude (star symbols); for this test we used \texttt{a-phot}, forcing the measurements within 2\arcsecond\ apertures (as for the definition of S/N adopted in this work) at the true input positions of the sources. Note that some distributions of points are overlapping (the three NIR bands have the same expected depth, and so do two of the LSST bands). The overall agreement with the expected values is very accurate. Finally, we proceeded to resample the maps along with the scientific images, again using \texttt{Swarp}, checking that the S/N values of the resampled images are correct when the RMS maps are used to estimate uncertanties.

\begin{figure*}[h!]
\centering
\includegraphics[width=0.99\textwidth]{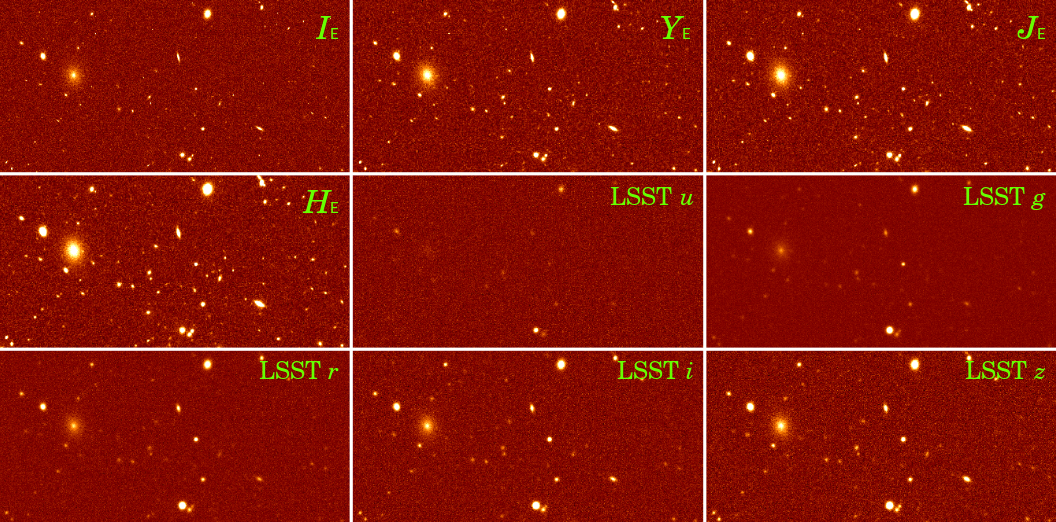}
\caption{A small region of the DS F0 realisation in the nine bands; all are shown with the same colour scale.}
\label{img1}
\end{figure*}

\begin{figure}[h!]
\centering
\includegraphics[width=0.49\textwidth]{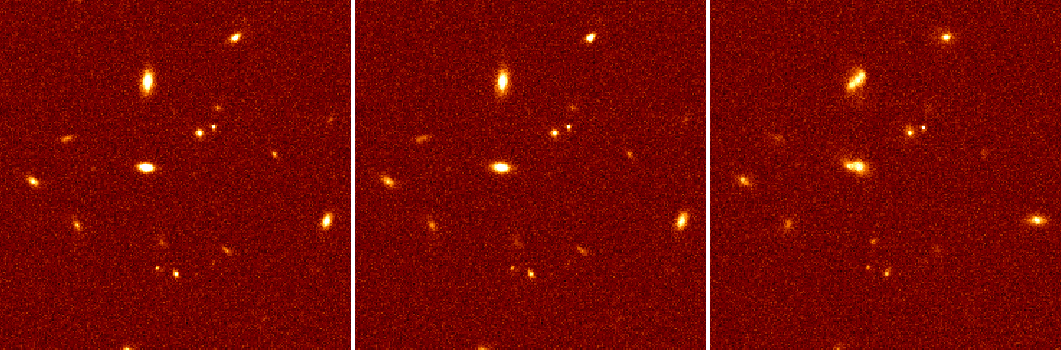}
\caption{A small crop of the three realisations of $I_{\scriptscriptstyle\rm E}$ images. Left to right: DS, SS, RM. Notice that in RM the orientations of the galaxies are not the same as in the other two realisations.}
\label{img2}
\end{figure}

\subsection{Challenge set-up}

The scientific and RMS images were finally uploaded to a private online repository for the participants to download, together with lists containing the IDs and input positions of the centroids of the simulated objects down to various nominal S/N levels ($100$, $10$, $5$, $1\sigma$). This was done to factor out possible inaccuracies in object detection and deblending, so that the challenge could be actually focused on the accuracy of fitting photometry and morphology without adding any further possible source of error. 
We point out that to obtain the lists we simply applied different cuts to the input $I_{\scriptscriptstyle\rm E}$ magnitudes, so they must be considered as coarse reference levels rather than accurate estimations of detection significance.

As a visual example, Fig.~\ref{img1} shows a small crop of the DS F0 images in all nine bands, while Fig.~\ref{img2} shows a small crop of the DS, SS and RM F0 $I_{\scriptscriptstyle\rm E}$ images. A sample of the ground-truth input values of the simulated galaxies was also provided to the participants (in particular, a small portion of the multi-band F0 data set and the whole F4 $I_{\scriptscriptstyle\rm E}$ data set), allowing for a check that their procedures were reasonably correct without evident errors. 

The output requested from each participant consisted of the estimates of (i) flux, (ii) \sersics index (in the SS and RM realisations), (iii) half-light semi-major axis, (iv) axis ratio, and (v) position angle, each with a corresponding 1$\sigma$ uncertainty, for each component of the simulated objects. In particular, while the SS and RM realisations only required a single fit with free $n$, for the DS realisation we asked for two estimates, namely: one fit with fixed indices ($n_{\rm disc}=1$ and $n_{\rm bulge}=4$, consistent with the way the images were simulated); and another with  $n_{\rm disc}=1$ and $n_{\rm bulge}$ left free to vary. 

As mentioned, the output estimates were required for the objects belonging to the list including only ${\rm S/N}>5$ (in $I_{\scriptscriptstyle\rm E}$) sources. Analysis of objects at lower S/N were explicitly mentioned as an optional output that would not influence the final comparison among the software packages.

%

\section{Model-fitting software packages} \label{codes}



Eight development teams of model-fitting software packages were invited to participate to the Challenge; of them, five (\texttt{DeepLeGATo}, \texttt{Galapagos-2}, \texttt{Morfometryka}, \texttt{ProFit}, and \texttt{SourceXtractor++}) provided at least partial results. All but one are based on parametric methods, using functional forms to fit the observed light distributions, the exception being \texttt{DeepLeGATo}, which exploits a convolutional neural network (CNN).

Here we briefly summarize the basic properties and features of these five software packages, and point out a few important details about the procedures each one followed. It is instructive and important to notice how various subtleties in the interpretation of the requests, usage of the provided input data sets, and processing methods used by each participant led to differences in the format and accuracy of the provided outputs.

\subsection{DeepLeGATo}
\texttt{DeepLeGATo} \citep{Tuccillo2018} is a software package for estimating galaxy structure based on a supervised deep-learning approach. The code uses CNNs to perform a simple regression between an image centered on a given galaxy and its structural parameters, providing results in very short times (see Appendix \ref{Dl}).
In the version used in this work, the training was performed with images of fixed size (128$\times$128 pixels) independently of the galaxy effective radius; this likely caused sub-optimal performance on the largest and brightest galaxies. The images used for training were not the ones provided with the EMC data-set; instead, they were idealised analytic 1- or 2-component \sersics profiles, convolved with the PSF of the corresponding band, to which noise having similar properties to that in the EMC data was added. A different training was performed for each structural parameter with slightly different architectures, which are variations of standard CNNs \citep[see][for more details]{Tuccillo2018}. The fits were performed at the positions of the sources as given in the input files.

Because the loss function used for training is a standard mean square error, only point estimates were provided; therefore in the implementation used for this work the uncertainty budget (i.e. the uncertainties on the estimates) was not computed. Noticeably, the 5$\sigma$ input source lists were subdivided into S/N bins using the other input catalogs (10 and 100$\sigma$), and different parameters were used for the fits in different bins; this can be seen in the distributions of the points in the plots (see Appendix \ref{AppCodes}).
While this makes the version of the software used in the Challenge not directly suitable for an implementation in the \Euclid pipeline, because the S/N of real sources cannot be known a priori, it is worth pointing out that the code remains under development, and more recent releases work without the need for this fine-tuning of the parameters for different input data.
Only $I_{\scriptscriptstyle\rm E}$ SS and DS fits were provided.

\subsection{Galapagos-2} 

\texttt{Galapagos-2} \citep{Haussler2022}\footnote{\url{https://github.com/MegaMorph/galapagos}} is an updated and enhanced version of \texttt{Galapagos} \citep{Barden2012}. It provides a wrapper around either \texttt{Galfit} \citep{Peng2002, Peng2010} for single-band fits, or \texttt{GalfitM} \citep{Haussler2013} for either single- or multi-band fits, and it is specifically designed to carry out fully automated fitting on all objects in a large survey.
Starting from the input images and a simple setup file, it employs \texttt{SExtractor} \citep{Bertin1996} for object detection, and then uses this information to automatically set up the fits. The postage stamp size used for each fit/object depends on the estimated size of the object, set up by the user. Using some limited input from the setup, e.g. enlargement factors to conservatively increase the size and shape of the object estimated by \texttt{SExtractor}, \texttt{Galapagos-2} takes care of neighbouring objects (deblending and fitting of bright nearby objects, masking of fainter and more distant objects), estimates the sky background level with a sophisticated and robust scheme \citep[see][for details]{Barden2012}, and sets up the fit using \texttt{SExtractor} values as initial guesses to run the fitting algorithm \texttt{GalfitM} for all objects, starting from the brightest objects and using the PSF provided. Once an object has been fit, it is merely subtracted from the fits of nearby objects. This significantly speeds up the process overall, as these bright, large objects take the longest to fit, but this only needs to be done once. In a fully automated pipeline, it then reads out the result and provides one final catalogue, which contains fitting information for all bands. \texttt{GalfitM} itself uses a Levenberg–Marquardt minimisation to derive the best-fit parameters and uncertainties. In the multi-band realisations of the EMC, all bands are connected via physically reasonable polynomials and fitted simultaneously, to reduce the degrees of freedom of the fit and make full use of the multi-wavelength information.
The software requires one additional input image compared to what was provided, namely a weight image to flag bad pixels. Since no bad pixels were in the data, this was trivially created as a uniform image of the correct size. 


All the requested outputs were provided; however, for DS F0 only a simultaneous fit was run, therefore including $I_{\scriptscriptstyle\rm E}$ in the multi-band fitting process -- in other words, there is no isolated fit for $I_{\scriptscriptstyle\rm E}$ DS F0. This causes the $I_{\scriptscriptstyle\rm E}$ fit for F0 to be substantially different from that of other fields; for this reason, it was decided not to include F0 in the analysis of $I_{\scriptscriptstyle\rm E}$-only results for all codes (see below). 

\subsection{Morfometryka}
\texttt{Morfometryka} \citep{Ferrari2015}, written in Python, was primarily designed to measure non-parametric morphological quantities, but as a bonus it performs single \sersics model fitting. The software takes as input a galaxy stamp (plus the PSF model), estimates the background with an iterative algorithm, segments the sources and defines the target. Then, it filters out external sources using the code \texttt{GalClean} \citep{Ferreira2018}. From the segmented region it calculates basic geometrical parameters (e.g. centroid, position angle, axis ratio) using light-profile moments. Then it performs photometry, measuring fluxes within ellipses with the aforementioned parameters (contextually masking out point sources over the ellipse annulus with a sigma clipping criterion). From the luminosity growth curve it establishes the Petrosian radius, inside which all the measurements are made. The \sersics fit is performed on the 1-D luminosity profile; for robustness, the 1-D outputs are used as inputs for a 2-D \sersics fit of the  galaxy pixels. Finally, it measures several morphometric parameters, e.g. concentrations, asymmetries, Gini and M20 \citep[the former is a coefficient quantifying the inequality among values of a frequency distribution, in this case of pixel values; the latter is the second order moment, i.e. the flux values weighted by the their square distance to the center, of the 20\% brightest pixels; see][]{Lotz2004}, entropy, spirality, curvature among others). In a forthcoming version, the luminosity profile curvature \citep{Lucatelli2019} will be used to provide a more robust input to a parametric model-based fit of the light profile, eventually replacing the 1-D \sersics fit as a metric, mainly to mitigate a long lasting problem of \sersics index determination (see discussion in EMC2022b). 
Only the $I_{\scriptscriptstyle\rm E}$ SS fit was provided.

\subsection{ProFit} 
\texttt{Profit} \citep{Robotham2017} is a software package designed to perform Bayesian two-dimensional photometric galaxy profile modelling. It consists of a low-level \texttt{C++} library accessible via a command-line interface and documented API, along with high-level \texttt{R} (v.3.6.1) and Python interfaces. The fitting process for each object starts running the source finder \texttt{ProFound} \citep{Robotham2018}\footnote{\url{https://github.com/asgr/ProFound}} on a $500\times500$ pixel cutout centred on each target, to create a segmentation map and find nearby sources requiring simultaneous modelling; the output also provides some reasonable initial guess for the profile solution. 
The actual fitting is then performed by the \texttt{Highlander} core software\footnote{\url{https://github.com/asgr/Highlander}}, which combines a genetic algorithm step with a CHARM \citep{Turchin1971, Smith1984} Markov chain Monte-Carlo (MCMC) process, repeated twice; each one is run for 100 steps (where model realisations are modified by the number of free parameters also). The CHARM algorithm is particularly useful on highly covariant parameter search, but it is computationally expensive, because a single iteration requires sampling all parameters. Since the kind of fitting used for \texttt{ProFit} is relatively low in the number of parameters, but sometimes quite highly covariant in the posterior, CHARM has proven to be a powerful exploration tool. The provided solution is the combination of parameters that generate the maximum likelihood given the per-pixel Data - Model residual.
The parameter priors are implicitly assumed to be uniform. Errors are estimated from the final MCMC run, with full covariance matrix information available. 

All requested data was provided.
Partially building upon the effort put in the EMC, the whole \texttt{ProFit} pipeline has recently been developed into a new package, \texttt{ProFuse} \citep{Robotham2022}.

\subsection{SourceXtractor++}
\texttt{SourceXtractor++} \citep{bertin_adassxxix,kuemmel_adassxxix}\footnote{\url{https://github.com/astrorama/SourceXtractorPlusPlus}} is a ground-up re-write of the widely used \texttt{SExtractor2} software \citep{Bertin1996}, written in \texttt{C++} with a strong focus on extensibility and model-fitting photometry; the software is under active development and the results submitted to the challenge represent a snapshot in this process (the version of \texttt{SourceXtractor++} used in the EMC is 0.12). 

Each \texttt{SourceXtractor++} run includes two stages, detection and measurement. The detection stage follows the same procedure as used in \texttt{SExtractor2}. 
Detection parameters need to be optimised for a compromise between the completeness of the true object list and the number of spurious objects extracted or deblended; over-extraction of sources impacts the performance of the run-time required, and may also reduce the accuracy of morphological measurements if objects are over-deblended. The parameters for the EMC were tuned aiming for good overall performance, and therefore not for reaching $100\%$ completeness of the input $5\sigma$ source list. 
The SS and DS simulations produce slightly different distributions in apparent extent and surface brightness for the galaxy images, and so the parameters governing detection were optimised separately for the different simulations.

Measurement in one or several bands is controlled via a Python configuration file with flexible model fitting at its core, which allows for the simultaneous source analysis over a large number of FITS files with different pixel grids. Various components (point source, exponential disc, free \sersics, etc.) can be used individually or in combination; reasonable priors must therefore be provided to the fitting engine in order to cover the range of parameter values and provide sensible fits. The chosen priors for the EMC are described in Appendix \ref{AppSEpriors}. 

Noticeably, the \texttt{SourceXtractor++} pipeline used for the EMC includes a pre-processing of the images, namely the extraction and usage of PSFs from images, performed using the \texttt{PSFEx} software \citep{Bertin2011}, and a re-downsampling to the original pixel scales of the NIR and LSST images ($\ang{;;0.3}$ and $\ang{;;0.2}$, respectively). This procedure was allowed by the guidelines of the EMC, given that no additional input data were used. However, given that no other participants did anything similar, we decided to also check the performance of the package on standard, non-pre-processed data, finding overall good agreement with a few differences that should be taken into account when considering the overall processing cost of the pipeline. We discuss this in Appendix \ref{Se} (see also EMC2022b).
Additionally, \texttt{SourceXtractor++} priors were obtained by comparing the output distribution of a given morphological parameter with the equivalent distribution for the provided samples of the input true catalogues; the priors on parameters were iteratively adjusted under the constraint that a simple analytical transfer function is required to map each distribution to a Gaussian. Each parameter was calibrated independently, without including covariances; only the statistical distributions were used (i.e. there was no object-by-object comparison in the process). A detailed description of the calibration of the priors is given in Appendix \ref{AppSEpriors}.
All requested data were provided, except for the multi-band DS fit with free $n_{\rm bulge}$.

\section{Diagnostic metric} \label{fom}

Given the high dimensionality of the output data, a straightforward comparison of the results was not feasible. In order to obtain a reasonably comprehensive overview of the quality of the performance, we defined an ad-hoc metric.
The participants provided catalogues that were matched to the input ones by means of the unique ID of each source. Then for each run we proceeded to estimate the difference between the input and the measured fluxes of each object, and computed averaged statistical diagnostics. 

Importantly, to compute such statistics we used a subset of sources from the $5\sigma$-limit list, including only those for which all software packages provided a meaningful fit. Recall that the default request was to provide a fit for all the sources in the nominal $5\sigma$ list; with some minor caveats and exceptions mentioned in Sect. \ref{codes}, all participants obliged to this, and some also provided results for lower S/N sources. However, not all the fits were successful (i.e. some were given as \texttt{NaN} or default values in the output catalogues), and some were flagged as `bad' or `unreliable' in one or more codes. In general, these sources were not included in the lists used to evaluate the accuracy of the results, which therefore only included the objects for which all the software packages had provided a reliable fit. 
An important exception is the case of \texttt{Galapagos-2}, which outputs several quality flags describing which component of a galaxy can be considered as reliably fit. In particular, for the DS runs a total of five flags were provided: \texttt{USE\_FLAG\_SS} indicates that the single \sersics fit is usable, as it did not run into fitting constraints; \texttt{USE\_FLAG\_BULGE\_CONSTR} and \texttt{USE\_FLAG\_DISK\_CONSTR} serve the same purpose for the bulge and disc components, respectively; in addition, \texttt{USE\_FLAG\_BULGE\_BRIGHT} and \texttt{USE\_FLAG\_DISK\_BRIGHT} indicate whether the bulge and the disc are relatively bright enough ($\btm>0.2$ and $\btm<0.8$, respectively), that their fit could in general be trusted, with the additional difficulty that \bts itself is defined via such a fit. However, all of these flags were ignored in our analysis, to avoid excessive complications in the definition of the common set of fitted sources within the submissions. While this choice certainly impacts the statistics of the results, because galaxies are taken into account that are known to violate fitting constraints (they are 4\% and 13\% of the total number of objects, for single and double component fits, respectively), we found that the effects on the overall analysis was marginal. We stress that a general user of \texttt{Galapagos-2} shall consider these flags, according to their purposes; a thorough description of the flags is provided in \citet{Haussler2022}. See the Appendix of EMC2022b for a comprehensive discussion on this topic.

To estimate the impact for each software tool, an additional term evaluating the completeness fraction of the output catalogues with respect to the full 5$\sigma$ list was included in the global metric, as described below. 


To build the metric, we started considering the relative flux difference of each object with respect to the input true flux, i.e. $\delta f=(f_{\rm meas}-f_{\rm true})/f_{\rm true}$. We then used $\delta f$ to evaluate three diagnostics: the bias $\mathcal{B}$; the dispersion $\mathcal{D}$; and the outlier fractions $\mathcal{O}$. In summary, the three diagnostics were first averaged over the sources belonging to bins of input magnitude (we used 15 bins to divide the full interval of simulated magnitudes, from 14 to 28), to quantify the impact of S/N; then these averages (normalised with weighting factors) were summed, and further combined with the completeness $\mathcal{C}$ diagnostic, finally yielding a global score $\mathcal{S}$ for each field and realisation. For $I_{\scriptscriptstyle\rm E}$, the values computed for each simulated field were finally averaged to obtain a single figure; while this is not strictly correct from a statistical point of view, given that the results in the different fields were very similar we assume that the outcome is sufficiently accurate. 
In more detail, the four quantities were defined as follows.

\begin{itemize} 

\item The bias $\mathcal{B_{\rm bin}}$ is the median of $\delta f$ in each bin of input magnitude, computed considering only the objects having $|\delta f|\leq5 \sigma_{\rm true}$, where $\sigma_{\rm true}$ is the standard deviation of $\delta f$ for an ideal distribution of fluxes, that we obtained by perturbing the input true values with a random realisation of observational Gaussian noise consistent with the expected depth of each image (we imposed a minimum value corresponding to $\delta f = 0.02$ (2\%), to avoid unrealistically small values of $\sigma_{\rm true}$ at the bright end). An unbiased measurement would yield $\mathcal{B_{\rm bin}}=0$. We then define the average value of the bias as the weighted mean of its values across the magnitude bins, $\mathcal{B} = \sum_{\rm bins} w_{\rm bin} \mathcal{|B_{\rm bin}|}$, where  $w_{\rm bin}$ is a weighting factor given by the fraction of objects in each bin of true magnitude (to give more weight to highly populated bins) multiplied by the logarithm of the median S/N in that bin (to give more weight to the fit of bright objects). Note that while the values of $\mathcal{B_{\rm bin}}$ can be positive or negative, $\mathcal{B}$ is defined to be positive.

\item The dispersion $\mathcal{D_{\rm bin}}$ is the ratio between $\sigma_{\rm meas}$, i.e. the standard deviation of the distribution of $\delta f$ (again only including objects within 5$\sigma_{\rm true}$) and $\sigma_{\rm true}$, in each magnitude bin. The average dispersion is defined as $\mathcal{D} = \sum_{\rm bins} w_{\rm bin} \mathcal{D_{\rm bin}}$.

\item The outlier fraction $\mathcal{O_{\rm bin}}$ is the number of objects having $|\delta f|>5\sigma_{\rm true}$ divided by the total number of fitted objects in each bin. These objects fall outside the expected distribution, and we assume that their large bias is due to some systematic error in their measurement (e.g. strong contamination from neighbours, or catastrophic failure of the fit). Therefore they were not included in the statistics of ``well-behaved'' sources, and were instead isolated into a separate diagnostic. The average outlier fraction is defined as $\mathcal{O} = \sum_{\rm bins} w_{\rm bin} \mathcal{O_{\rm bin}}$.

\item The completeness $\mathcal{C}$ is simply the number of objects for which a successful fit was provided, divided by the total number of objects in the input list of ${\rm S/N}>5$ sources (we do not weight this quantity by the magnitude bins). A galaxy is considered not to be fit if there is no entry in the provided output catalogue, or if the challenge participant flagged that galaxy as a `bad fit' (see discussions above). Each software package has different ways of identifying unreliable fits, and we refer the reader to the publications describing each code for additional information. Here, we simply trusted the participants' verdict on the reliability of their fits. 

\end{itemize}

We point out that the definitions used in EMC2022b are very similar, but since it is difficult to construct  meaningful expectations for ideal perturbed distributions of morphological parameters (corresponding to the $\sigma_{\rm true}$ we use here for the fluxes), some differences were introduced. The interested reader should therefore pay attention to these details.

Finally, the global diagnostic $\mathcal{S}$ for each run is defined as 
\begin{equation}
 \mathcal{S}= (1-\mathcal{C}) + k_{\mathcal{B}} \mathcal{B}  +  k_{\mathcal{D}} (\mathcal{D}-1)  + k_{\mathcal{O}} \mathcal{O} 
\label{Svalue}
\end{equation}



\noindent where we subtract 1 from $\mathcal{D}$ when computing the final global statistics, because when the dispersion is `ideal' the ratio with $\sigma_{\rm true}$ is 1, and we want the value of all diagnostics to be close to zero for ideal fits. In this expression, the $k$ factors are multiplicative constants assigned to each of the three diagnostics in an attempt to reasonably weight their relative contributions. 

We chose $k_{\mathcal{B}}=20.0$, $k_{\mathcal{D}}=0.6667$ and $k_{\mathcal{O}}=5.0$. While these choices are to some extent arbitrary, it is worth pointing out that the large differences in these values do not reflect the actual weight given to each diagnostic; on the contrary, they were chosen exactly to try and reach a reasonable balance between the three weighted quantities. We argue that a fit might be defined as `optimal' if it has e.g. $\mathcal{C}=1.0$ (100\% completeness above 5$\sigma$), $\mathcal{B}<0.015$ (1.5\% median bias), $\mathcal{D}<1.33$ (dispersion no larger than 4/3 of that from the perturbed true fluxes), and $\mathcal{O}<0.1$ (10\% of outliers) in all bins of magnitude; in the case of these exact values, applying the chosen weights one gets $\mathcal{S}\simeq0.3+0.22+0.5=1.02$ (the bin weights $w_{\rm bin}$ are not relevant here). So, we see that if $\mathcal{S}\leq1$ the fit can be considered as optimal; $\mathcal{S}\leq1.33$ is very good; $1.33<\mathcal{S}\leq1.67$ is good; and $1.667<\mathcal{S}\leq2.0$ is acceptable. Finally, with this metric values of $\mathcal{S}$ much larger than 2 indicate a bad overall fit.
Note that when marginalizing the contributions for a 100\% complete fit, $\mathcal{S}=2.0$ can be due to: a 10\% overall offset; a 3$\sigma_{\rm true}$ standard deviation; or a 40\% outlier fraction. 

\begin{figure}[h!]
\centering
\includegraphics[width=0.49\textwidth]{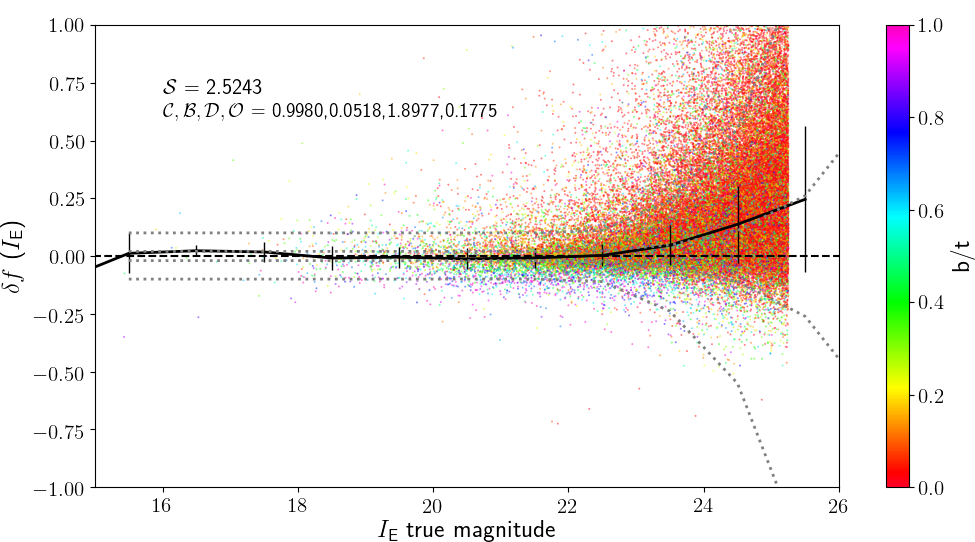}
\caption{Example of a diagnostic plot used to analyse each output catalogue provided by the participants to the EMC. Each dot shows the $\delta f$ value of a fitted galaxy, colour-coded by the \bts value from the original \texttt{Egg} catalogue (colour-bar on the right), as a function of the input true magnitude in the considered band. The black solid line is the running median of the distribution, which should be identically zero for a perfect fit. The dotted lines show the (positive and negative) 1 and 5 $\sigma_{\rm true}$ levels, used to compute the diagnostics described in the text and reported in the top left corner of the plot (see Sect. \ref{fom}).}
\label{trumpet}
\end{figure}

The diagnostics were evaluated automatically by means of Python scripts, but the results were also visualized graphically, to allow for sanity checks and for a quick grasp of any particular features. Figure \ref{trumpet} shows an example of a diagnostic plot that we used to analyse one of the provided output catalogues; similar plots were created for the outputs for each field and realisation that each participant provided. Each dot is a single fitted galaxy, and its $\delta f$ is plotted against its true input magnitude in the considered band. The dots are colour-coded by the true bulge-to-total ratio (which for the SS and RM realisations is a proxy for the \sersics index, $n=3\; \btm +1$). For each bin of magnitude, the median, standard deviation, and outlier fraction of the distribution were computed, and the values were then used to compute the diagnostics described above; the dotted lines show the $1\sigma_{\rm true}$ and $5\sigma_{\rm true}$ levels. Specific examples are given in Appendix \ref{AppCodes}.

\begin{figure}[h!]
\centering
\includegraphics[width=0.45\textwidth]{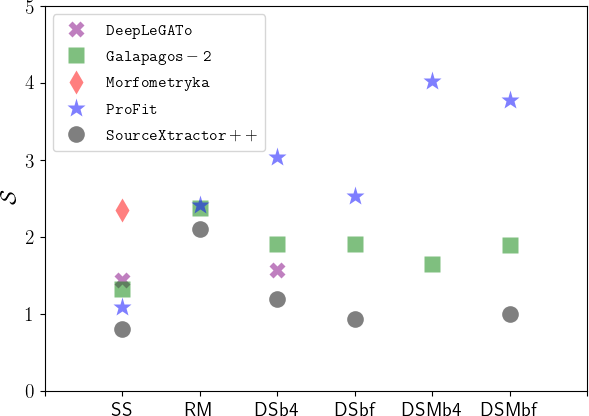}
\caption{Visual summary of the $\mathcal{S}$-values listed in Table \ref{codestab}.}
\label{summary}
\end{figure}

\begin{figure*}[h!]
\centering
\includegraphics[width=0.33\textwidth]{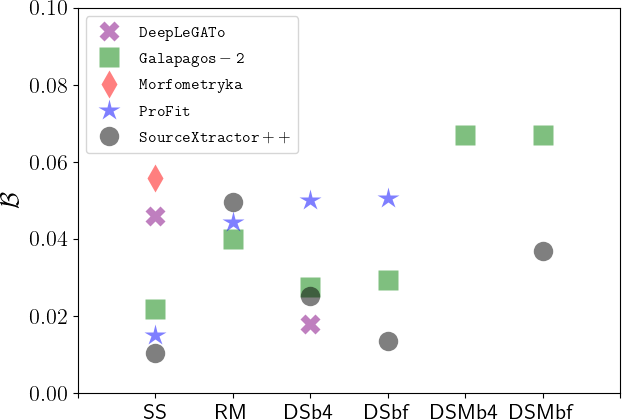}
\includegraphics[width=0.33\textwidth]{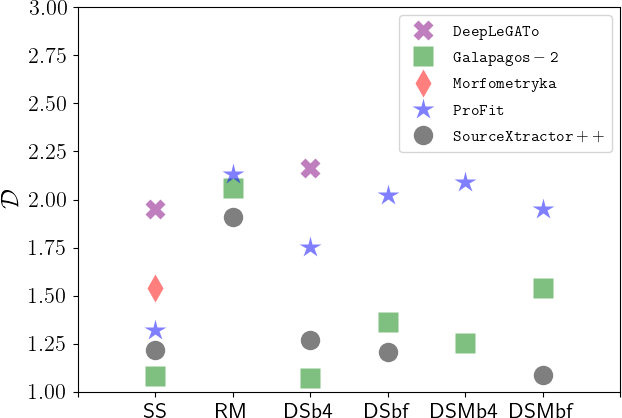}
\includegraphics[width=0.33\textwidth]{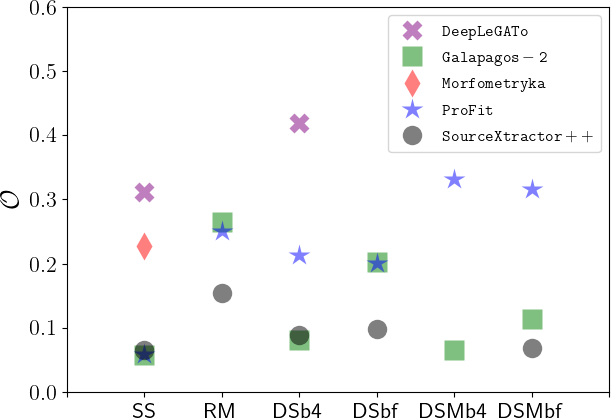}
\caption{Visual summary of the diagnostic quantities: (left to right) absolute value of the median bias ($\mathcal{B}$), average ratio of $\sigma_{\rm meas}$ and $\sigma_{\rm true}$ ($\mathcal{D}$), and outlier fraction ($\mathcal{O}$). In the bias panel, the points corresponding to \texttt{ProFit} DS multi-band runs are not shown, being off the scale (their values are 0.49 and 0.44, respectively). See text for details.}
\label{summaryb}
\end{figure*}

Because of a technical problem not related to the performance of the code and gone unnoticed during the run, the processing of F1 by \texttt{DeepLeGATo} was interrupted before the end of the input list, and the corresponding catalogue was therefore incomplete. To ease the comparisons, and considering that the problem was not caused by a bug in the code (since the processing of all the other fields ended smoothly), we decided to remove F1 entirely from the analysis process, after checking that this would not favour one of the codes with respect to the others.

Finally, as already mentioned the \texttt{Galapagos-2} runs on DS F0 were performed in a simultaneous multi-band fit, including $I_{\scriptscriptstyle\rm E}$ with the other bands; this caused the results to be significantly different from those in the other four $I_{\scriptscriptstyle\rm E}$ fields. To avoid any impact on the evaluation, and considering the many different approaches of the other participants (\texttt{DeepLeGATo} only provided the $I_{\scriptscriptstyle\rm E}$ fits, \texttt{Morfometryka} did not provide the fit, \texttt{ProFit} only provided non-simultaneous fits, and \texttt{SourceXtractor++} provided both a simultaneous and a separated fit), we decided to remove F0 from the analysis of $I_{\scriptscriptstyle\rm E}$ DS.

In summary, $I_{\scriptscriptstyle\rm E}$ SS and RM were analysed by averaging F0, F2, F3, and F4 (resulting in 212\,000 objects for SS, and 204\,229 for RM, where the bright galaxies were not simulated); $I_{\scriptscriptstyle\rm E}$ DS by averaging F2, F3 and F4 (207\,064 objects); and the other bands on DS F0 alone (because it was the only field simulated with a multi-band data set; it contains 70\,700 objects). 


\section{Results} \label{results}

\begin{table*}
\label{codestab}
\centering       
\begin{tabular}{|l| c| c| c| c| c| c|} 
\hline
Software package & SS & RM &  DSb4 & DSbf & DSMb4 & DSMbf \\
\hline

\texttt{DeepLeGATo} & 1.44 (0.97) & -- & 1.57 (0.96) & -- & -- & -- \\
\texttt{Galapagos-2} & 1.33 (0.90) & 2.37 (0.75) & 1.91/2.03 (0.95) & 1.91/2.51 (0.95) & 1.65/1.90 (0.54) & 1.90/2.00 (0.95) \\
\texttt{Morfometryka} & 2.35 (0.84) & -- & -- & -- & -- & -- \\
\texttt{ProFit} & 1.09 (1.00) & 2.42 (0.94) & 3.04 (1.00) & 2.53 (1.00) & 4.02 (0.73) & 3.78 (0.73) \\
\texttt{SourceXtractor++} & 0.81 (0.96) & 2.11 (0.87) & 1.19 (0.97) & 0.95 (0.97) & -- & 0.99 (0.97)\\
\hline




\hline
\end{tabular}
\caption{Software package, runs and $\mathcal{S}$-values considering the common list of sources (see text for details). Lower $\mathcal{S}$ means better performance. The acronyms in the columns refer to the different realisations described in Sect. \ref{sims}, with the following additional specifications for the DS runs: b4 = fixed \sersics index for bulge ($n=4$); bf = free \sersics index for bulge; M = multi-band (i.e. NIR and LSST bands, excluding $I_{\scriptscriptstyle\rm E}$; no M is for $I_{\scriptscriptstyle\rm E}$ band only). For the \texttt{Galapagos-2} DS runs, we give the numbers of the best and  worst performance (first and second number, respectively), corresponding to either the double \sersics fit or a single \sersics fit, which the software produces in all cases (see Appendix \ref{G2}). The numbers in parenthesis are the completeness $\mathcal{C}$, i.e. the fraction of sources from the input 5$\sigma$ list having a successful measurement in the output catalogue. } 
\end{table*}

In this section we discuss the results obtained with the different software packages. First of all, it is worth pointing out again that the complexity of the challenge caused a significant scatter in the interpretation of its goals and spirit by the participants. This caused substantial differences in the adopted approaches, level of processing and output formats between them. Together with the high dimensionality of the data set, this makes a direct and comprehensive comparison of the results very challenging. 
In other words, the strategies and techniques adopted by the participants influenced the overall accuracy of the provided output, and this must be taken into account in the analysis, to ensure a fair overview of each code's capabilities and limitations. Nevertheless, we believe it is possible to draw some interesting general conclusions from the comparison. We will describe the overall outcomes in the following, with some particular cases discussed in more detail, when necessary.

Individual diagnostic plots for all the different runs are available in an on-line interactive tool.\footnote{\url{https://share.streamlit.io/hbretonniere/euclid_morphology_challenge}} Further discussions on the results provided by each participating team are provided in Appendix \ref{AppCodes}, together with a summary of the computational times and memory workload required by each software package. 

\subsection{Global outcome}

In the following we separately analyse the three realisations SS, DS, and RM. We separate the multi-band data set from the $I_{\scriptscriptstyle\rm E}$-only DS fits, since the results are significantly different (we identify the multi-band case with the addition of the letter `M' to the acronym DS whenever necessary). The values of the global metric $\mathcal{S}$ for each realisation are listed in Table \ref{codestab} (where the values of the completeness factor $\mathcal{C}$ are also reported, to allow for a better evaluation of its impact), and shown in Fig.~\ref{summary}. A visual summary of the values of each diagnostic used to compute $\mathcal{S}$ (i.e. $\mathcal{B}$, $\mathcal{D}$ and $\mathcal{O}$), averaged over the magnitude bins and fields, is given in Fig.~\ref{summaryb}. For the multi-band case, these values are the average of the ones obtained in each of the nine individual bands. Note that in these plots the values are shown before the weighting by the $k$ factors in Eq. \ref{Svalue}, so the relative difference between the values obtained by any two codes for the three diagnostics does not straightforwardly reflect their final difference in $\mathcal{S}$. 

\begin{figure}[h!]
\centering
\includegraphics[width=0.47\textwidth]{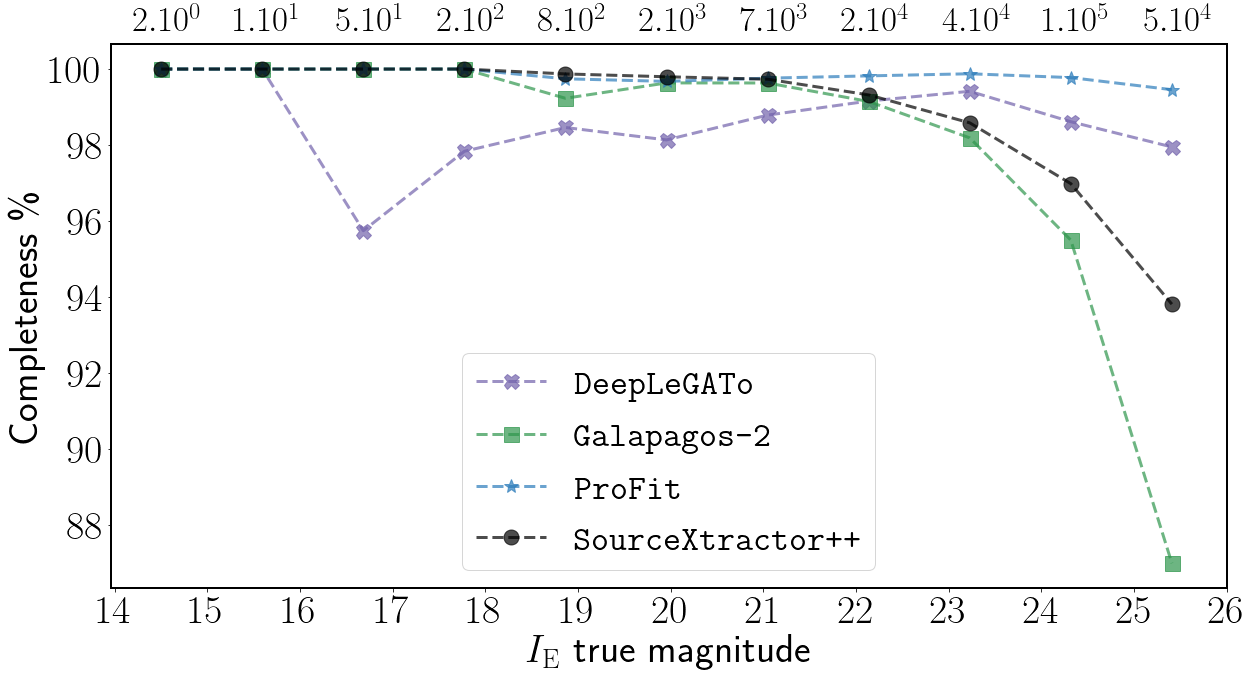}

\vspace{0.3cm}

\includegraphics[width=0.47\textwidth]{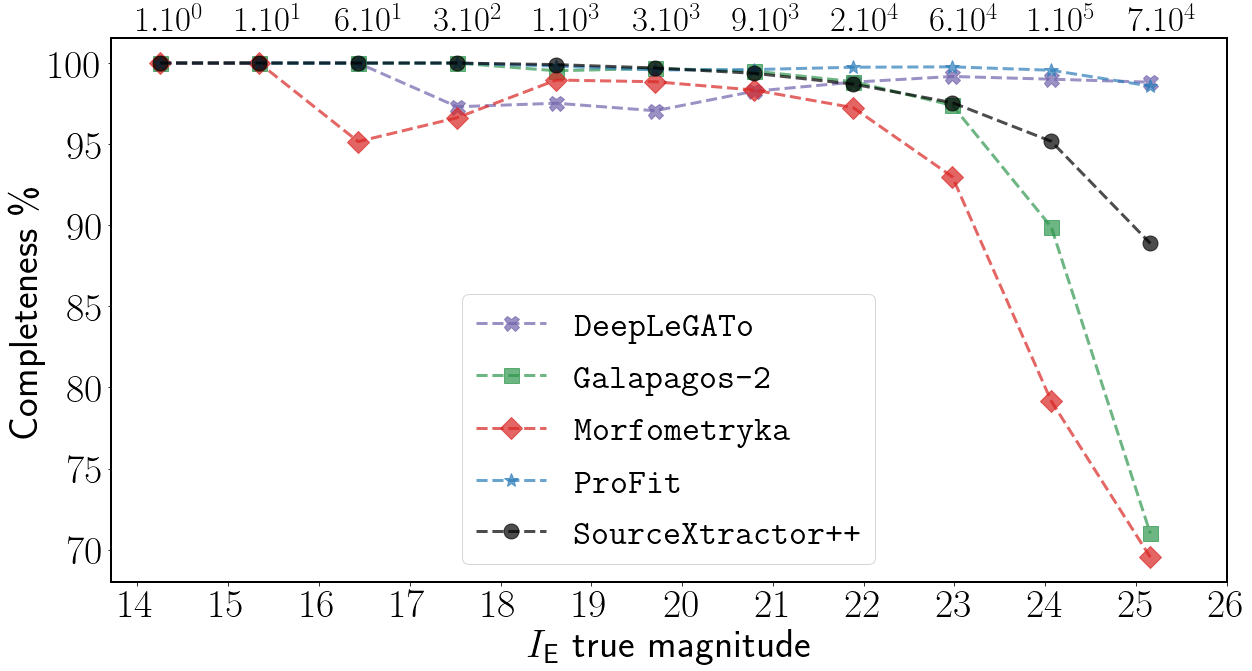}

\vspace{0.3cm}

\includegraphics[width=0.47\textwidth]{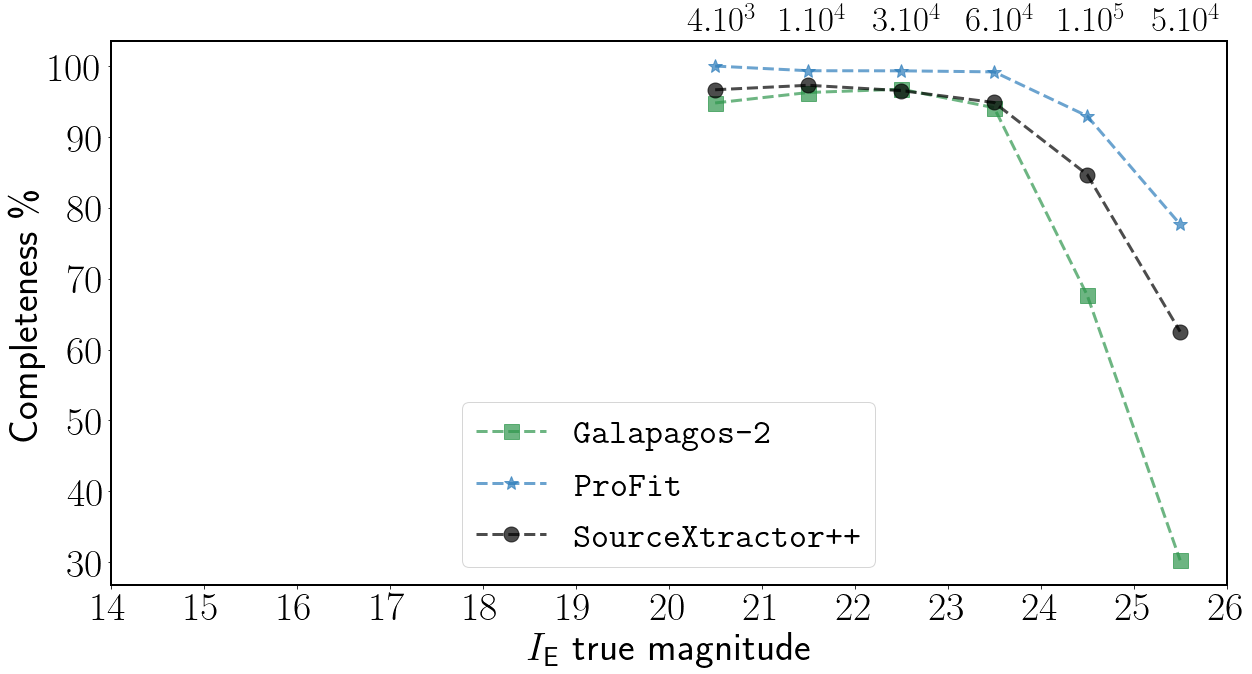}
\caption{Trends of the completeness $\mathcal{C}$ with input $I_{\scriptscriptstyle\rm E}$ magnitude of the output provided by the five participants to the challenge, for (top to bottom) DS, SS, and RM realisations (in the latter, points start at $I_{\scriptscriptstyle\rm E}=20.5$ because bright objects were not included in the simulated images).}
\label{compl}
\end{figure}

Because the global diagnostic quantities only provide a crude overview of the results, being averages over the input parameter space obtained with arbitrary weights, in Sect. \ref{trends} we also present a collection of summary plots (Figs. \ref{compl} to \ref{summ2}), showing the trends of the diagnostics as a function of the input magnitudes in all the cases of interest. 

We want to begin emphasizing that each code proved to have points of strength and of weakness, so the comparison of the global score $\mathcal{S}$ and of its factors is only intended as a quick overview, and should by no means be taken as a rigorous evaluation and ranking of the software packages. We fully acknowledge that this is a simplified view and therefore alternative metrics, tailored to specific science cases or assigning different weights to the considered diagnostics, could result in different conclusions.
That said, we can claim that with our metric all software packages provided acceptable to good results in at least some of the realisations. Some differences are present in a few cases (e.g. a particular realisation or band, the faint end of the simulated distribution of galaxies, etc.), but the outputs provided are typically fairly accurate. Unsurprisingly, in the DS realisation the best results were obtained in the $I_{\scriptscriptstyle\rm E}$ runs, for most of the software packages, given its high resolution and depth. The multi-band data set proved to be more demanding, because of the lower S/N and resolution of the images, and also of the resampling procedure which introduced noise correlations. 
All the participants reached at least 95\% global completeness in the $I_{\scriptscriptstyle\rm E}$ DS data set; in the other realisations, some lower scores were obtained (see the numbers in parenthesis in Table \ref{codestab}). 

In the $I_{\scriptscriptstyle\rm E}$-only runs, the best global results by all codes were obtained on the SS realisation, with all software packages reaching values of $\mathcal{S}$ between 1.0 and 1.7, with the exception of \texttt{Morphometryka} ($\mathcal{S}=2.54$), penalised by a significant bias $\mathcal{B}$ caused by systematic underestimation of fluxes of bright bulge-dominated galaxies, and of all objects at faint magnitudes (see discussions in Sect. \ref{trends} and in Appendix \ref{Mm}). 
On the DS realisation of the $I_{\scriptscriptstyle\rm E}$ band, \texttt{SourceXtractor++} and \texttt{DeepLeGATo} reached $\mathcal{S}<2.0$, although it must be recalled that the scheme adopted by the latter (dividing the sources according to their nominal S/N bin) introduces peculiar features in the distribution of $\delta f$ (see Appendix \ref{Dl}); \texttt{Galapagos-2} reached $\mathcal{S} \simeq 2.0$, while \texttt{ProFit} was penalised by a large $\mathcal{B}$ caused by a strong overestimation of fluxes at faint magnitudes (see Appendix \ref{Pr}). Interestingly, the fits with free $n_{\rm bulge}$ were in general slightly better than those with fixed $n_{\rm bulge} = 4$ (except for \texttt{Galapagos-2}), despite the fact that the simulated galaxies had indeed $n_{\rm bulge} = 4$.

Finally, the results for the RM realisation were less accurate than the ones for the other $I_{\scriptscriptstyle\rm E}$ images. This should be expected, given the inherently more difficult task of fitting analytical profiles on complex realistic morphologies. Interestingly, here the three codes that provided results have very similar trends; this seems to imply that when dealing with realistic galaxy shapes, the impact of prior calibrations, pre-processing of the images, and robustness of the algorithm become of secondary importance with respect to the inherent difficulty of the task.

The results for the multi-band data set show the evident impact of the different strategies followed by the participants. \texttt{SourceXtractor++} (which included  an image and PSF pre-processing pipeline, calibrated its priors on the provided example ground-truth data, and fitted all bands simultaneously) obtained optimal results, comparable to those for the $I_{\scriptscriptstyle\rm E}$ band fitted alone. \texttt{Galapagos-2} did perform a simultaneous fit, but without the image and PSF pre-processing obtained sub-optimal results. \texttt{ProFit}, which performed a separate fit on each band, had weaker performance due to not using information from the $I_{\scriptscriptstyle\rm E}$ image in the multiband fits: faint galaxies detected in $I_{\scriptscriptstyle\rm E}$ are likely to have very low S/N or could even be undetected in most NIR and LSST bands, and without the $I_{\scriptscriptstyle\rm E}$ parameter to constrain the fit, it is very difficult to properly model their light profiles in these bands. This outcome is important for highlighting how the synergy with \Euclid can significantly improve the accuracy of Rubin/LSST measurements, as already pointed out by several studies \citep[see e.g.][]{Rhodes2017,Capak2019}.

It is worth mentioning here that any statistical result showing a dependence on the bulge fraction of the sources is biased by the low fraction of simulated bulge-dominated galaxies with respect to the real Universe distribution (see Sect. \ref{sims}), so the overall performance on real data might be worse. 

\subsection{Trends of the diagnostics with input magnitudes} \label{trends}

In Fig.~\ref{compl} the trends of completeness $\mathcal{C}$ are shown for each software package as a function of $I_{\scriptscriptstyle\rm E}$ in bins of magnitude, for the three realisations DS, SS and RM (we remind the reader that for the latter only objects with $I_{\scriptscriptstyle\rm E}>20.5$ were simulated).

As one can expect, the fraction of successfully fitted objects decreases with increasing magnitude for all codes, since fainter objects are generally harder to detect and fit; the only exception is \texttt{DeepLeGATo}, for which bright objects are more often prone to failure. This is a consequence of the fixed stamp sizes that are used in the current version of the software, so that very bright (and large) objects may be larger than the stamp size.

Overall, down to $I_{\scriptscriptstyle\rm E}\simeq 23.5$ all codes successfully fit more than $95\%$ of the galaxies; the completeness then typically decreases, with the noticeable exception of \texttt{ProFit} which always stays close to $100\%$ for the SS and DS realisations. This is likely due to the different thresholds used to perform detection in the software packages (we recall that \texttt{Galapagos-2} uses \texttt{SExtractor} and \texttt{ProFit} uses \texttt{ProFound}), leading to different efficiencies in the actual detection of sources.

In each panel of Figs. \ref{summ0} to \ref{summ2}, the trends of one of the diagnostics $\mathcal{B_{\rm bin}}$, $\mathcal{D_{\rm bin}}$, and $\mathcal{O}_{\rm bin}$ are shown as a function of magnitude in the considered band, for all the software packages that have provided a fit in the relevant realisations. Some individual cases of particularly notable behaviour are described in more detail in Appendix \ref{AppCodes}. 

\subsubsection{Bias in $I_{\scriptscriptstyle\rm E}$ runs}

In the $I_{\scriptscriptstyle\rm E}$ DSb4 runs, for \texttt{Galapagos-2} and \texttt{ProFit}, the typical absolute bias in the measured flux $\mathcal{B_{\rm bin}}$ is below $1\%$ for bright sources ($I_{\scriptscriptstyle\rm E}<23$), increasing to $10$--$15$\% at the faint end. This is likely due to contamination from nearby brighter sources, and/or to the inherent difficulty in fitting low S/N objects with analytical profiles. \texttt{DeepLeGATo} has a less stable trend, with slightly larger values of bias for intermediate and bright sources, but a lower bias at the faint end. Finally, \texttt{SourceXtractor++} shows the most stable trend, without a strong overestimation at the faint end, despite a more pronounced average bias of $2$--$3\%$.

In the SS runs, in most cases we see more stable trends, with  monotonic trends towards overestimation of fluxes with decreasing brightness reaching about 5\% bias at the faint end. Noticeably, \texttt{ProFit} behaves differently, slightly underestimating intermediate magnitude sources before starting to overestimate at the faint end. An even more striking exception is given by \texttt{Morphometryka} (SS was the only provided output data set), which has a clearly declining monotonic trend, reaching $\delta f\simeq-15\%$ at $I_{\scriptscriptstyle\rm E}\simeq25.3$. 

The situation is completely different in the RM realisation, where all three of the codes that provided results have similar declining trends in the bias $\mathcal{B}$, with faint sources typically having fluxes underestimated by about $10$--$15$\% at $I_{\scriptscriptstyle\rm E}=24$--$25$. This is at odds with the results from the other realisations. We checked that this is not an issue in the simulations: while minor inconsistencies between the input true fluxes in the catalogue and the actual realisations of the sources in the images can be present because of the simulation method (see Sect. \ref{sims}), we found that the impact of this is negligible, with a typical mean offset of about 0.05\% and some scatter in the values that is not sufficient to explain the global trends of the measurements. We postpone further investigation on this topic to future work, given that it does not strongly impact the present analysis; the trend is almost identical for the three considered software packages, leaving the comparison among them essentially unchanged.

\subsubsection{Dispersion in $I_{\scriptscriptstyle\rm E}$ runs}
The dispersion $\mathcal{D_{\rm bin}}$ for all codes is typically comparable to $\sigma_{\rm true}$ at the faint end in all realisations. In DSb4, at $I_{\scriptscriptstyle\rm E}<23$ there is a hierarchy of performance with \texttt{DeepLeGATo} reaching values around 3.0 (again probably because of the limited dimensions of the stamps), \texttt{Profit} reaching around 2.0, \texttt{SourceXtractor++} 1.5 and \texttt{Galapagos-2} going from 0.5 to 1.5.  In SS this hierarchy is less pronounced with all codes, including \texttt{Morfometryka}, staying below $2.0$ at all magnitudes; the evident exception is again \texttt{DeepLeGATo}. In the RM case again we see similar (and sub-optimal) trends for the three codes that provided results.

\subsubsection{Outliers in $I_{\scriptscriptstyle\rm E}$ runs}

The outlier fraction $\mathcal{O_{\rm bin}}$ in DSb4 stays below 10\% at all magnitudes for \texttt{SourceXtractor}, and goes from very low values to 20\% at faint magnitudes for \texttt{Galapagos-2}. \texttt{Profit} reaches 30\%, while again \texttt{DeepLeGATo} suffers from the limited dimensions of the stamps, reaching 100\% outliers at the bright end, while remaining close to zero at $I_{\scriptscriptstyle\rm E}>23$. In SS, \texttt{SourceXtractor} and \texttt{Galapagos-2} stay below $10\%$ at all magnitudes, while \texttt{Morfometryka} has large values (around 50\%) at the bright end, likely because of a sub-optimal estimate of bulge-dominated sources (see Appendix \ref{Mm}). Again in the RM case there are no major differences between the quality of the performance, except for  \texttt{SourceXtractor} performing better on $I_{\scriptscriptstyle\rm E}<22$ objects.


\subsubsection{The free S\'ersic bulge case}

There is no substantial difference between the fixed and free $n_{\rm bulge}$ cases in the DS realisation. We do not show the trends for the free $n_{\rm bulge}$ case; as mentioned, the latter yields slightly better results than the fixed $n_{\rm bulge}=4$ case for \texttt{SourceXtractor++}, reducing the bias at all magnitudes, and for \texttt{ProFit}, which has better trends for all diagnostics; for \texttt{Galapagos-2}, the bias is almost identical in the two runs, while the dispersion and the outlier fraction have opposite trends (the free case having many more bright outliers but, simultaneously, a lower dispersion ratio for the few ``well-behaved'' sources), resulting in a similar final score (see Fig.~\ref{summaryb}).

\subsubsection{The multi-band data set}
\begin{figure}[h!]
\centering
\includegraphics[width=0.49\textwidth]{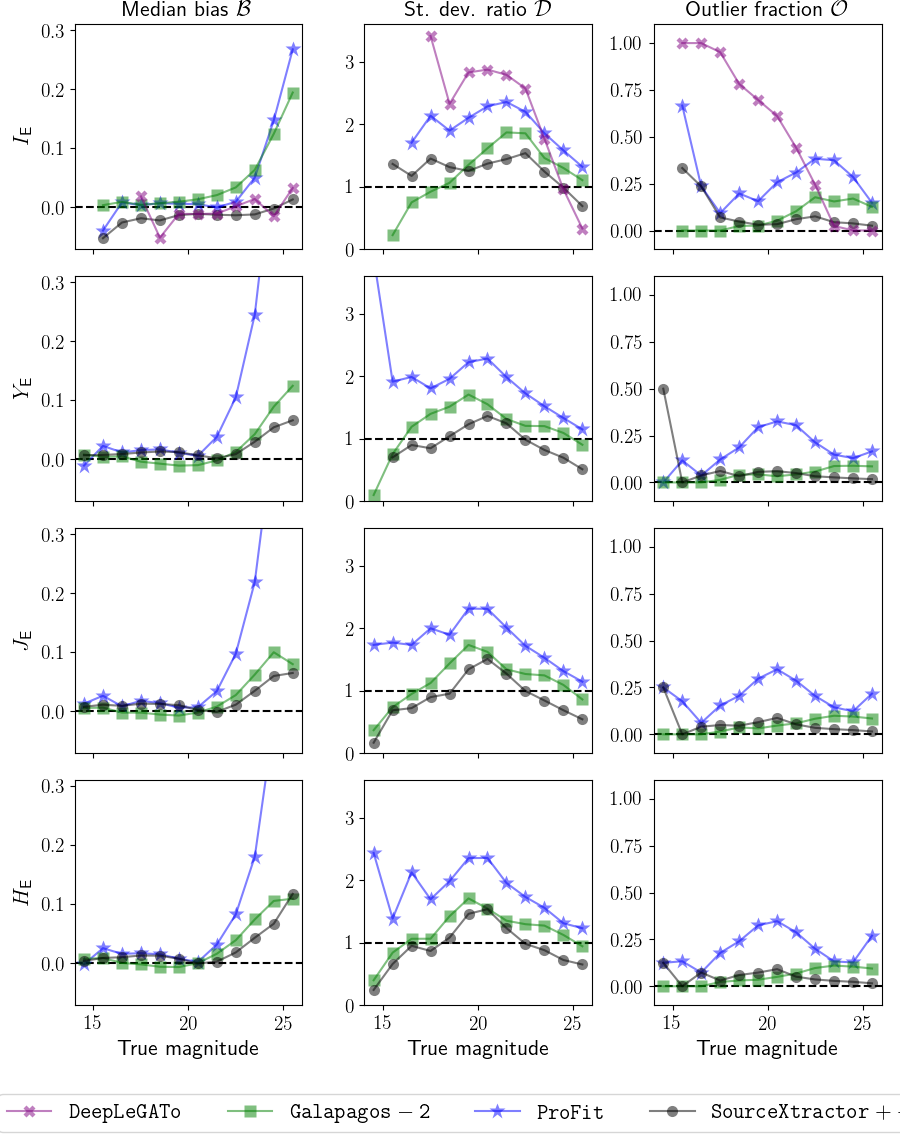}
\caption{Multi-band summary plots for $I_{\scriptscriptstyle\rm E}$ and the NIR bands. Left to right: median bias $\mathcal{B}$; dispersion (standard deviation) $\mathcal{D}$; and fraction of outliers $\mathcal{O}$. Top to bottom: $I_{\scriptscriptstyle\rm E}$, $Y_{\scriptscriptstyle\rm E}$, $J_{\scriptscriptstyle\rm E}$, and $H_{\scriptscriptstyle\rm E}$. Each line and colour correspond to a different code as indicated in the legends; note the different $y$-axis scales.}
\label{summ1}
\end{figure}

\begin{figure}[h!]
\centering
\includegraphics[width=0.49\textwidth]{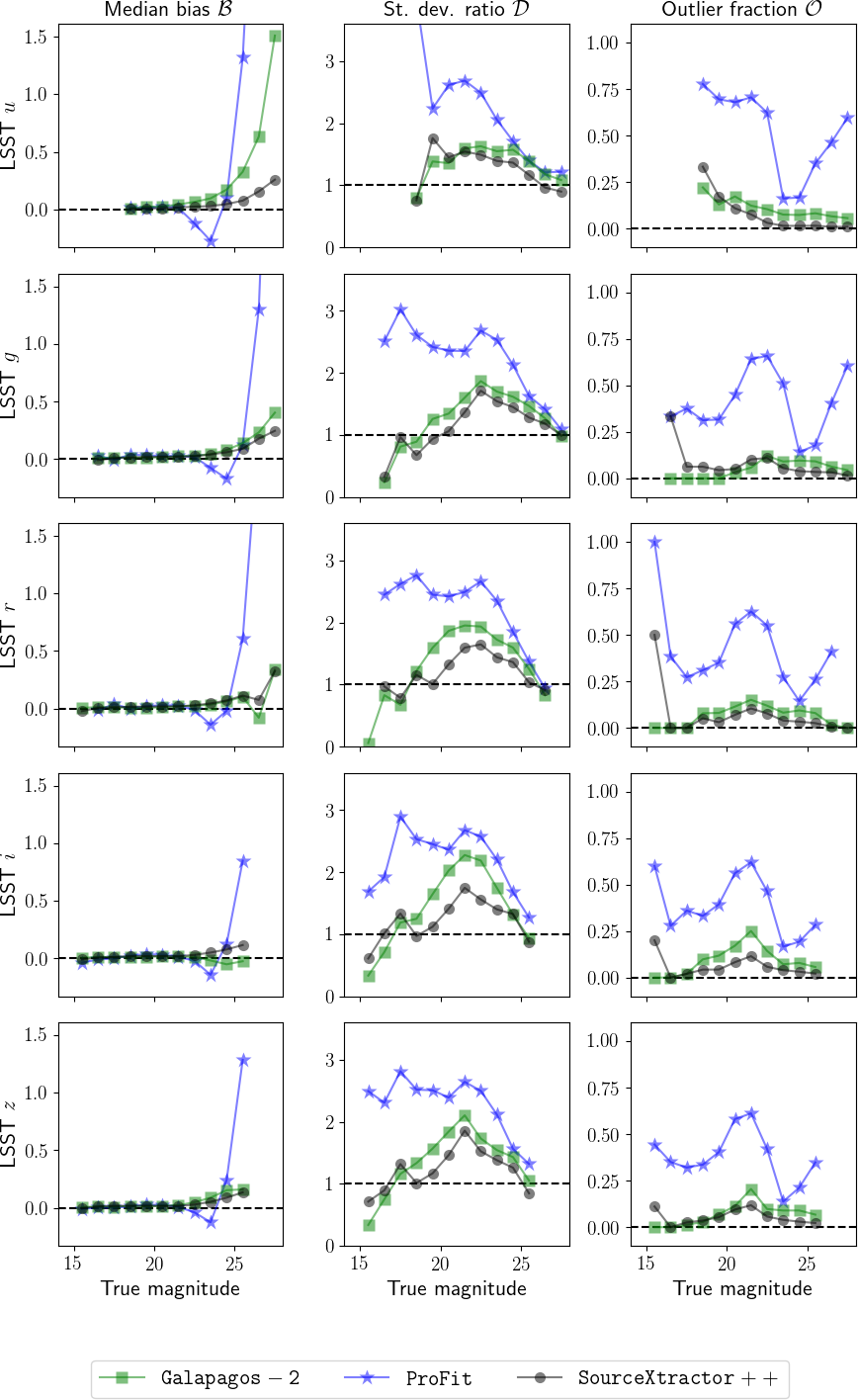}
\caption{Same as Fig.~\ref{summ1}, for LSST bands. Top to bottom: $u$, $g$, $r$, $i$, $z$. Note the different $y$-- axis scales.}
\label{summ2}
\end{figure}

In the multi-band data set (for which only \texttt{Galapagos-2}, \texttt{ProFit}, and \texttt{SourceXtractor++} provided results; see Figs. \ref{summ1} and \ref{summ2}) the trends are generally similar to those of the $I_{\scriptscriptstyle\rm E}$ case, with larger values of bias at the faint end for all codes in the NIR and LSST bands. In particular, \texttt{Profit} reaches $\mathcal{B_{\rm bin}}\simeq 1.5$ in the NIR bands, and approximately 4 in the LSST $u$ and $g$ bands, likely because of the independent fits performed on low S/N sources (we do not show the corresponding points in the plots, for readability); there is also a particular trend in the LSST bands, with sources being underestimated at intermediate magnitudes before turning to a strong overestimation at the faint end (we were not able to find an easy explanation for this).
\texttt{Galapagos-2} and \texttt{SourceXtractor} obtain better (and similar) results in NIR thanks to the simultanous fit, reaching $\mathcal{B_{\rm bin}}\simeq0.1$; in LSST they behave similarly as well, with \texttt{SourceXtractor++} performing clearly better only in the $u$ band.
Here the free $n_{\rm bulge}$ fit generally yields slightly worse results in dispersion and outlier fraction for \texttt{Galapagos-2}, and slightly better results for \texttt{ProFit}, than the fixed  $n_{\rm bulge}$ fit (not shown in these plots; see again Fig.~\ref{summaryb}). 

\begin{figure}[h!]
\centering
\includegraphics[width=0.49\textwidth]{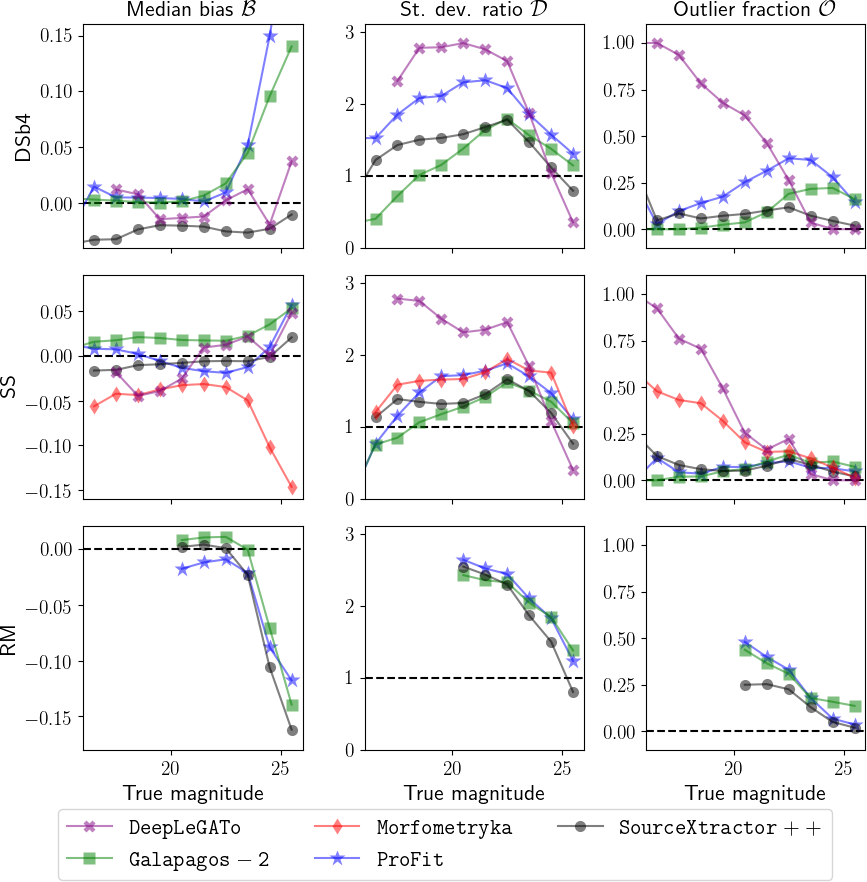}
\caption{DSb4 (top), SS (center), and RM (bottom) summary plots for $I_{\scriptscriptstyle\rm E}$. Left to right: median bias $\mathcal{B}$; dispersion (standard deviation) $\mathcal{D}$; and fraction of outliers $\mathcal{O}$. Each line and colour correspond to a different code as indicated in the legend; note the different $y$-axis scales.}
\label{summ0}
\end{figure}

\subsection{Separated bulge and disc estimates}

\begin{figure}[h!]
\centering
\includegraphics[width=0.49\textwidth]{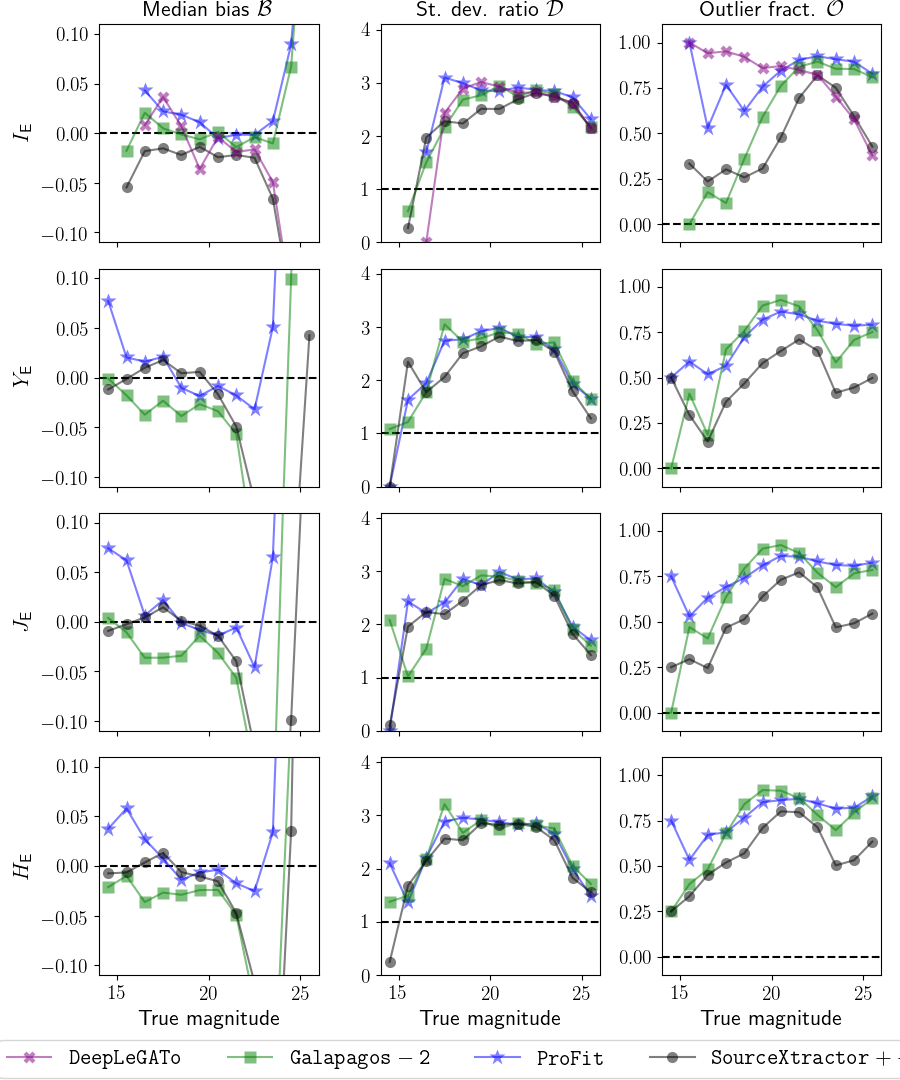}
\caption{Trends for bulges only, in the multi-band DSb4 fit, for $I_{\scriptscriptstyle\rm E}$ and the NIR bands. Left to right: median bias $\mathcal{B}$; dispersion (standard deviation) $\mathcal{D}$; and fraction of outliers $\mathcal{O}$. Top to bottom: $I_{\scriptscriptstyle\rm E}$, $Y_{\scriptscriptstyle\rm E}$, $J_{\scriptscriptstyle\rm E}$, and $H_{\scriptscriptstyle\rm E}$. Each line and colour correspond to a different code as indicated in the legends; note the different $y$-axis scales.}
\label{summ3a}
\end{figure}

\begin{figure}[h!]
\centering
\includegraphics[width=0.49\textwidth]{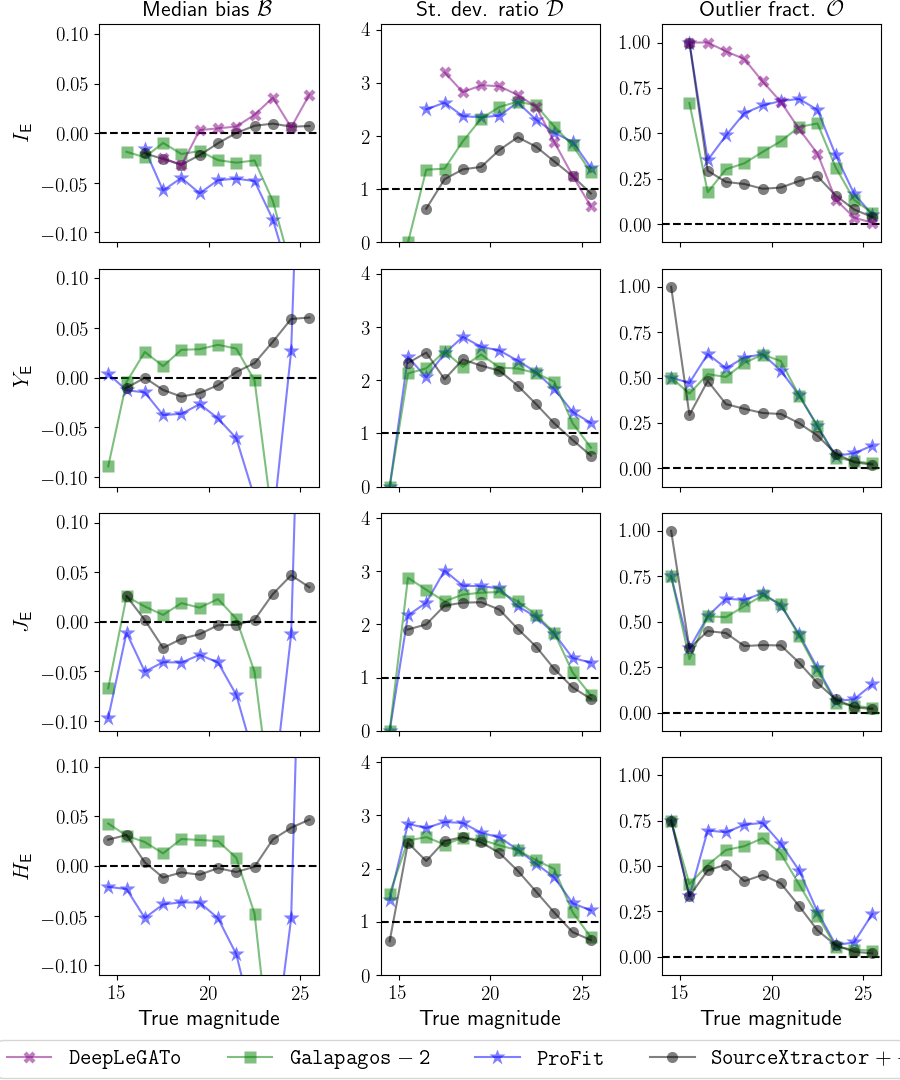}
\caption{Same as Fig. \ref{summ3a}, but fot discs only.} 
\label{summ4a}
\end{figure}

\begin{figure}[h!]
\centering
\includegraphics[width=0.49\textwidth]{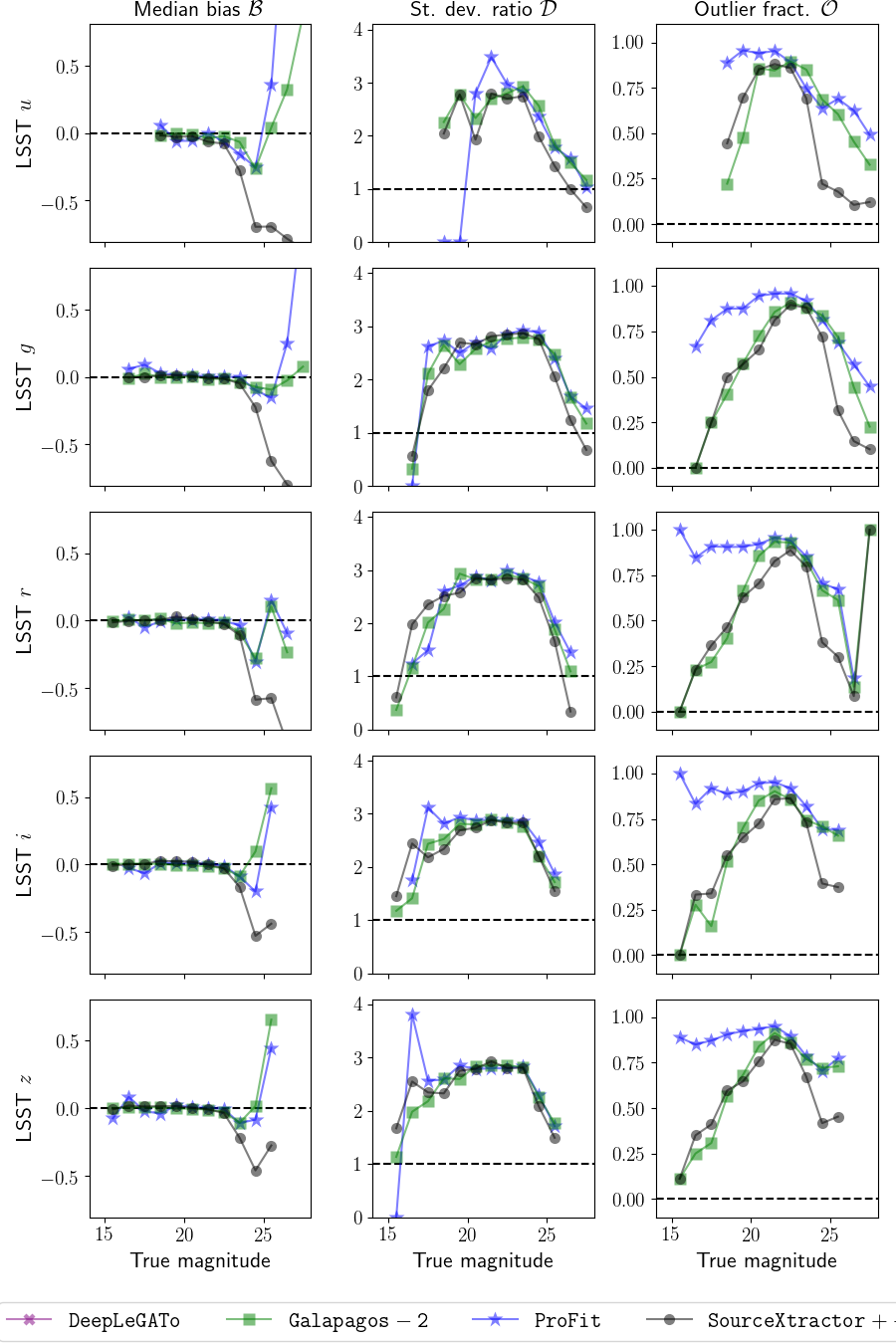}
\caption{Trends for bulges only, in the multi-band DSb4 fit (as in Fig.~\ref{summ3a}), for LSST bands. Top to bottom: $u$, $g$, $r$, $i$, and $z$.}
\label{summ3b}
\end{figure}

\begin{figure}[h!]
\centering
\includegraphics[width=0.49\textwidth]{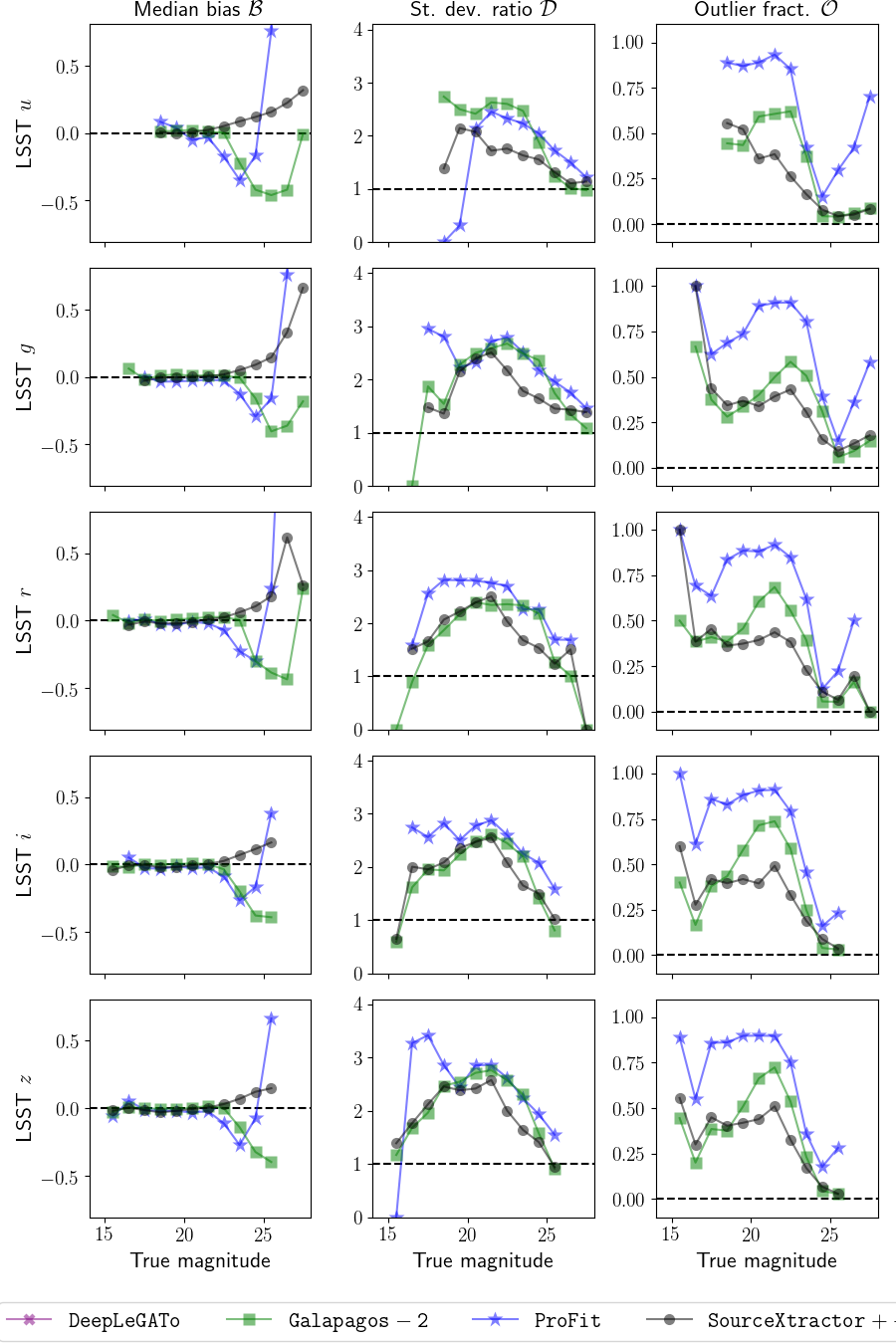}
\caption{Same as Fig. \ref{summ3b}, but for discs only.} 
\label{summ4b}
\end{figure}

So far we have considered total fluxes, but it is also instructive to investigate how the software packages performed in the separate flux measurements of the two components of each galaxy (bulge and disc) in the DS realisation. It is worth stressing that both estimates can be individually worse than the total flux one, if their sum is close to the true total value, but the partition among bulge and disc is not well recovered; this is linked to the accuracy of the morphological parameter estimates, discussed in EMC2022b.

In Figs. \ref{summ3a} to \ref{summ4b} we show the summary plots for bulges and discs separately. 
In $I_{\scriptscriptstyle\rm E}$, evident features are the bias trends for \texttt{Galapagos-2} and \texttt{ProFit}, both overestimating the flux of bulges at the faint end and underestimating that of discs at all magnitudes, by approximately 5\% down to $I_{\scriptscriptstyle\rm E}\simeq22$ and then getting worse; \texttt{DeepLeGATo} and \texttt{SourceXtractor++} show a strong underestimating trend for faint bulges, while showing a reasonable accuracy for discs at all magnitudes (\texttt{DeepLeGATo} overestimating their flux by a few percent). The dispersion and the outlier fraction are generally larger for bulges (often reaching very high values) than for discs, in particular at intermediate and faint magnitudes, where discs show nearly optimal values. 
For bulges, the dispersion has very similar trends for all codes, while for discs \texttt{SourceXtractor++} performs better, followed by \texttt{Galapagos-2}, while \texttt{ProFit} and \texttt{DeepLeGATo} suffer at the bright end. \texttt{SourceXtractor++} also yields the best performance concerning the outlier fraction, both for bulges and for discs.

Similar observations can be made for the other bands. In the NIR bands, the bias of bulges for \texttt{Galapagos-2} and \texttt{SourceXtractor++} show a similar underestimation of 3--5\% at intermediate magnitudes, and all codes show a strong overestimation at the faint end, particularly dramatic for \texttt{ProFit}. For discs, \texttt{SourceXtractor++} stays close to zero quite firmly, \texttt{Galapagos-2} shows an underestimation at the faint end balancing out the bulge overestimation, while \texttt{ProFit} again shows an overestimation. In LSST bands, \texttt{Galapagos-2} shows the same opposite trend for bulges and discs at faint magnitudes; interestingly, \texttt{SourceXtractor++} tends towards the reverse (underestimating bulges and counterbalancing with an overestimation of discs). \texttt{ProFit} still shows the strong rising trend in all cases.

The trend of the dispersion for bulges is very similar for the three codes, all having $\mathcal{D}\simeq2.0$--$3.0$ for the bulk of the bins in all bands; for discs, better values are found towards the faint end, again with \texttt{SourceXtractor++} performing slightly better than the other two codes. Finally, bulges typically have very high outlier fractions, typically above $50\%$ in all bins in the NIR bands, while having lower values at the bright end in the LSST bands, but always peaking at values above $75\%$ at intermediate magnitudes; on the other hand, discs have a lower fraction of outliers at the faint end.

\subsection{Colours in the multi-band data set}

\begin{figure*}[h!]
\centering
\includegraphics[width=0.9\textwidth]{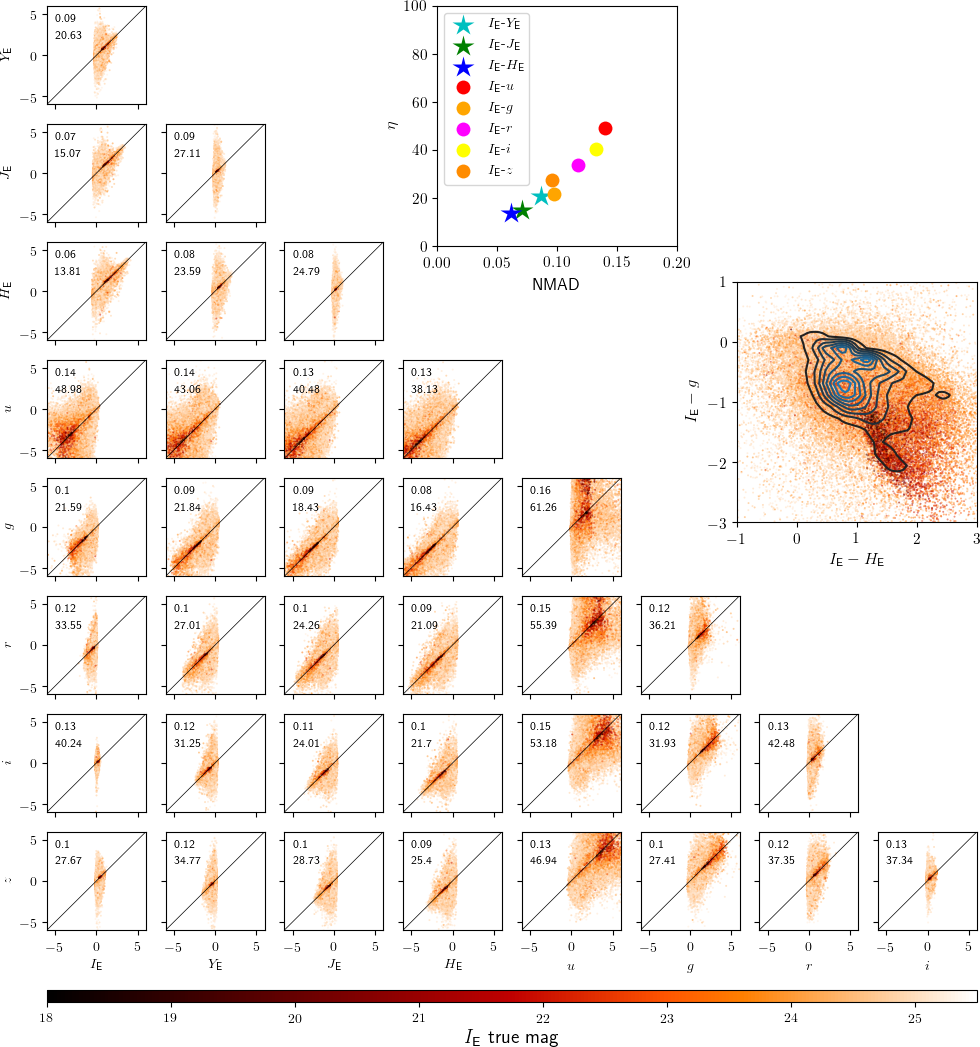}
\caption{Colours estimates for \texttt{Galapagos-2}. Each small sub-panel on the left part of the plot shows the measured versus true colour (magnitude difference) for a pair of bands: the colour is always $x$-axis label $-$ $y$-axis label. The numbers in each panel are the NMAD and $\eta$, as defined in the text. The upper larger panel shows the values  of NMAD and $\eta$ for the colours of all bands with respect to $I_{\scriptscriptstyle\rm E}$, as reported in the legend. The rightmost larger panel shows the $I_{\scriptscriptstyle\rm E}-g$ versus $I_{\scriptscriptstyle\rm E}-H_{\scriptscriptstyle\rm E}$ colour-colour diagram; points are the measured colours for individual galaxies, while the black lines are density contour levels of the corresponding true colours distribution.}
\label{colors1}
\end{figure*}
\begin{figure*}[h!]
\centering
\includegraphics[width=0.9\textwidth]{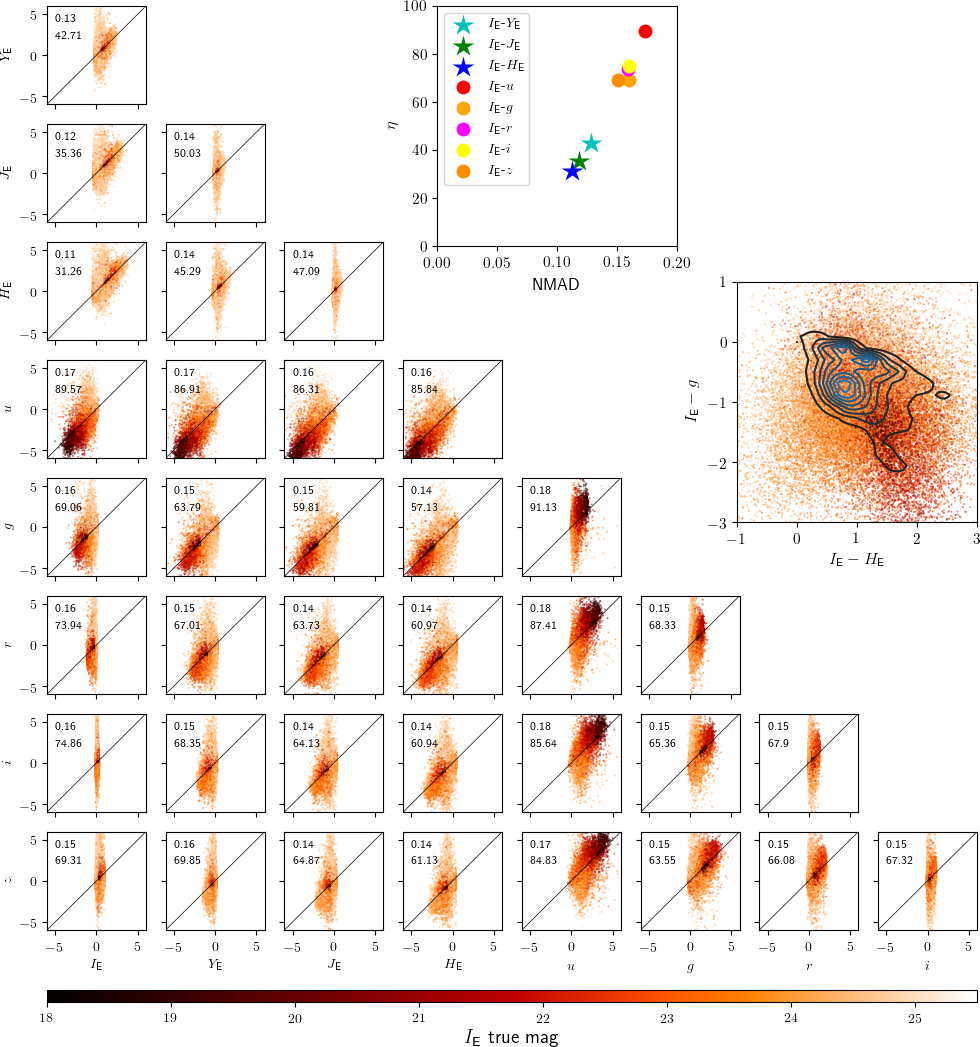}
\caption{Same as Fig.~\ref{colors1}, for \texttt{ProFit}.}
\label{colors2}
\end{figure*}
\begin{figure*}[h!]
\centering
\includegraphics[width=0.9\textwidth]{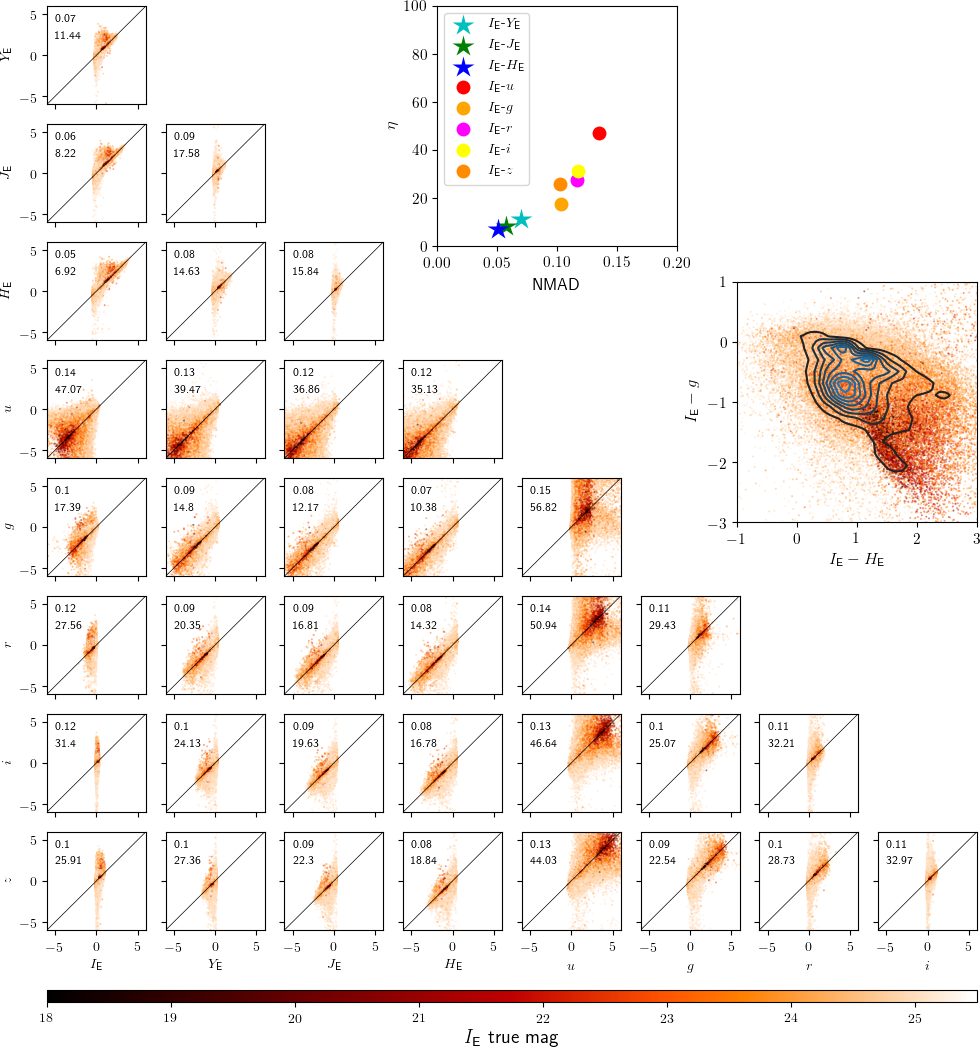}
\caption{Same as Fig.~\ref{colors1}, for \texttt{SourceXtractor++}.}
\label{colors3}
\end{figure*}


Thus far, we have focused on the accuracy of the flux estimates. However, for the multi-band data set it is also interesting to examine how accurately the three software packages that provided results were able to recover flux ratios among the bands, i.e. colours. Here too, the multi-dimensionality of the data requires some effort to summarize the results in a few informative plots. In each small sub-panel of Figs. \ref{colors1} (\texttt{Galapagos-2}), \ref{colors2} (\texttt{ProFit}), and \ref{colors3} (\texttt{SourceXtractor++}), we show the comparison between a measured colour and its corresponding true value. For each fitted galaxy (colour-coded by its true $I_{\scriptscriptstyle\rm E}$ magnitude) the measured colour is the $y$-axis coordinate, and the true colour is the $x$-axis coordinate (so a perfect colour estimate would result in a diagonal line from the bottom left to the top right corners; for reference, we plot this line in each panel). The figures show all possible combinations between the nine bands, giving an overview of the results. In each small sub-panel, the considered colour is given by the $x$-axis band label and the $y$-axis band labels, so for example the first sub-panel in the upper left corner is the $I_{\scriptscriptstyle\rm E}-Y_{\scriptscriptstyle\rm E}$ colour, the one below it is the $I_{\scriptscriptstyle\rm E}-J_{\scriptscriptstyle\rm E}$ colour, and so on. A distribution of points with a narrow vertical strip, which is often seen (e.g. in colours including $I_{\scriptscriptstyle\rm E}$), imply that while the true colours are close to zero, the measured ones have a large dispersion; if the distribution of points follows the diagonal the colour is estimated with good accuracy.

In each panel the values of two statistical diagnostics are also shown. Defining $\delta c=|{\rm colour}_{\rm meas}-{\rm colour}_{\rm true}|/(1+{\rm colour}_{\rm true})$, the first number is the normalised median absolute deviation, NMAD $=1.48 \mathcal{M}(\delta c)$ considering the sources with $\delta c \leq 0.2$ ($\mathcal{M}$ is the median of the distribution); and the second one is $\eta$, the percentage of outliers having $\delta c>0.2$. The upper bigger panel in the figures shows these two values for the colours including $I_{\scriptscriptstyle\rm E}$, as reported in the legend. As a further example, the lower bigger panels show a colour-colour diagram, $I_{\scriptscriptstyle\rm E}-g$ versus $I_{\scriptscriptstyle\rm E}-H_{\scriptscriptstyle\rm E}$; again, plotted are all measured colours of individual galaxies, colour-coded by their input $I_{\scriptscriptstyle\rm E}$ magnitudes; the density contours shown as solid lines represent the true colour distribution.

Obviously, fainter sources have more scattered distributions. In all cases, the colours including the $u$ band are clearly the least accurate, with large NMAD and $\eta$; this is easily understandable given the fact that deep-sky galaxies are typically fainter in blue bands, because of intrinsic properties, dust absorption, and cosmological redshifting; also, the band itself is quite shallow ($m_{\rm lim}=23.6$ at 10$\sigma$).

These diagrams again show how the impact of the different approaches used by the three participants affect the results. \texttt{Profit}, using independent fits in each band, obtains the least accurate results both in terms of NMAD and $\eta$. \texttt{Galapagos-2} and \texttt{SourceXtractor++} perform significantly better, probably because of the multi-band simultaneous fit. Between the two, \texttt{SourceXtractor++} yields slightly better results, most likely due to the careful prior calibration used (see Appendix \ref{AppSEpriors}).


\subsection{Uncertainty budgets} \label{errors}

\begin{figure}[h!]
\centering
\includegraphics[width=0.49\textwidth]{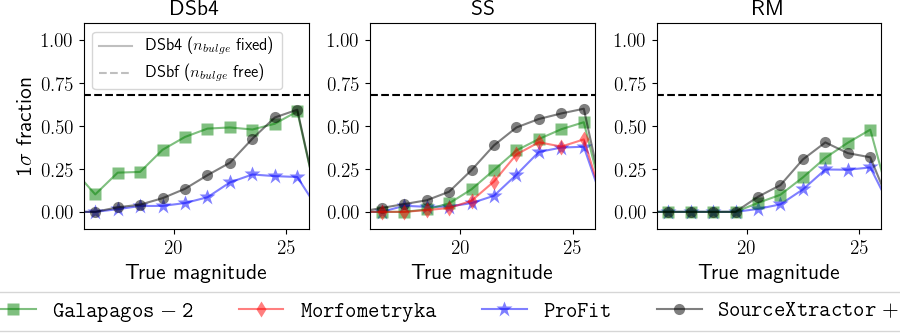}
\caption{Summary plots showing the fraction of sources satisfying the relation $|f_{\rm meas}-f_{\rm true}|<\sigma_{\rm meas}$ in each magnitude bin, for the $I_{\scriptscriptstyle\rm E}$ band runs. Expected fractions should be equal to 0.683 (horizontal dashed line). Left to right: DSb4, SS, RM. Each symbol and colour correspond to a different code, as indicated in the legends.}
\label{errors1}
\end{figure}

\begin{figure}[h!]
\centering
\includegraphics[width=0.49\textwidth]{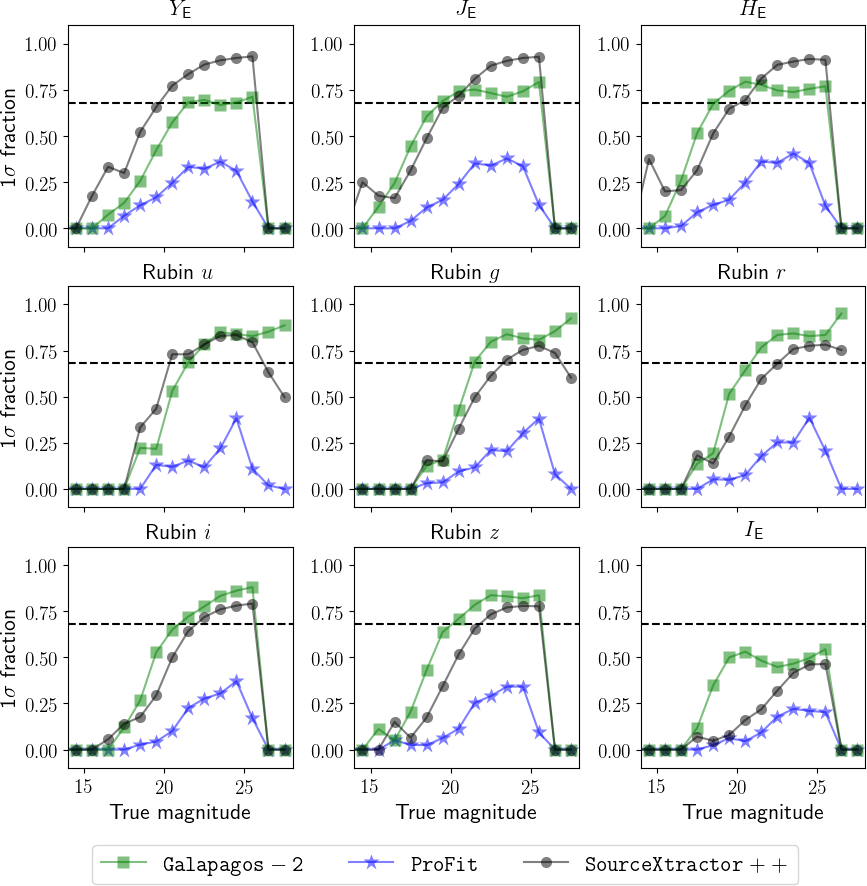}
\caption{Same as Fig.~\ref{errors1}, for the DSb4 multi-band run. Each panel refers to a band, as indicated in the titles (note that $I_{\scriptscriptstyle\rm E}$ is the last panel, for readability of the plot).}
\label{errors2}
\end{figure}

We computed a final quantity of interest, the fraction of sources that have a measurement of flux consistent with the true value within their nominal uncertainty budget, as provided in the output catalogues: i.e., $|f_{\rm meas}-f_{\rm true}|<\sigma_{\rm meas}$ \citep[e.g.][]{Haussler2007}. This fraction should in principle be 68.3\% (assuming a Gaussian distribution of the measurements, a reasonable assumption given that the noise map is predominantly Gaussian at least when the photon noise is not relevant, i.e. for intermediate and faint galaxies).

For the codes that provided magnitudes, we computed the errors on fluxes as $\sigma_f = 0.921  f\, \sigma_m$. For the double component runs, we estimated a global uncertainty with the usual approach, i.e. $\sigma_{\rm tot} = \btm\; \sigma_{\rm bulge} + (1-\btm)\, \sigma_{\rm disc}$, since we cannot consider the errors on the two components as independent.
We show the results in Figs. \ref{errors1} ($I_{\scriptscriptstyle\rm E}$) and \ref{errors2} (multi-band DSb4). It is evident that uncertainties are typically underestimated for bright sources despite their per pixel RMS being enhanced by the contribution of photon noise, while for faint sources they are mostly correct, in a statistical sense. 

A possible explanation for these trends is that bright sources are fit with less uncertainty because of their high S/N, but the uncertainty budget (which is based on the information in the RMS map) does not account for systematic errors that cause the actual offsets in the measurements. Note that in the evaluation of the bias $\mathcal{B}$ we considered the \textit{relative} offset $\delta f$, which is typically small for bright sources even when the absolute offset $|f_{\rm meas}-f_{\rm true}|$ is large. 

At the faint end, \texttt{ProFit} yields substantially less accurate results than \texttt{Galapagos-2} and \texttt{SourceXtractor++}, in all bands. It must be pointed out that the version of \texttt{DeepLeGATo} that was used for the EMC did not include a consistent algorithm for the uncertainty budget estimation. Therefore, we have not included \texttt{DeepLeGATo} in this analysis. A new version including error estimation is under development. The uncertainty estimation in \texttt{Morfometryka} takes into account the RMS map, however it only represents the fit parameters uncertainty, since this is extracted from the covariance matrix of the fit. For fluxes and magnitudes, the quoted uncertainty is only due to the fitted radius and underlying region used to integrate the flux.

\section{Conclusions} \label{conclusions}

This is the first of two papers presenting and discussing the Euclid Morphology Challenge, a project aimed at comparing the results of model-fitting software packages on a set of simulated \Euclid observations, with the aim of providing a first quantitative study to define the best-suited algorithm to be included in the \Euclid Science Ground Segment processing pipeline. Five model-fitting tools (\texttt{DeepLeGATo}, \texttt{Galapagos-2}, \texttt{Morfometryka}, \texttt{ProFit}, and \texttt{SourceXtractor++}) were tested in the challenge, providing partial or complete results. It is worth stressing that the results obtained for the EMC and discussed in this paper and in EMC2022b were used by the developers to improve and in some cases significantly upgrade the software packages, which is \textit{per se} a relevant outcome of this work.

While the companion paper (Euclid Collaboration: Bretonni\`ere et al. 2022) focuses on results concerning the morphological parameters, in this work we have presented the simulated data set, and focused on the results concerning photometric estimates only. 

We used the code \texttt{Egg} \citep{Schreiber2017} to produce mock galaxy catalogues, which were then exploited to create synthetic images in three different realisations: two using \texttt{GalSim} \citep{Rowe2015}, to generate single and double component \sersics analytical profiles (SS and DS respectively, the latter with $n_{\rm bulge}$=4 and $n_{\rm disc}$=1 and varying bulge-to-total luminosity ratios); and one with the method described in \citet{Bretonniere2022} and \citet{Lanusse2021} to obtain realistic morphologies by means of a CNN technique (RM). We did not simulate irregular galaxies, which constitute a substantial fraction of the high-redshift population, and we did not include AGNs. Furthermore, the distribution of the bulge-to-total ratios of the simulated galaxies is more skewed towards disc-dominated objects than in the real Universe (see Fig. \ref{distr}); this implies that any result showing a dependence on this feature must be considered as statistically optimistic.

We created five fields of view in the \Euclid $I_{\scriptscriptstyle\rm E}$ band with an area of 0.482 deg$^2$ each, and for one of them we also simulated eight additional bands in the double \sersics realisation (\Euclid NIR $Y_{\scriptscriptstyle\rm E}$, $J_{\scriptscriptstyle\rm E}$, and $H_{\scriptscriptstyle\rm E}$, plus five Rubin/LSST bands $u$, $g$, $r$, $i$ and $z$). The images were produced with zero background flux (i.e., already `backgroud subtracted'). Gaussian noise realisations mimicking the expected depths of the \Euclid Wide Survey were added to the noiseless images. The analytical profiles of the galaxies were finally randomized, following a consistent Poissonian distribution. RMS maps including photon noise were also simulated, and provided to the participants along with the PSFs of each field, and lists of objects having nominal ${\rm S/N}>5$ in $I_{\scriptscriptstyle\rm E}$, for which the actual centroid positions in pixel coordinates were given. The requested output consisted of the photometric (and morphological) estimates for all the objects in these lists. 

To allow for a quantitative analysis of the results and for a comparison of the accuracy of the codes, we defined a metric as described in Sect. \ref{fom}, computing three quantities related to, respectively, the median bias in the flux measurements ($\mathcal{B}$), the mean dispersion of the measured flux distribution ($\mathcal{D}$), and the outlier fraction ($\mathcal{O}$); these quantities were evaluated on a subset of the data set including only the sources for which all participants provided a fit, and we also computed a further value quantifying the fraction of sources of the input list for which a software package succeeded to provide a successful fit ($\mathcal{C}$). Finally, we combined them in a single figure of merit $\mathcal{S}$; the lower the value of $\mathcal{S}$, the better the global fit, with $\mathcal{S}\leq1$ being considered optimal performance. The global results are summarised in Table \ref{codestab} and Figs. \ref{summary} to \ref{summ2}.
We also analysed how accurately colours were retrieved, and the reliability of the uncertainty estimates. The computational times and some in-depth analysis for each individual software package are discussed in the Appendix.

In general, all the participants provided acceptable to good results in at least some of the tests, with a few differences from case to case. A thorough comparison could only be performed with the results of \texttt{ProFit}, \texttt{Galapagos-2}, and \texttt{SourceXtractor++} (the first two provided all the requested outputs, and the third only lacked one part of a set, namely the free $n_{\rm bulge}$ run in the double \sersics multi-band realisation). \texttt{DeepLeGATo} provided fits only for the $I_{\scriptscriptstyle\rm E}$ analytical realisations, yielding good results with the caveat of a S/N-dependent calibration of the models, while \texttt{Morfometryka} provided data only for the single \sersics set; therefore, their contribution to the challenge can only be considered as indicative of their potential.

Considering the $\mathcal{S}$ values, among the three codes that provided the complete output the best performance was obtained by \texttt{SourceXtractor++}, in all realisations. It is worth stressing again that the pipeline it adopted for the EMC was the most tuned on the data set, with the inclusion of a substantial pre-processing of the data, and the priors being modeled using the samples of true values that were provided along with the simulated images.
In the analytical realisations (SS and DS), \texttt{SourceXtractor++} and \texttt{Galapagos-2} reached typical values in the bias of the measured fluxes with respect to the true values below 1\% at S/N above 10 in all bands, and above 5 in $I_{\scriptscriptstyle\rm E}$ (in at least one of the possible configurations of the codes for which results were provided). For these two codes, the dispersion is typically slightly larger than, but comparable to, the one expected from a theoretical distribution obtained by perturbing the input true fluxes with simulated observational noise consistent with the nominal depths of the images; a small fraction of outliers (objects having a relative error in the fitted flux larger than 5 times the expected theoretical distribution) is always present.
\texttt{ProFit} also yielded good results in SS, but was less accurate on the DS realisation (\texttt{Galapagos-2} also being sub-optimal at the faint end in $I_{\scriptscriptstyle\rm E}$ for this realisation). 
The three software packages performed similarly on the RM data set,
and in all cases results were poorer with respect to the analytical realisations; this seems to imply that the overall quality of the fits might be sub-optimal when dealing with real data. However, further investigation is needed to assess to what extent the quality of the simulation might have impacted the results.

All codes tend to underestimate the uncertainties at the bright end, but statistically provide a reasonable estimate for intermediate and faint sources.
Finally, a relevant outcome is the importance of the simultaneous fitting of the multi-band data set: using the information provided by the deep and high resolution $I_{\scriptscriptstyle\rm E}$ band helps to obtain more accurate measurements on the shallower NIR bands, and noticeably also on the ground-based LSST bands, both in terms of absolute fluxes and colours. Once again this highlights how the synergy between the two surveys will be of paramount important for the success of both.

We did not attempt to investigate the performance of the model-fitting algorithms in the context of the \Euclid Deep Survey \citep[see][]{Laureijs2011}. It is reasonable to expect that the results should be similar to those obtained for the Wide Survey at equivalent S/N, although the increased level of contamination and blending might worsen the performance. Dedicated simulations would be required to quantitatively confirm this hypothesis, but this is beyond the scope of the present study.

The results of the EMC set the baseline to decide which algorithm is to be implemented in the \Euclid pipeline. One possible way to proceed might be to use the computational power of \texttt{DeepLeGATo} to provide fast and reliable priors to \texttt{SourceXtractor++}; we will investigate this option in the next future. However, given the many different approaches and techniques adopted by the participants (different use of priors, different pre-processing strategies, different approaches for multi-band fitting, etc.) and the other parameters not included in the metric that should be taken into account (such as the computing time, required resources, compatibility with the current pipeline, accuracy of uncertainty budget estimates etc.), it is important to stress that the results presented in this work and in EMC2022b must be interpreted with caution. We followed one possible approach to analyse in a compact way a very complex and multi-layered data set; and we emphasise that some of the software packages might be better suited than the ones obtaining the best score here, for other specific science cases. We invite the interested reader to check the full set of results using the online tool. 

Future work will include testing at least some of the software packages in more realistic environments using \Euclid official simulations, and finally the implementation in the \Euclid pipeline of the chosen algorithms.

\begin{acknowledgements} 
\AckEC
H. Hildebrandt is supported by a Heisenberg grant of the Deutsche Forschungsgemeinschaft (Hi 1495/5-1) as well as an ERC Consolidator Grant (No. 770935).
\end{acknowledgements}

\bibliographystyle{aa}
\bibliography{biblio} 


\appendix

\section{Technical information} \label{AppTec}

Here we provide detailed information about the technical realisation of the noise background and of the RMS maps for the data set.

\subsection{Noise maps and zero-points} \label{AppZP}

We created Gaussian background noise maps using \texttt{SkyMaker} \citep{Bertin2009}. The software requires the value of the expected background surface brightness (${\rm SB}_{\rm bkg}$, in magnitude arcsec$^{-2}$) and the observational ZP at 1 second of exposure as inputs, to produce a map at the desired depth. The conceptual steps to compute the ZP, following \citet{Martinet2019}, are the following.
Given a desired limiting magnitude $m_{\rm lim}$ within a given area $A$ (e.g. in square arcseconds), the corresponding limiting flux for the total exposure time is

\begin{equation}
        S=f_{{\rm lim},A}=\frac{t_{\rm exp}}{g}10^{({\rm ZP}-m_{\rm lim})/2.5}\;,
\end{equation}

\noindent where $g$ is the gain of the detector. The uncertainty per pixel from the background for the same source is 

\begin{equation}
        \sigma_{\rm bkg,pixel}=\sqrt{f_{\rm bkg,pixel}+\sigma_{\rm RON}^2}\simeq\sqrt{l^2 \frac{t_{\rm exp}}{g} 10^{({\rm ZP}-{\rm SB}_{\rm bkg})/2.5}}\;,
\end{equation}    

\noindent where $l$ is the pixel-scale, and for the second step we assume that the contribution of the read-out noise $\sigma_{\rm RON}$ is negligible. 
Assuming that $\sigma_{\rm bkg,pixel}$ is constant, the total noise within the aperture is

\begin{equation}
N=\sqrt{\sum_{{\rm pixels},A} \sigma_{\rm bkg,pixel}^2} = \sqrt{A} \sigma_{\rm bkg,pixel}\;;  
\end{equation}

\noindent so if $r$ is the radius of the aperture relative to $A$, $r=\sqrt{A/\pi}$, the background S/N reads

\begin{equation}
        {\rm S/N}=\frac{f_{{\rm lim},A}}{\sqrt{A}\sigma_{\rm bkg,pixel}}=\frac{t_{\rm exp}10^{({\rm ZP}-m_{\rm lim})/2.5}}{g\sqrt{\frac{\pi r^2}{l^2} }\sqrt{l^2 \frac{t_{\rm exp}}{g} 10^{({\rm ZP}-{\rm SB}_{\rm bkg})/2.5}}}\;.
\end{equation}

Finally, the formula can be inverted to make the expression for ZP explicit:

\begin{equation}
        ZP=2.5   \logten \left[ \frac{(S/N)^2 \pi r g}{t_{\rm exp}} \right] - 2m_{\rm lim} - SB_{\rm bkg}\;,
\end{equation}

\noindent where it can be noticed that the pixel-scale terms $l^2$ have canceled out. 

We computed the ZP of each image putting $\mbox{S/N}=10$ and considering the corresponding $m_{\rm lim}$ computed in 2\arcsecond\ apertures (so $r=1$\arcsecond\ in the above formulas), SB$_{\rm bkg}$, and exposure times as given in Table \ref{depths}. We point out again that this method is used to compute the ZP for 1 second of exposure (to obtain the ZP at the given exposure time, a term $2.5 \logten t_{\rm exp}$ should be added).

\subsection{RMS maps and photon noise} \label{AppRMS}

To obtain the RMS map for each scientific image, first of all we produced a map with a constant value equal to the standard deviation of the original Gaussian background noise image, which only provides information on the noise due to the unresolved sky background and undetected faint sources.

The flux of the photons arriving from galaxies is a Poissonian distribution and contributes to the uncertainty on the actual flux estimate. This contribution (photon noise) was added to the background noise map with the following procedure. Given an arbitrary exposure time $t$, the pixel values $S_{t}=C_{t}+{\rm sky}_{t}$ are due to the sum of the counts coming from sources ($C_{t}$) and from the background (${\rm sky}_{t}$). The noise per pixel $N_{t}$ is found by summing in quadrature the sky contribution ${\rm RMS}_{{\rm sky},t}$ and the photon noise contribution ${\rm RMS}_{{\rm source},t} = \sqrt{C_t})$:

\begin{equation}
\label{eq1}
N_{t}=\sqrt{{\rm RMS}^2_{{\rm sky},t}+\left( \sqrt{C_t} \right) ^2}=\sqrt{{\rm RMS}^2_{{\rm sky},t}+C_t}\;.
\end{equation}

Considering a science image normalised to $t=1$ second, in which pixels have values $S_{1}=S_{t}/t$, with $N_{1}$ being the corresponding RMS map pixel values, the S/N of each pixel must be conserved, i.e.
\begin{equation}
\label{eq2}
\frac{S_{t}}{N_{t}}=\frac{S_{1}}{N_{1}}\;.
\end{equation}

This relation leads to the procedure to build the noise map $N_{1}$ of the normalised image as

\begin{equation}
\label{eq3}
N_{1}=\frac{N_{t} S_{1}}{S_{t}}=\frac{N_{t}}{t}\;.
\end{equation}

By combining Eqs. \ref{eq1} and \ref{eq3}, and considering that $C_t=C_1 t$ we finally obtain
\begin{equation}
\label{eq4}
N_{1}=\sqrt{\frac{{\rm RMS}^2_{{\rm sky},t}+C_t}{t^2}}=\sqrt{{\rm RMS}^2_{{\rm sky},1}+\frac{C_1}{t}}\;.
\end{equation}

\section{Computational times and separate analysis of the performance of each software package} \label{AppCodes}

In this section we provide information about the computational times and memory workload required by each software package to complete the runs, and analyse in more detail some particular cases in which the results were especially interesting, adding some final remarks where appropriate.

\subsection{DeepLeGATo} \label{Dl}

\begin{figure}[h!]
\centering
\includegraphics[width=0.49\textwidth]{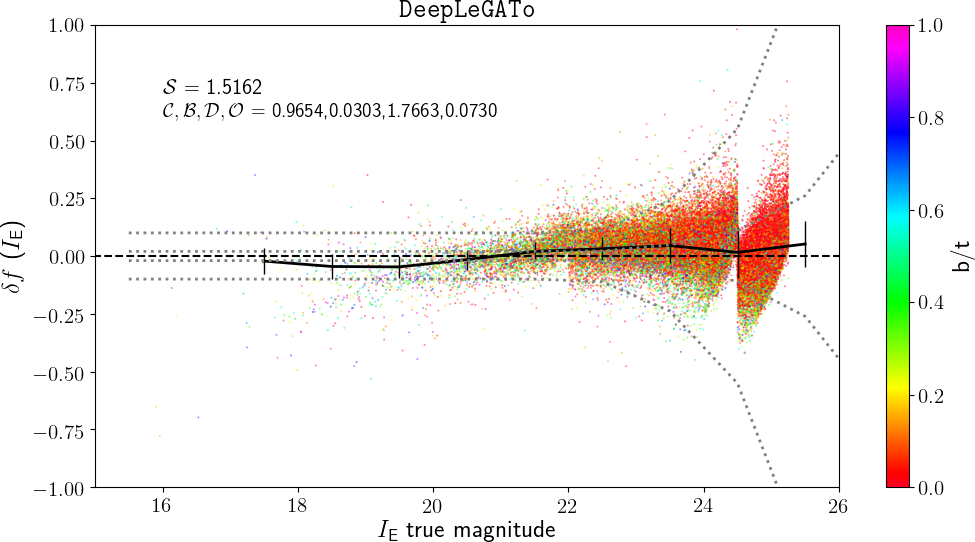}

\vspace{0.3cm}

\includegraphics[width=0.49\textwidth]{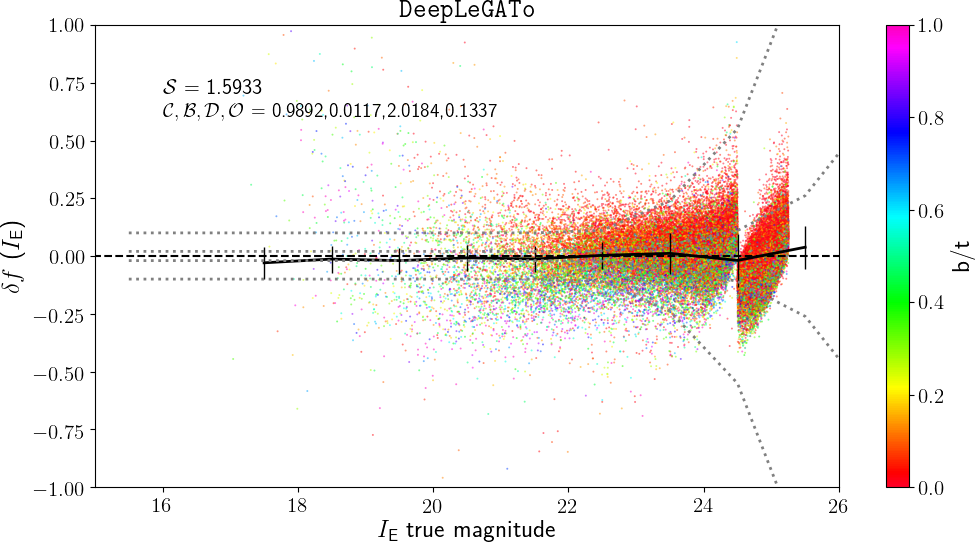}
\caption{Examples of diagnostic plots for \texttt{DeepLEGaTo}. Top, $I_{\scriptscriptstyle\rm E}$ SS F4 run; bottom, $I_{\scriptscriptstyle\rm E}$ DS F4 run. See Fig.~\ref{trumpet} for a description of this and the subsequent similar plots. Note that the colour coding is given by the value of the bulge fraction \bts in the original \texttt{Egg} double-Sersic catalogue; it can be considered as a proxy for the value of the \sersics index, $n=3\; \btm +1$.} 
\label{dleg}
\end{figure}

\texttt{DeepLeGATo} exploits neural network models, that for the EMC were trained on a single Tesla V100 GPU.
The output was provided in three separate catalogues of different minimum S/N, according to the input source lists that were distributed to the participants; a different model was used for each S/N section, and the training time was of the order of 2 hours per model. At inference, it then took a few minutes to ``fit'' all galaxies in a simulated field; therefore, concerning computational times \texttt{DeepLeGATo} has by far the best performance, due to the very different fitting technique.

We stress again that the subdivision of the catalogue into subsamples fitted with different models caused evident features in the global distribution of $\delta f$, though the average trends and the final $\mathcal{S}$ values were generally quite good. In particular, the bright sources in each section were systematically underestimated while the faint ones were overestimated. There were also many bright outliers especially in the DS runs, with bulges typically being underestimated and discs being overestimated; this is likely to be a consequence of the CNN fitting technique (fixed dimensions of the stamps, and less training at the bright end). Figure~\ref{dleg} provides an example of these features.

\subsection{Galapagos-2}  \label{G2}

\texttt{Galapagos-2} employs the down-hill gradient minimisation method of \texttt{Galfit}/\texttt{GalfitM}, so it is typically fast, and can be reasonably parallelized to use 16 cores (due to an \texttt{IDL} limitation; ways around this limit exist, as it can also run several sessions in parallel). For the EMC, the code was run on a machine with 48 cores (2.3 GHz each) at 96 threads and ample (512 GB) memory, using 12 threads (at 1.15 GHz) at a time (and 4 sessions in parallel) for multi-band fits. 
The run times of the \texttt{Galfit} fits themselves were measured and stored, and account for the majority of the time used, with the wrapper not adding significant computational time. The typical \texttt{GalfitM} fitting times were approximately as follows:
\begin{itemize}
\item DS F1 single-band single \sersics fit, 160 hours (about 5.7 seconds per galaxy);
\item DS F1 single-band double \sersics fit, 150 hours (about 5.4 seconds per galaxy);
\item DS F0 multi-band single \sersics fit, 8000 hours (about 288 seconds per galaxy);
\item DS F0 multi-band double \sersics fit, 10\,000 hours (about 360 seconds per galaxy).
\end{itemize}

Therefore the multi-band fitting, while providing the results for nine bands in parallel, took significantly more than the time elapsed to fit individual bands. However, as shown, it clearly provides better results for the shallower bands. 

\begin{figure}[h!]
\centering
\includegraphics[width=0.49\textwidth]{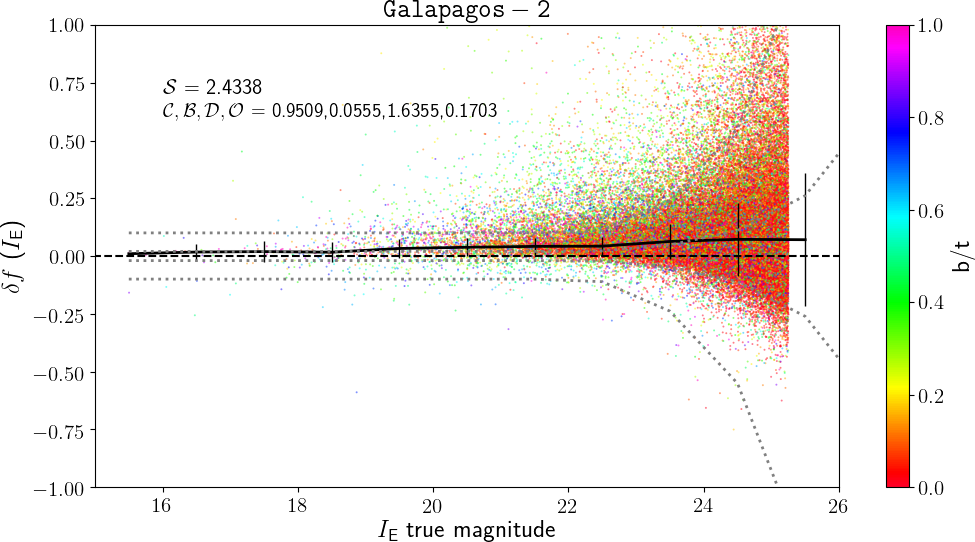}

\vspace{0.3cm}

\includegraphics[width=0.49\textwidth]{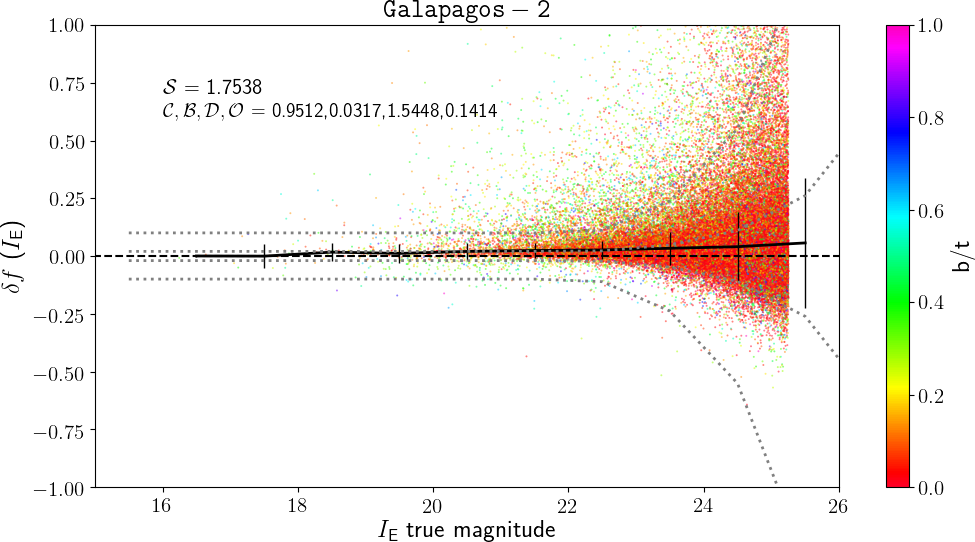}
\caption{\texttt{Galapagos-2} results on F0 (top) and F1 (bottom) DS $I_{\scriptscriptstyle\rm E}$ band (one component fits). The fit on F1, performed on the $I_{\scriptscriptstyle\rm E}$ image only rather than in multi-band mode, is more accurate. See text for details.}
\label{boris_3}
\end{figure}

The overall results were good, with no evident pathological behaviour. The performance on $I_{\scriptscriptstyle\rm E}$ in the multi-band fit DS F0, i.e. obtained simultaneously with the other bands, is significantly worse than those on other fields, where $I_{\scriptscriptstyle\rm E}$ is measured alone (see Fig.~\ref{boris_3}); this is not surprising, because when fitting one band individually all the parameters get optimised to minimise that band, while fitting more bands in parallel the structural parameters are minimized averaging over the wavelengths (and this is particularly true if the band is the deepest and the one with the higher resolution, as is the case for $I_{\scriptscriptstyle\rm E}$). However, it should be pointed out that multi-band fits ``fail'' less often than the combination of several single-band fits, which have independent constraints \citep[see][]{Haussler2013}.

\begin{figure}[h!]
\centering
\includegraphics[width=0.49\textwidth]{figs/results/Boris/EUCLID1_vis_ds_ssfit_and_bdfit_nb4_all_match0.30_1comp.png}

\vspace{0.3cm}

\includegraphics[width=0.49\textwidth]{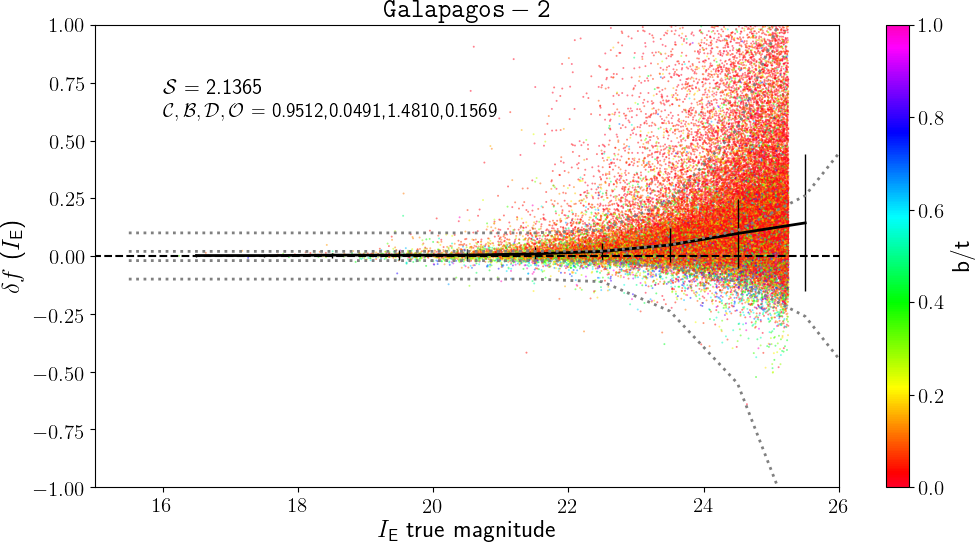}
\caption{\texttt{Galapagos-2} results on F1 DS $I_{\scriptscriptstyle\rm E}$ band,  one (top) and two (bottom; $n_{\rm bulge}$ fixed) component fits. The one-component (i.e. single \sersic) fit is more accurate in $I_{\scriptscriptstyle\rm E}$-only runs. See text for details.}
\label{boris_1}
\end{figure}

\begin{figure}[h!]
\centering
\includegraphics[width=0.49\textwidth]{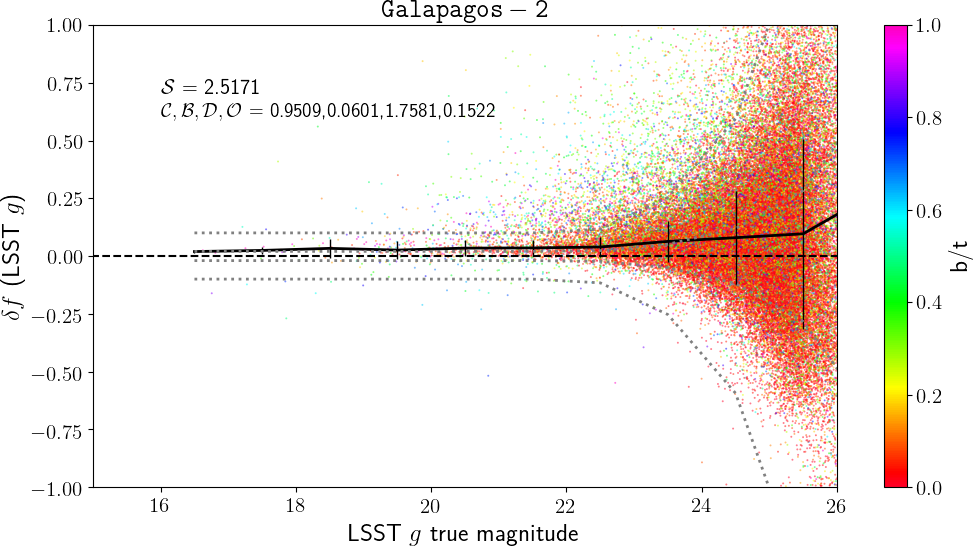}

\vspace{0.3cm}

\includegraphics[width=0.49\textwidth]{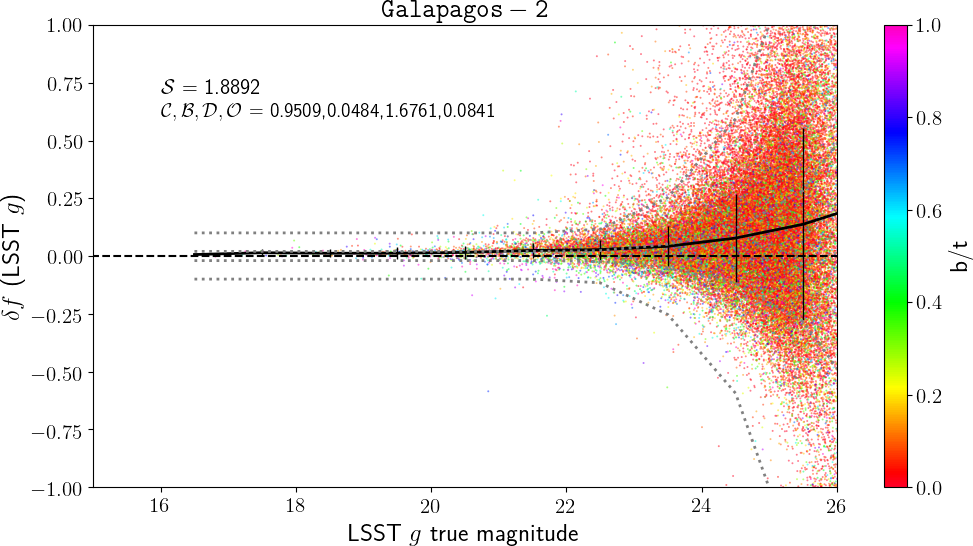}
\caption{\texttt{Galapagos-2} results on F0 DS LSST $g$ band, one (top) and two (bottom; $n_{\rm bulge}$ fixed) component fits. The two-component (double \sersic) fit is more accurate for non-$I_{\scriptscriptstyle\rm E}$ bands in the multi-band simultaneous fit. See text for details.}
\label{boris_2}
\end{figure}

The total magnitude estimated with a single \sersics fit (which is always provided by the code) is typically better than the one from the double \sersics in $I_{\scriptscriptstyle\rm E}$, even on the DS simulation (Fig.~\ref{boris_1}). On the contrary, the double \sersics magnitude estimate is better for the other bands (Fig.~\ref{boris_2}). While the latter result is expected, given that the code is fitting a double profile with two separate analytical functions (and, the shallower the band the more the multi-band fit should improve the results, so the NIR and LSST bands should benefit more than $I_{\scriptscriptstyle\rm E}$ from it), the opposite outcome found on $I_{\scriptscriptstyle\rm E}$ images is more surprising. We point out that since the other teams did not provide single \sersics flux estimates for DS runs, we could not check whether this particular result is a general one, or only concerns \texttt{Galapagos-2}.

\begin{figure}[h!]
\centering
\includegraphics[width=0.48\textwidth]{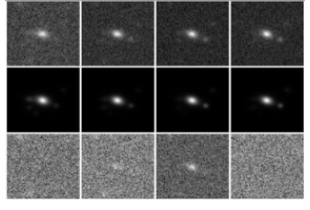}
\vspace{1.0cm}
\includegraphics[width=0.48\textwidth]{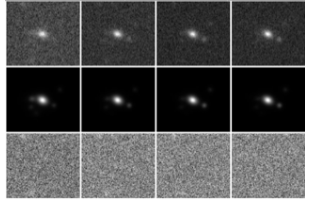}
\caption{Test on the \texttt{Galapagos-2} LSST $i$ band issue. In both sets of panels, for a chosen badly fitted source (ID 2158 in F0), top to bottom we show scientific image; \texttt{Galapagos-2} models; and residuals. Left to right are LSST $g$, $r$, $i$, and $z$ bands. Top set: reference run including $I_{\scriptscriptstyle\rm E}$ information in the fit. Bottom set: test run without including $I_{\scriptscriptstyle\rm E}$ information in the fit. The residual in the $i$ band, evident in the upper panel, vanishes in the bottom panel, when $I_{\scriptscriptstyle\rm E}$ is not included. See text for details.}
\label{boris_res}
\end{figure}

\begin{figure}[h!]
\centering
\includegraphics[width=0.49\textwidth]{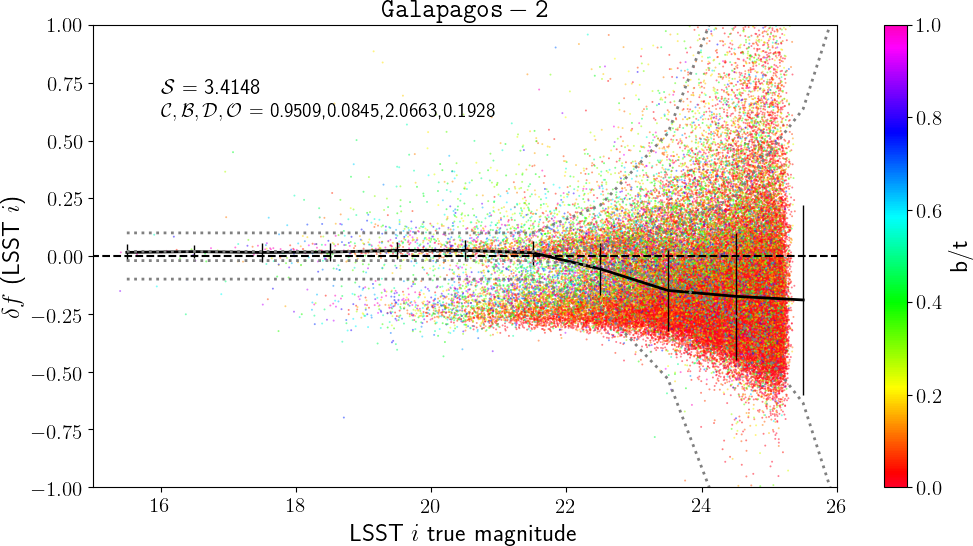}
\caption{\texttt{Galapagos-2} results on F0 LSST $i$ band. The evident double trend is due to the multi-band simultaneous fitting (see Sect. \ref{G2}).}
\label{boris_i}
\end{figure}

Interestingly, the fit of the LSST $i$ band on the DS F0 data set by \texttt{Galapagos-2} is substantially worse than the fits provided for the other bands. After an in-depth analysis, we concluded that this is actually due to the polynomial fitting method used in the software. In short, \texttt{Galapagos-2} works with effective central wavelengths, and when performing a multi-band fit on a data set including two bands with a very close central wavelength, the shallower of the two bands might get a sub-optimal fit with a systematic offset. This is exactly the case here, with the LSST $i$ band having a central wavelength very close to the $I_{\scriptscriptstyle\rm E}$ one (${\rm \lambda}_{I_{\scriptscriptstyle\rm E}}=710$ nm  and ${\rm \lambda}_{i}=754$ nm), and very different filter widths; see Sect. \ref{sims} and Fig.~\ref{filters}. Noticeably, this does not affect the whole catalogue, but only a fraction of the sources. Despite trying to find a correlation with some other input or output parameter of the data set, we were not able to identify the criteria resulting in a good or a bad fit for a given source. Nevertheless, we verified the reliability of our conclusion by picking one source having a bad fit, and looking at the residual images produced by \texttt{Galapagos-2} in the full multi-band run, and in a test run in which the $I_{\scriptscriptstyle\rm E}$ band information was not used to fit the $i$ band. The result of the test is shown in Fig.~\ref{boris_res}: while the other LSST bands are well-fitted in both cases, when the $I_{\scriptscriptstyle\rm E}$ band is used the $i$ band has an evident residual, which flattens out when excluding $I_{\scriptscriptstyle\rm E}$. Of course, this cannot be considered a viable, definitive `solution' to the issue, since $I_{\scriptscriptstyle\rm E}$ is the most important and deepest band, so it cannot be excluded from the fit to improve the fit in another band; on the contrary, one might want to exclude the $i$ band from the fit, although this would imply entirely discarding a band and its information. Therefore, for the sake of comparison with the other software packages we agreed to keep the delivered data set including this issue, although it non-negligibly worsens the statistics. Finally, we note that the issue is more evident when using the fluxes of the single \sersics fit, while it is somehow mitigated if the double \sersics fit is considered (\texttt{Galapagos-2} provides both for the DS realisation, which is the only one including LSST bands).

\subsection{Morfometryka} \label{Mm}

\begin{figure}[h!]
\centering
\includegraphics[width=0.49\textwidth]{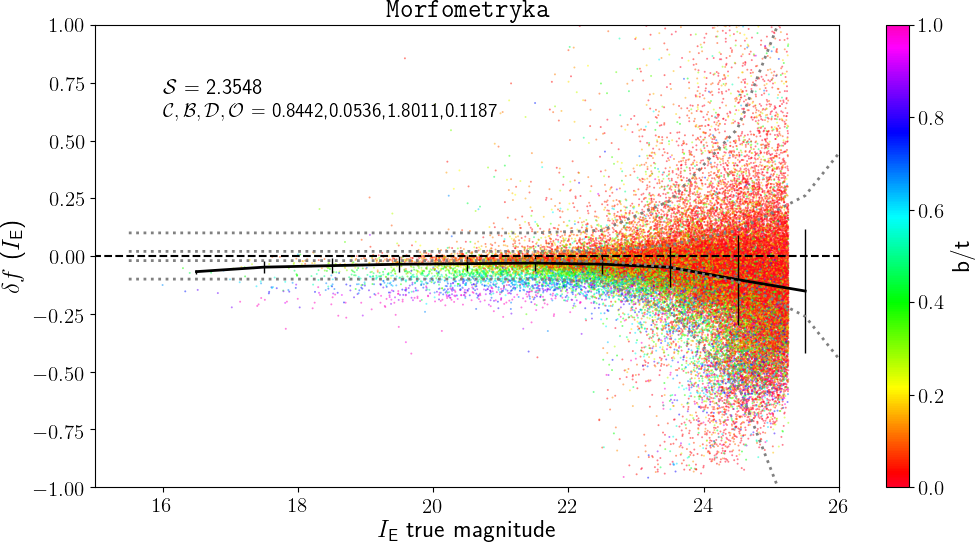}
\caption{Example of diagnostic plot for \texttt{Morfometryka}. This plot refers to the $I_{\scriptscriptstyle\rm E}$ SS F2 run. The colour coding clearly shows that the bimodal distribution of points correlates with the input bulge-to-total fraction, with bulge-dominated galaxies (bluer points) being fitted with less accuracy than disc-dominated ones (redder points).} 
\label{metryka}
\end{figure}

\texttt{Morfometryka} runs were performed on an Intel(R) Xeon(R) CPU E5-2640 v3 2.60 GHz shared 32-core workstation, running 16 jobs in parallel at any given time. The wall time was roughly 920 minutes per field, which is roughly 10 seconds per source. For each run, \texttt{Morfometryka} spends around 30\% of its running time for the \sersics profile fits, and the remaining time is dedicated to other measurements; however, runtimes strongly depend on the input image size.

There is an evident dependence on morphological type in the results. Bulge-dominated galaxies (having high \sersics index, which in the SS realisation is obtained from the bulge-to-total ratio given in the \texttt{Egg} catalogue, see Sect. \ref{sims}) are strongly biased ($\mathcal{B}\simeq 0.15$), and therefore are considered outliers in the computation of $\mathcal{S}$, thus significantly affecting the accuracy and resulting in a sub-optimal overall performance (see Fig.~\ref{metryka}). 

\subsection{ProFit} \label{Pr}

\texttt{ProFit} ran at a mean single-core time of 144 minutes per run, resulting in about 93\,000 hours of total CPU time; the runs were performed on the Magnus super computer (operated by Pawsey in Western Australia), which is a multi-node cluster where each node is 24 core (48 thread), comprised of two Intel Xeon E5-2690 v3 (Haswell) 12-core CPUs.

The typical single \sersics profile fit took a little under a minute per object, and a double \sersics fit around two minutes on average; this included the whole processing, but in practice the vast majority of the time was spent doing the \texttt{ProFit} optimisation, with the \texttt{ProFound} detection stage usually taking about one second. This was deemed enough time to find a good global solution per fit, but it was not long enough to thoroughly explore uncertainties, as shown in Sect. \ref{errors}. 

\begin{figure}[h!]
\centering
\includegraphics[width=0.49\textwidth]{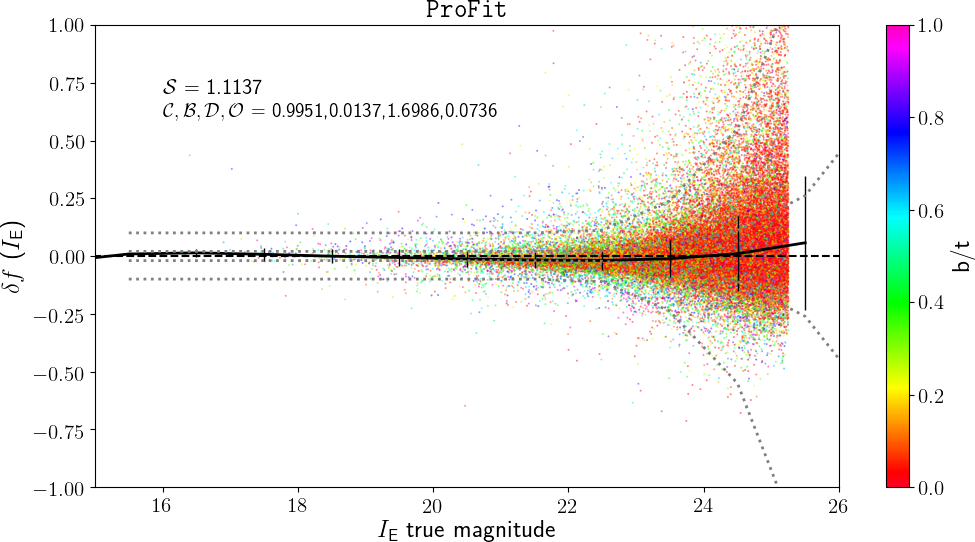}

\vspace{0.3cm}

\includegraphics[width=0.49\textwidth]{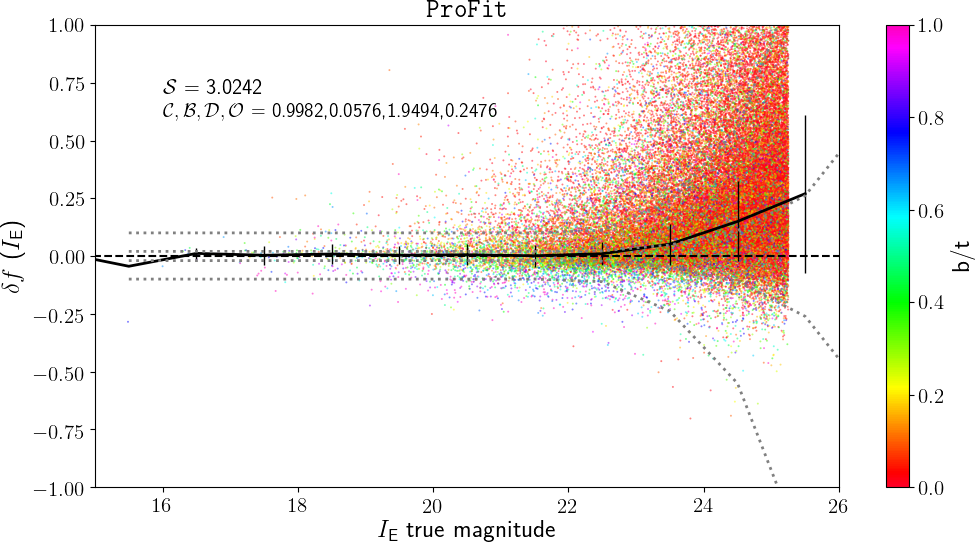}

\vspace{0.3cm}

\includegraphics[width=0.49\textwidth]{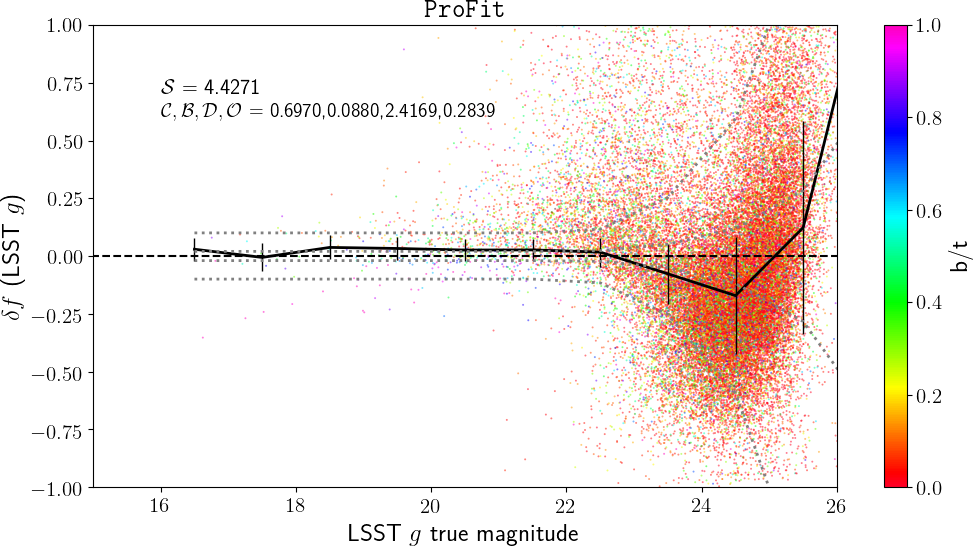}
\caption{Example of diagnostic plot for \texttt{ProFit}. Top to bottom: $I_{\scriptscriptstyle\rm E}$ SS F2; $I_{\scriptscriptstyle\rm E}$ DS F2 (fixed bulge); and LSST $g$ F0. The $I_{\scriptscriptstyle\rm E}$ SS fit is substantially better than the DS fit, and there is a `U-shaped' trend in the LSST bands.} 
\label{profit}
\end{figure}


Good results were provided for the SS realisation, but much less so on the DS and RM ones. The faint end is typically strongly biased (fluxes are overestimated); LSST bands also have a `U-shaped' trend (see Figs. \ref{profit}), for which we were not able to find a simple explanation.

\subsection{SourceXtractor++} \label{Se}

Single-band runs (SS, RM, DS on $I_{\scriptscriptstyle\rm E}$) were performed on a small cluster at LMU Munich, with 56 cores and 2 GB of RAM per core. Computational times were of about 50 minutes per field for single \sersics runs, and 2 hours per field for double \sersics runs.

Multi-band runs were performed on a local server at the Institut Astrophysique de Paris, with 24 AMD EPYC 7402P cores (ignoring hyperthreading) running at 2.8 GHz, using an improved version of the code with increased parallelisation efficiency. The computational time was 45 hours (it is worth recalling that the NIR and LSST bands were rebinned to their original pixel scales, so their sizes were smaller than the $I_{\scriptscriptstyle\rm E}$ one -- respectively 12\,500 and 8333 pixels per side).

\begin{figure}[h!]
\centering
\includegraphics[width=0.49\textwidth]{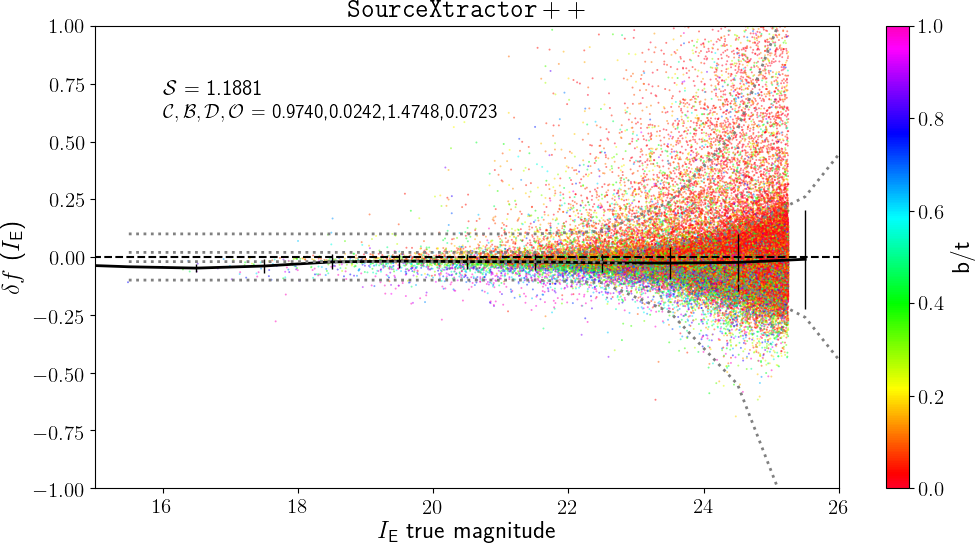}

\vspace{0.3cm}

\includegraphics[width=0.49\textwidth]{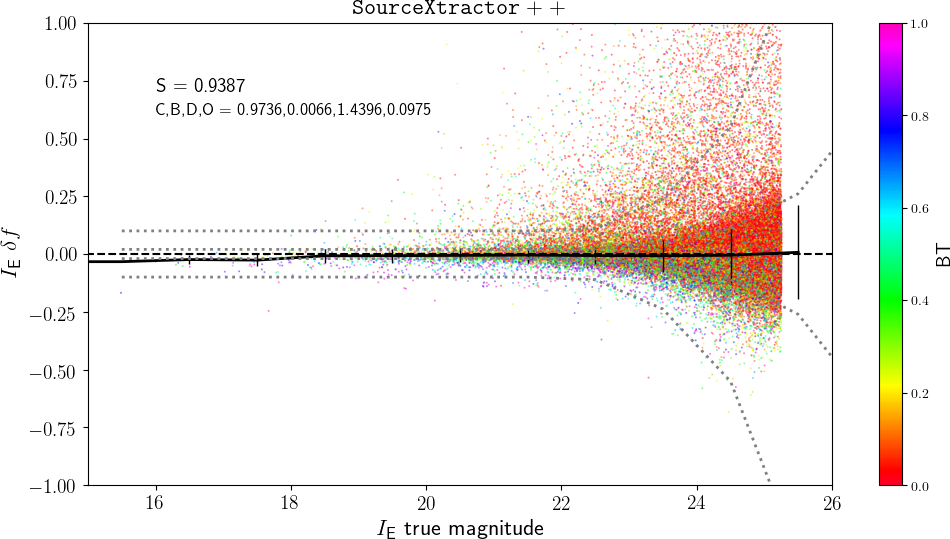}
\caption{Example of diagnostic plot for \texttt{SourceXtractor++}. These plots refers to the $I_{\scriptscriptstyle\rm E}$ DS F2 run; the upper panel shows the fit with $n_{\rm bulge}=4$, while the bottom panel shows the one with $n_{\rm bulge}$ left free to vary.} 
\label{se++}
\end{figure}

The results were generally good, with no evident trends or pathological behaviour. Among the provided output, \texttt{SourceXtractor++} typically obtained the best values for the global diagnostics. Interestingly, in the DS realisation the runs with free $n_{\rm bulge}$ typically yielded slightly better results than those with fixed $n_{\rm bulge}=4$, in terms of average bias (see Fig.~\ref{se++}).

As mentioned, the processing pipeline adopted for the EMC included a substantial pre-processing of the data set, and the priors used for the measurements were calibrated on the provided samples of true values, albeit only in a statistical sense (see below).

\subsubsection{Re-runs on raw data}

\begin{figure}[h!]
\centering
\includegraphics[width=0.49\textwidth]{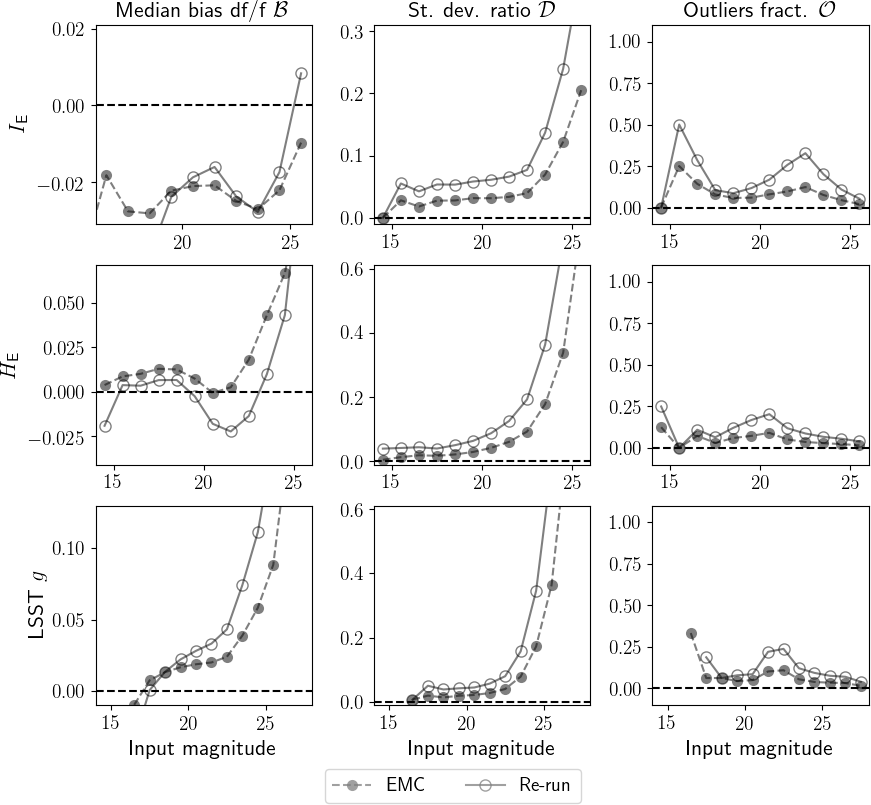}
\caption{\texttt{SourceXtractor++} re-runs on non-pre-processed (top to bottom) $I_{\scriptscriptstyle\rm E}$, $H_{\scriptscriptstyle\rm E}$, and LSST $g$ F0 DS images (empty symbols, solid lines), compared to output provided for the EMC, obtained with pre-processing of the images (full symbols, dashed lines). Note the different limits on the axes.}
\label{SErerun}
\end{figure}

The NIR and LSST images were rebinned back to their original pixel scales (while the ones provided in the data set shared with the participants had been rebinned to the $I_{\scriptscriptstyle\rm E}$ pixel scale of 0.1\arcsecond). Additionally, the PSFs used in the fitting process were not those provided in the data set; instead, they were extracted from the images using \texttt{PSFEx}. This did not require any additional input, and therefore it was not outside the guidelines of the challenge. Still, we felt it was necessary to check the performance of the software on the non-processed images and using the official PSFs, for a fairer comparison with the other participants. To this end, a few additional runs were performed by the \texttt{SourceXtractor++} team (using version 0.16 of the code instead of 0.12), and we compared the results with the ones officially provided for the EMC. The values of the four diagnostics $\mathcal{C}$, $\mathcal{B}$, $\mathcal{D}$, and $\mathcal{O}$, and of the global $\mathcal{S}$, for both the EMC runs and the re-run, are given in Table \ref{tabrer}. Some trends for the DSb4 multi-band realisation are also shown in Fig.~\ref{SErerun} as examples (namely the $I_{\scriptscriptstyle\rm E}$, $H_{\scriptscriptstyle\rm E}$, and $g$ trends, since the aim was to check the impact of the re-extraction of PSFs and of the re-rebinning of images with native pixel scale different from the $I_{\scriptscriptstyle\rm E}$ one). The results do not seem to significantly improve because of the pre-processing; most of the values and trends look similar, and in some cases even slightly better in the re-run. The evident exception is given by the NIR bands, in particular concerning the median bias; including the weighting factors, the values of $\mathcal{S}$ are significantly worse in the re-run, albeit still reasonably good (i.e. close to 1).

These re-runs required 40 minutes per field for the single \sersics runs (SS and RM), and 1 hour and 20 minutes per field for the double \sersics $I_{\scriptscriptstyle\rm E}$ runs (using 28 cores). Finally, the multi-band re-run was performed on a different set of nodes, with 32 AMD EPYC 7302 16-Cores at 3.2 GHz, and required 67 hours.



\begin{table}
\label{tabrer}
\centering       
\begin{tabular}{|l| c| c| c| c| c|} 
\hline

\hline
Run &  $\mathcal{S}$ &  $\mathcal{C}$ &  $\mathcal{B}$ &  $\mathcal{D}$ &  $\mathcal{O}$\\
\hline

SS F4 $I_{\scriptscriptstyle\rm E}$ &0.81&0.94&0.01&1.44&0.07\\
&0.87&0.94&0.01&1.40&0.08  \\ \hline

DSb4 F4 $I_{\scriptscriptstyle\rm E}$ &1.20&0.96&0.02&1.48&0.07\\
&0.92&0.96&0.00&1.41&0.11\\\hline

DSbf F0 $I_{\scriptscriptstyle\rm E}$ &0.95&0.96&0.01&1.46&0.10\\
&0.94&0.96&0.00&1.42&0.11\\\hline

RM F4 $I_{\scriptscriptstyle\rm E}$ &2.13&0.83&0.03&2.00&0.14\\
&2.23&0.83&0.03&2.00&0.15\\\hline

DSb4 F0 $Y_{\scriptscriptstyle\rm E}$ &0.51&0.97&0.01&1.07&0.04\\
&1.20&0.96&0.04&1.23&0.05\\\hline

DSb4 F0 $J_{\scriptscriptstyle\rm E}$ &0.58&0.97&0.01&1.12&0.05\\
&1.18&0.96&0.03&1.26&0.06\\\hline

DSb4 F0 $H_{\scriptscriptstyle\rm E}$ &0.67&0.97&0.01&1.15&0.05\\
&1.02&0.96&0.02&1.26&0.06\\\hline

DSb4 F0 $u$ &1.01&0.97&0.03&1.43&0.03\\
&0.99&0.96&0.03&1.42&0.03\\\hline

DSb4 F0 $g$ &1.46&0.97&0.04&1.50&0.06\\
&1.45&0.96&0.04&1.53&0.07\\\hline

DSb4 F0 $r$ &1.40&0.97&0.04&1.47&0.05\\
&1.39&0.96&0.03&1.49&0.07\\\hline

DSb4 F0 $i$  &1.34&0.97&0.04&1.48&0.06\\
&1.35&0.96&0.03&1.52&0.73\\\hline

DSb4 F0 $z$  &1.32&0.97&0.03&1.51&0.06\\
&1.32&0.96&0.03&1.55&0.08\\\hline













\end{tabular}
\caption{Example values of the diagnostics for the \texttt{SourceXtractor++} runs in the output catalogues, provided for the EMC (top value in each box) and in corresponding runs with non-pre-processed images and PSFs (i.e., using those provided for the Challenge; bottom value in each box).} 
\end{table}

\subsubsection{SourceXtractor++ priors} \label{AppSEpriors}

As mentioned, \texttt{SourceXtractor++} priors were obtained by determining an appropriate transfer function to map each prior to a Gaussian mimicking the distributions of the provided samples of the input true catalogues. Each parameter was calibrated independently, without including covariances; only the statistical distributions were used (i.e. there was no object-by-object comparison in the process). We have shown how this choice likely had a significant impact on the accuracy of the results.

Even though \texttt{SourceXtractor++} models source ellipticity internally, with the standard axis ratio $q$ and position angle $\varphi$ variables, it was found useful to express the priors needed for these two degrees of freedom in terms of two $e_1$, $e_2$ Cartesian ellipticity parameters (that usually enter weak lensing studies). More specifically, in complex notation, $e= \frac{1-q}{1+q} \exp (2 {\rm i} \varphi) \equiv e_1 + {\rm i} e_2 $. Hence, a Normal prior on both $e_1$ and $e_2$ was set, centred on zero and with $\sigma=0.3$. This was in broad agreement with the population distribution of axis ratios in the SS F4 simulation, for which ground truth was provided. A similar prior was built for the \sersics index, as well as for the effective radius. 

In multi-component models, priors were set under the assumption that the bulge and disc of each source share a common position angle, but axis ratios are not correlated (this is not actually the case, since axis ratios do not vary across the spectrum in the \texttt{Egg} catalogues). Therefore the default $q$ and $\varphi$ variables were used instead of the ($e_1$, $e_2$) ellipticities. A single fit across all bands was run for bulge and disc effective radii, ellipticity, and position angle; the amplitudes of the bulge and disc components were left free to vary. Here again, the fitting procedure prefers the definition of a set of total magnitudes and a set of \bts (with their own associated priors), instead of using the flux of each component in each band as a set; in practice, \bts is expressed via the transform $X_{\btm} = \logten (\btm +0.01 ) /( 1.01- \btm)$. The priors on $X_{\btm}$ were set to be band dependent, but with little variation. 

\end{document}